\documentclass[aps,prd,10pt,superscriptaddress,twocolumn,nofootinbib,dvipsnames,longbibliography]{revtex4-1}

\usepackage[pdftex]{color}
\usepackage[sort&compress]{natbib}
\usepackage[colorlinks=true,linkcolor=purple,filecolor=orange,urlcolor=blue,citecolor=purple,pdftex,plainpages=false]{hyperref}
\usepackage{url}
\pdfoutput=1
\usepackage{ifpdf}
\usepackage[utf8]{inputenc}
\usepackage{color,hhline,psfrag,rotating}
\usepackage{dcolumn}
\usepackage{verbatim}
\usepackage{lipsum}
\usepackage{bm}
\usepackage{slashed}
\usepackage{braket}
\usepackage{rotating}

\usepackage{graphicx}
\usepackage[dvipsnames]{xcolor}
\usepackage{amsfonts,amssymb,amsmath}
\usepackage{booktabs}  
\newcommand\brachat[1]{\mathord{\mathop{#1}\limits^{\scriptscriptstyle(\wedge)}}}  

\mathchardef\mhyphen="2D

\begin{document}

\global\let\newpage\relax

\title{Improved $V_{cs}$ determination using precise lattice QCD form factors for $D \rightarrow K \ell \nu$}
\author{Bipasha~Chakraborty}
\email[]{bc335@cam.ac.uk}
\affiliation{DAMTP, Centre for Mathematical Sciences, University of Cambridge, Wilberforce Road, Cambridge, CB3 0WA}
\author{W.~G.~Parrott}
\email[]{w.parrott.1@research.gla.ac.uk}
\affiliation{SUPA, School of Physics and Astronomy, University of Glasgow, Glasgow, G12 8QQ, UK}
\author{C.~Bouchard}
\affiliation{SUPA, School of Physics and Astronomy, University of Glasgow, Glasgow, G12 8QQ, UK}
\author{C.~T.~H.~Davies}
\email[]{christine.davies@glasgow.ac.uk}
\affiliation{SUPA, School of Physics and Astronomy, University of Glasgow, Glasgow, G12 8QQ, UK}
\author{J.~Koponen}
\affiliation{Helmholtz Institute Mainz, Johannes-Gutenberg-Universit\"{a}t Mainz, 55099 Mainz, Germany}
\author{G.~P.~Lepage}
\affiliation{Laboratory of Elementary-Particle Physics, Cornell University, Ithaca, New York 14853, USA}
\collaboration{HPQCD collaboration}
\homepage{http://www.physics.gla.ac.uk/HPQCD}
\noaffiliation

\date{\today}

\begin{abstract}
We provide a 0.8\%-accurate determination of $V_{cs}$ from combining experimental results 
for the differential rate of $D \rightarrow K$ semileptonic decays with precise 
form factors that we determine from lattice QCD. This is the first time that $V_{cs}$ has been 
determined with an accuracy that allows its difference from 1 to be seen.   
Our lattice QCD calculation uses the Highly Improved Staggered Quark (HISQ) 
action for all valence quarks 
on gluon field configurations 
generated by the MILC collaboration that include the effect of $u$, $d$, $s$ and $c$ 
HISQ quarks in the sea.  
We use eight gluon field ensembles with five values of the lattice spacing ranging 
from 0.15 fm to 0.045 fm and 
include results with physical $u/d$ quarks for the first time. 
Our calculated form factors cover the full $q^2$ range of the physical decay 
process and enable a Standard Model test of the shape of the differential decay rate as 
well as the determination of $V_{cs}$ from a correlated weighted average 
over $q^2$ bins. 
We obtain $|V_{cs}|= 0.9663(53)_{\text{latt}}(39)_{\text{exp}}(19)_{\eta_{EW}}(40)_{\text{EM}}$, 
where the uncertainties come from lattice QCD, experiment, short-distance electroweak and
electromagnetic corrections, respectively. This last uncertainty, 
neglected for $D \rightarrow K \ell \nu$ hitherto, now needs attention if the uncertainty 
on $V_{cs}$ is to be reduced further.    
We also determine $V_{cs}$ values in good agreement using the measured total branching 
fraction and the rates extrapolated to $q^2=0$. Our form factors enable tests of lepton flavour 
universality violation. We find the ratio of branching fractions for $D^0 \rightarrow K^-$ 
with $\mu$ and $e$ in the final 
state to be $R_{\mu/e}=0.9779(2)_{\text{latt}}(50)_{\mathrm{EM}}$ in the Standard Model,
with the uncertainty dominated by that from electromagnetic corrections. 
\end{abstract}

\maketitle

\section{Introduction} \label{sec:intro}

The flavour changing weak interactions between quarks via emission of $W$ 
bosons can be parameterised in terms of the unitary Cabbibo-Kobayashi-Maskawa 
(CKM) matrix in the Standard Model, given by~\cite{Cabibbo:1963yz,Kobayashi:1973fv}
\begin{equation}
V_{\mathrm{CKM}} = \begin{bmatrix} V_{ud}&V_{us}&V_{ub}\\V_{cd}&V_{cs}&V_{cb}\\V_{td}&V_{ts}&V_{tb} \end{bmatrix}.
\end{equation}
Precise and independent determination of each of the CKM matrix elements 
from multiple processes is crucial to test the Standard Model stringently. 
Current accuracy varies from 0.014\% for $V_{ud}$ to 6\% for $V_{ub}$ with 
several reviews in~\cite{pdg} discussing different aspects of their determination. 
For a recent review of the impact of lattice QCD on this endeavour see~\cite{mw2020}. 
Here we will focus on the determination of $V_{cs}$ and provide a significant improvement 
in its accuracy that expands the range of tests we can perform of the CKM matrix. 

Any significant deviation from unitarity of the CKM matrix would signal the existence of 
physics beyond the Standard Model, but the accuracy with which unitarity tests can be 
performed varies substantially across the matrix. 
The unitarity of the first row has been tested to a precision of 0.05\%.
A result of
\begin{equation}
\label{eq:firstrowunitarity}
|V_{ud}|^2 + |V_{us}|^2 + |V_{ub}|^2 = 0.9985(3)_{V_{ud}}(4)_{V_{us}}
\end{equation}
is quoted in `$V_{ud}$, $V_{us}$, the Cabibbo Angle and CKM Unitarity' in~\cite{pdg}, noting 
that the value of $V_{ub}$ is too small to affect this relation. 
$V_{ud}$ here is determined from super-allowed nuclear $\beta$ decay with a 0.01\% experimental 
accuracy that requires careful treatment of electroweak radiative corrections 
(see the review for a discussion of this). 
$V_{us}$ is determined from a weighted average of results from combining experimental results 
for $K$ meson leptonic and semileptonic decays with lattice QCD calculations 
of the appropriate hadronic parameters~\cite{Bazavov:2017lyh,Bazavov:2018kjg,Aoki:2019cca}. 
A determination with uncertainty better than 0.3\% is possible in both cases, paying 
attention to various sources of electroweak radiative corrections. 
The value quoted in Eq.~(\ref{eq:firstrowunitarity}), with its 3$\sigma$ hint of a discrepancy 
with unitarity,
results from a weighted average 
of leptonic and semileptonic $V_{us}$ values with an uncertainty 
increased by a factor of two to allow for
the tension between them. 

Tests of unitarity for other rows and columns of the CKM matrix are much less stringent, either 
because of larger experimental uncertainties, larger theoretical uncertainties or both. 
Our aim here is to improve $V_{cs}$. 
Since $V_{cs}$ is close to 1 in value, it needs to have a small uncertainty to 
avoid ruining any CKM unitarity test that it appears in. 
The determination of $V_{cs}$
proceeds most directly, as for $V_{us}$, either through a study 
of leptonic decays of the $D_s$ meson or through $D$ semileptonic decay to $K\ell\nu$. 
We summarise its current status below before outlining our new determination. 

We will not discuss the determination of $V_{cs}$ from other semileptonic 
channels such as $D_s \rightarrow \phi$~\cite{Donald:2013pea} 
or $\Lambda_c \rightarrow \Lambda$~\cite{Meinel:2016dqj}. These 
are not currently competitive but do provide further checks on $V_{cs}$; the baryon 
channel is particularly important to provide constraints on new physics 
complementary to those available from meson decays.

\subsection{Current situation on $V_{cs}$}
\label{sec:introvcs}
The experimental measurement of the branching fraction for $D_s$ leptonic decay has been 
challenging, with the average drifting downwards slowly with time as newer results are added. 
The current situation is reviewed 
in `Leptonic decays of charged pseudoscalar mesons' in~\cite{pdg} 
(we will refer to this as RSV). 
See also results from the Heavy Flavor Averaging Group (HFLAV)~\cite{Amhis:2019ckw}. 
There are now experimental results from BaBar, Belle,
BES III and CLEO-c with either $\mu$ or $\tau$ in the final state. 
The experimental branching fraction 
$D_s \rightarrow \ell \overline{\nu}$
is obtained after removing the 
effect of QED bremsstrahlung at leading-log order using PHOTOS~\cite{Golonka:2005pn}. 
The measured width is then given by 
\begin{equation}
\label{eq:leptrate}
\Gamma = \frac{G_F^2m^2_{\ell}M_{D_s}}{8\pi} (\eta_{EW}f_{D_s}|V_{cs}|)^2\left(1-\frac{m^2_{\ell}}{M^2_{D_s}}\right)
\end{equation}
up to remaining QED effects (RSV apply a 1\% correction to BaBar and Belle $\mu$ results 
to account for contamination 
from $D_s \rightarrow (D_s^*\rightarrow \ell\overline{\nu})\gamma$~\cite{Dobrescu:2008er}). 
$\eta_{EW}$ accounts for short-distance electroweak 
corrections to the value of $G_F$ obtained from the $\mu$ lifetime~\cite{Sirlin:1981ie}, a 
correction applied as standard in the $K$ leptonic and semileptonic decays discussed above. 
The experimental width then yields a result for the combination $\eta_{EW}f_{D_s}|V_{cs}|$, 
where $f_{D_s}$ is the decay constant of the $D_s$ meson, the hadronic parameter that 
determines the amplitude for annihilation to a $W$ boson.
RSV take $\eta_{EW}=1.009$ and obtain an average from experiment of 
\begin{equation}
\label{eq:pdglept}
|V_{cs}|f_{D_s} = 245.7(3.1)(3.4) \ \mathrm{MeV}.
\end{equation}
The first error here comes from the experimental branching fractions and the second error 
takes a 100\% uncertainty from the applied radiative corrections 
($\eta_{EW}$ and the additional 1\% on the rate to $\mu$ above). 
The average from HFLAV~\cite{Amhis:2019ckw} (included in the review  
`CKM Quark-Mixing Matrix' in~\cite{pdg}) has a larger central value because 
they take $\eta_{EW}=1$, and a smaller uncertainty since they do not include the 
second uncertainty above. The total experimental uncertainty then ranges from 
1.3\% from HFLAV~\cite{Amhis:2019ckw} to 1.9\% from Eq.~\eqref{eq:pdglept}.  

Early full lattice QCD calculations~\cite{Aubin:2005ar} of the $D_s$ decay 
constant were undertaken 
before the experimental results were obtained. They had rather large (6\%) 
systematic uncertainties from 
discretisation effects associated with the relatively heavy $c$ quark mass and 
uncertainties from matching the normalisation of the 
lattice representation of the $c\overline{s}$ weak current 
to that in the continuum. 
A step-change in accuracy was made possible by the development of HPQCD's Highly Improved 
Staggered Quark (HISQ) action~\cite{Follana:2006rc}. 
This has good control of discretisation effects (going beyond 
$\mathcal{O}((ma)^2)$) and a partially conserved axial current relation that 
enables the decay constant to be absolutely normalised. 
HPQCD used this to obtain a 1\% accurate result for $f_{D_s}$~\cite{Follana:2007uv, Davies:2010ip} 
back in 2010. Combined with the higher experimental average for the branching fraction  
at that time it led to 
a $V_{cs}$ result with a central value above 1. 
More recent results from the Fermilab/MILC 
collaboration~\cite{Bazavov:2017lyh} using HISQ give 
a 0.2\% uncertainty on $f_{D_s}$. RSV then give a leptonic determination 
\begin{equation}
\label{eq:introvcsl}
|V_{cs}|_{\mathrm{lept}} = 0.983(13)(14)(2) 
\end{equation}  
where the first uncertainty is from experiment, the second from radiative corrections 
and the third from $f_{D_s}$. We see that the current picture for $V_{cs}$ from leptonic 
decays is one in which the experimental uncertainty dominates that from lattice QCD ($f_{D_s}$), 
which is now almost negligible here. 
When radiative corrections are considered, as in RSV, they also have a sizeable uncertainty.  
The value obtained for $V_{cs}$ is consistent with 1. 

The situation with semileptonic $D \rightarrow K$ decays is somewhat different. 
Smaller experimental uncertainties have been available for some time 
but lattice QCD 
calculations are harder to do, with less accurate results to date. 
The hadronic quantities that parameterise the amplitude for the $c \rightarrow s$ 
transition within the meson are form factors, functions of the squared 4-momentum 
transfer, $q^2$, from the initial $D$ to final $K$ meson. The only form factor that 
contributes here, for light leptons in the final state, is the vector form factor, $f_+(q^2)$.  
Here we will improve substantially on previous lattice QCD uncertainties 
for the $D \rightarrow K$ form
factor and 
demonstrate the improvement in accuracy of $V_{cs}$ that results. 
This will inevitably mean, as discussed above, that uncertainties from 
electroweak radiative corrections will 
rear their heads. 

Experimental results for $D \rightarrow K \ell \overline{\nu}$ 
are available from BaBar, Belle, BES III and 
CLEO-c~\cite{Amhis:2019ckw} and will be discussed in more detail later. 
Results exist for both charged and neutral $D$ mesons and with both 
$e$ and $\mu$ in the final state. They are either given in the form of 
a differential distribution in bins of $q^2$ or, following a fit to the distribution 
combined with an analysis of radiative bremsstrahlung corrections 
using PHOTOS, a value 
for $\eta_{EW}|V_{cs}|f_+(0)$. 
HFLAV~\cite{Amhis:2019ckw} quote an average for this latter quantity 
with a 0.5\% uncertainty from experiment. 
Note that $\eta_{EW}$ is taken to be 1 in 
these analyses and does not appear as a factor.  

Full lattice QCD calculations of the $D \rightarrow K$ form factors again 
began before experimental results were available~\cite{Aubin:2004ej} but 
were limited in accuracy (to 10\%) by systematic effects from the discretisation of the 
quark action.  
The use of the HISQ action by HPQCD brought a big improvement~\cite{Na:2010uf}
coupled with the fact that the scalar form factor $f_0$ (equal to $f_+$ at $q^2=0$) 
can be determined with absolute normalisation. HPQCD extended this 
to a determination 
of the vector form factor across the full physical $q^2$ range 
in~\cite{Koponen:2013tua} with nonperturbative normalisation of the vector current.   
This allowed a 1.6\%-accurate determination of $|V_{cs}|$ using a bin-by-bin 
comparison of the differential distribution with experiment, 
thus providing also a Standard Model test of the shape 
of the distribution. Recently the European Twisted Mass Collaboration (ETMC) 
determined the full shape of the $D \rightarrow K$ form 
factors~\cite{Lubicz:2017syv, Riggio:2017zwh} using the twisted mass formalism 
and combined that with experimental results to obtain a 3.5\% accurate result for 
$|V_{cs}|$. Work is also underway by other groups; see, 
for example,~\cite{Kaneko:2017sct, Li:2019phv}.

The ETMC result for $f_+^{D\rightarrow K}(0)$  
is used in the `CKM Quark-Mixing Matrix' review 
in~\cite{pdg} (quoting~\cite{Aoki:2019cca}) 
to give a semileptonic determination of $V_{cs}$ as  
\begin{equation}
\label{eq:introvcs-sl}
|V_{cs}|_{\mathrm{semi}} = 0.939(38). 
\end{equation}  
The uncertainty here is strongly dominated by that from lattice QCD. The result
takes $\eta_{EW}=1$ and does not include additional uncertainties to allow for 
possible missing QED corrections. 
Combining their results over the full range of $q^2$ with experiment, 
ETMC~\cite{Riggio:2017zwh}
instead obtains 
\begin{equation}
\label{eq:introvcs-sl-etm}
|V_{cs}|_{\mathrm{semi}} = 0.978(35), 
\end{equation}  
with similar uncertainty. Both results above are consistent with the value 1 within 
2$\sigma$ because 
of the large uncertainty. They also agree with the expectation 
$V_{cs}=V_{ud}=0.97370(14)$~\cite{pdg} to $\mathcal{O}((\lambda=V_{us})^4)$.  

The results for $|V_{cs}|$ in Eqs~(\ref{eq:introvcsl}),~(\ref{eq:introvcs-sl}) 
and~(\ref{eq:introvcs-sl-etm}) contribute 4--7\% uncertainties to 
CKM second row or column unitarity, i.e. two orders of magnitude worse than that for 
the first row discussed earlier (Eq.~(\ref{eq:firstrowunitarity})). 
This precludes picking up hints of new physics.  

\vspace{2mm}

Here we provide a substantial improvement to the lattice QCD determination of 
these form factors using the HISQ action 
on gluon field configurations that include $u$, $d$, $s$ and $c$ quarks in the sea. 
We build on~\cite{Koponen:2013tua} (although using a method for normalising the vector 
current suggested but not 
implemented there) to determine the scalar and vector form factor across the full physical 
$q^2$ range for the decay. 
This enables us to compare to experimental results in each 
$q^2$ bin as well as at $q^2=0$, 
as in~\cite{Koponen:2013tua, Lubicz:2017syv, Riggio:2017zwh}, to determine $V_{cs}$. 
We include results over a larger range of lattice spacing values than 
in~\cite{Koponen:2013tua} and with
sea light quark masses going down to physical values of the $u/d$ mass. 
Our work will also provide form factors for the improved experimental determinations 
to come in future, for example from Belle II~\cite{Kou:2018nap}. 

The paper is laid out as follows: Section~\ref{sec:formalism} 
lays out our formalism and then Section~\ref{sec:lattice} describes our 
lattice QCD calculation. This includes details of the gluon ensembles used 
and the correlation functions calculated followed by a 
description of how the calculated 
lattice correlation functions are fitted and values for the form 
factors extracted. We then describe how the form factor results are extrapolated to the physical 
continuum limit. Section~\ref{sec:lattice} can be omitted by anyone who is not interested in 
the lattice QCD details. Section~\ref{sec:results} gives our results for the 
physical form factors, with instructions on how to reconstruct them from the parameters given. 
We compare the shape of the vector form factor to that obtained from the differential
decay rate by experiment. We also give the ratio of branching fractions for a 
muon in the final state to that for an electron as a function of $q^2$ for tests of 
lepton flavour universality. 
Section~\ref{sec:vcs} gives three different methods for determining $V_{cs}$ using our 
results and experimental measurements of the $D \rightarrow K \ell \nu$ decay rate.  
Our preferred method is to use a bin-by-bin comparison with the differential decay rate 
but we also give values determined from the total branching fraction and from the 
rate at $q^2=0$. Section~\ref{sec:vcsdiscuss} puts our improved results for $V_{cs}$ into context with 
previous results and other CKM elements in tests of unitarity of the CKM matrix. 
Finally, Section~\ref{sec:conclusions} gives our conclusions. 

\section{Formalism}
\label{sec:formalism}

We write the differential decay rate for $D\rightarrow K \ell \overline{\nu}$ (inclusive of photons) as: 
\begin{eqnarray}
\label{eq:diffdecayrate}
\frac{d\Gamma}{dq^2} &=& \frac{G^2_F}{24\pi^3}(\eta_{EW}|V_{cs}|)^2(1-\epsilon)^2(1+\delta_{EM})\times \nonumber \\ 
&& \left[ |\vec{p}_K|^3(1+\frac{\epsilon}{2})|f_+(q^2)|^2 + \right. \nonumber \\
&&\hspace{2.0em}\left. |\vec{p}_K|M_D^2\left(1-\frac{M_K^2}{M_D^2}\right)^2\frac{3\epsilon}{8}|f_0(q^2)|^2 \right]  
\end{eqnarray}
where $\epsilon=m^2_{\ell}/q^2$, $m_{\ell}$ being the lepton mass, 
and $\vec{p}_K$ is the 3-momentum of the $K$ in the $D$ rest frame. 
Note that the contribution of $f_0$ to the differential rate is suppressed by $\epsilon$.
$\eta_{EW}$ accounts for universal short-distance  
corrections to $G_F$ from box diagrams in the Standard Model~\cite{Sirlin:1981ie}. 
We take 
\begin{equation}
\label{eq:eta-ew-value}
\eta_{EW} = 1+\frac{\alpha_{QED}}{\pi}\log\left(\frac{M_Z}{M_D}\right) = 1.009(2) 
\end{equation}
where the uncertainty allows for a factor of two variation in the lower scale from the 
central value of $M_D$. $\delta_{EM}$ accounts for QED corrections to the leading-order formula.  
Some of these corrections may be $q^2$-dependent. We will handle $\delta_{EM}$ by taking an 
overall uncertainty for it, rather than making an explicit correction (see Section~\ref{sec:vcs}).  

In Eq.~(\ref{eq:diffdecayrate}) $f_+$ and $f_0$  
are the vector and scalar form factors for the process, respectively. 
They are defined from the matrix element of the vector part of the weak current between 
$D$ and $K$, since that is the only part that contributes for a 
pseudoscalar meson to pseudoscalar meson 
semileptonic decay. 
The parameterisation of the matrix element of the vector current, $V^{\mu}=\overline{\psi}_s\gamma^{\mu}\psi_c$,  in the continuum can be written as
\begin{eqnarray}
\langle K|V^\mu|D\rangle &=& f_+^{D \rightarrow K}(q^2) \left[p_D^\mu + p_K^\mu - \frac{M^2_D - M^2_K}{q^2}q^\mu\right] \nonumber\\
           & & + f_0^{D \rightarrow K}(q^2) \frac{M^2_D - M^2_K}{q^2}q^\mu,  \label{eq:D2Kff} 
\end{eqnarray} 
where $M_D$ and $M_K$ are the masses of the $D$ and $K$ mesons 
(charged or neutral, as appropriate) respectively. 
The momentum transfer, $q^{\mu}=p^{\mu}_D - p^{\mu}_K$. 

Application of the partially conserved vector current (PCVC) relation shows that the scalar 
form factor can also be obtained from the matrix element of the scalar current, $S=\overline{\psi}_s\psi_c$ : 
\begin{equation}
\langle K|S|D\rangle = \frac{M^2_D - M^2_K}{m_{c} - m_{s}}f_0^{D \rightarrow K}(q^2).\label{eq:scff}
\end{equation}   
The PCVC relation also holds in lattice QCD for the HISQ discretisation~\cite{Follana:2006rc} 
of the quark action that we use. 
This means that $(m_c-m_s)\langle K | S | D \rangle$ is not renormalised and $f_0$ is obtained 
from the HISQ lattice QCD calculation with absolute normalisation~\cite{Na:2010uf}. 
Eq.~(\ref{eq:D2Kff}) requires that $f_+(0)=f_0(0)$ and hence a determination of the scalar form 
factor obtained at $q^2=0$ is sufficient to determine the vector form factor there. This can then be combined with experimental results, if they are given in the form of a determination of 
$|V_{cs}|f_+(0)$, to yield a value for $V_{cs}$~\cite{Na:2010uf}. We will make use of this 
as one method to obtain $|V_{cs}|$. 

Here we also determine $f_+$ over the full range of physical $q^2$ for the decay so that 
we can compare to the differential rate from experiment using Eq.~(\ref{eq:diffdecayrate}). 
Although the HISQ action has a conserved vector current that is well understood~\cite{Hatton:2019gha}, 
it is a complicated operator with several different multilink point-split components. 
Instead we use here a much simpler local vector current but this must be renormalised to 
match the (partially) conserved current. We do this by writing 
$V^{\mu}=Z_V V^{\mu}_{\mathrm{latt}}$ and determine $Z_V$ by comparing scalar and temporal 
vector matrix elements in the `zero recoil' configuration where 
the $D$ and $K$ are both at rest and $q^2\equiv q^2_{\mathrm{max}}=(M_D-M_K)^2$.  
Then, from Eq.~\eqref{eq:D2Kff}
\begin{equation}
\label{eq:zvcalc1}
Z_V\langle K|V^0_{\mathrm{latt}}|D\rangle = 
            f_0^{D \rightarrow K}(q^2_{\mathrm{max}}) (M_D+M_K)
\end{equation} 
so that $Z_V$ can be determined at this kinematic point~\cite{Koponen:2013tua} from 
\begin{equation}
\label{eq:zvcalc2}
(M_D-M_K)Z_V\langle K|V^0_{\mathrm{latt}}|D\rangle_{q^2_{\mathrm{max}}} = 
           (m_c-m_s)\langle K | S | D\rangle_{q^2_{\mathrm{max}}}.  
\end{equation} 
Note that $m_c$ and $m_s$ here are the HISQ lattice quark masses for $c$ and $s$.
This provides a self-consistent normalisation for the matrix elements in Eqs.~(\ref{eq:D2Kff}) and~(\ref{eq:scff}) that matches that in the continuum. 

\begin{table*}
  \caption{Parameters for the $N_f=2+1+1$ gluon field configurations used in this work. 
The Wilson flow parameter~\cite{Borsanyi:2012zs} is used to 
determine the lattice spacing, $a$, via the values for $w_0/a$. 
We take $w_0=0.1715(9)\text{fm}$, as determined in \cite{Dowdall:2013rya} 
from $f_{\pi}$. 
Column 4 gives approximate values for $a$ in fm for each set, and column 5 gives the 
approximate value for the ratio of the light quark mass to that of strange in the sea 
(the physical value is close to 0.036~\cite{Bazavov:2017lyh}). 
Column 6 gives the spatial ($N_x$) and temporal ($N_t$) dimensions of each lattice in lattice units and 
column 7 the number of configurations and time origins used in our calculation. 
Columns 8--12 give the masses 
of the valence and sea quarks in lattice units. For the light ($u/d$) quark the valence and sea masses are the same. 
Column 13 shows values for the normalisation $Z_{\text{disc}}$, defined in~\cite{Monahan:2012dq} and appearing in Eq.~\eqref{Eq:fitnormalisation}.}
  \begin{center} 
    \begin{tabular}{l l l l l l l l l l l l l}
      \hline
      Set & $\beta$ & $w_0/a$ & $a$ (fm) & $(m_l/m_s)^{\mathrm{sea}}$ & $N_x^3\times N_t$  &$n_{\mathrm{cfg}}\times n_{\mathrm{src}}$ &    $am_{l}^{\mathrm{sea/val}}$ & $am_{s}^{\mathrm{sea}}$ & $am_c^{\mathrm{sea}}$& $am_{s}^{\mathrm{val}}$ & $am_c^{\mathrm{val}}$& $Z_{\mathrm{disc}}$\\
  \hline
  \hline
    1   & 5.8    & 1.1367(5) & 0.15 & \ 0.036 &  $32^3\times 48$    & $998\times 16$ &     0.00235        &  0.0647        &  0.831&  0.0678 &  0.8605& 0.99197\\
    \hline
    2    & 6.0    & 1.4149(6) & 0.12 & \ 0.036 &  $48^3\times 64$    & $985\times 16$&   0.00184       &   0.0507        &  0.628  &0.0527  & 0.643& 0.99718\\
    \hline
    3   &   6.3  & 1.9518(7) & 0.09 & \ 0.033 & $64^3\times 96$    & $620\times 8$&  0.00120       &  0.0363         &  0.432&   0.036  &  0.433& 0.99938\\
    \hline
    4    & 5.8    & 1.1119(10) & 0.15 & \ 0.20 & $16^3\times 48$    & $1020\times 16$&     0.013        &  0.065        &  0.838 &  0.0705 &  0.888 & 0.99105\\
    \hline
    5    & 6.0    &  1.3826(11) & 0.12 & \ 0.20 & $24^3\times 64$   & $1053\times 16$ &     0.0102        &  0.0509        &  0.635  &  0.0545  & 0.664 & 0.99683\\
    \hline
    6    & 6.3    &  1.9006(20)  & 0.09 & \ 0.20 & $32^3\times 96$  & $499\times 16$  &   0.0074       &   0.037        &  0.440&  0.0376  &  0.449& 0.99892\\
    \hline
    7   & 6.72    &  2.896(6) & 0.06 & \ 0.20 & $48^3\times 144$    & $415\times 8$&  0.0048       &  0.024         &  0.286& 0.0234 &  0.274 & 0.99990\\
    \hline
    8   & 7.0    &   3.892(12) & 0.044 & \ 0.20 & $64^3\times 192$    & $375\times 4$&  0.00316       &  0.0158         &  0.188& 0.0165 &  0.194 & 0.99997 \\
    \hline
    \hline
    \end{tabular}
  \end{center}
  \label{tab:ensembles}
\end{table*}

\begin{table}
  \caption{Details of the $T$ values and $K$ meson momenta used on each ensemble. Momenta can be obtained from twist, $\theta$, via $\theta = |a\vec{p}_K|N_x/(\sqrt{3}\pi)$, where $N_x$ is the spatial dimension of the lattice, given in Table~\ref{tab:ensembles}.}
  \begin{center} 
    \begin{tabular}{c c c}
      \hline
      Set & $\theta$ & $T/a$  \\
  \hline
  \hline
    1    & 0, 2.013, 3.050, 3.969 &9, 12, 15, 18   \\
    \hline
    2    & 0, 2.405, 3.641, 4.735 &12, 15, 18, 21\\
    \hline
    3   & 0, 0.8563, 2.998, 5.140 & 14,17,20 \\
    \hline
    4    & 0, 0.3665, 1.097, 1.828 & 9, 12, 15, 18 \\
    \hline
    5    & 0,  0.441, 1.323, 2.205, 2.646 & 12, 15, 18, 21 \\
    \hline
    6    & 0, 0.4281, 1.282, 2.141, 2.570& 14, 17, 20 \\
    \hline
    7   & 0, 1.261, 2.108, 2.946, 3.624 & 20, 25, 30  \\
    \hline
    8   &0, 0.706, 1.529, 2.235, 4.705 & 24, 33, 40 \\
    \hline
    \hline
    \end{tabular}
  \end{center}
  \label{tab:twists}
\end{table}

\section{Lattice calculation}
\label{sec:lattice}

\subsection{Simulation details}
The calculation used gluon ensembles generated by the MILC collaboration~\cite{Bazavov:2012xda}. 
The gluon action is improved through $\mathcal{O}(\alpha_sa^2)$~\cite{Hart:2008sq} 
and includes the 
effect of four flavours of sea quarks ($N_f=2+1+1$) using the HISQ action~\cite{Follana:2006rc}. 
The $u$ and $d$ sea quark masses are taken to be the same, with value denoted $m_l^{\text{sea}}$. 
The eight ensembles used have parameters listed in Table~\ref{tab:ensembles}. 
Sets 1, 2 and 3 have $m_l^{\text{sea}}$ set to the physical average value of $m_u$ and $m_d$, 
whilst sets 4-8 have $m_l^{\mathrm{sea/val}}=0.2m_s^{\mathrm{sea}}$. 
These `second-generation' gluon field configurations are a significant improvement over the 
`first-generation' $N_f = 2+1$ Asqtad configurations used in~\cite{Koponen:2013tua}. 
We also have results for a bigger range of lattice spacing values and going 
down to smaller values, from $a = 0.15$ fm 
to $a = 0.045$ fm. Although~\cite{Koponen:2013tua} discussed the use of the local temporal 
vector current, the results were obtained using a one-link-split spatial vector current. 
We believe that the approach using the local temporal current that we adopt here gives
improved statistical and systematic uncertainties. 

\subsection{Lattice correlation functions}

\begin{figure}
  \includegraphics[width=0.48\textwidth]{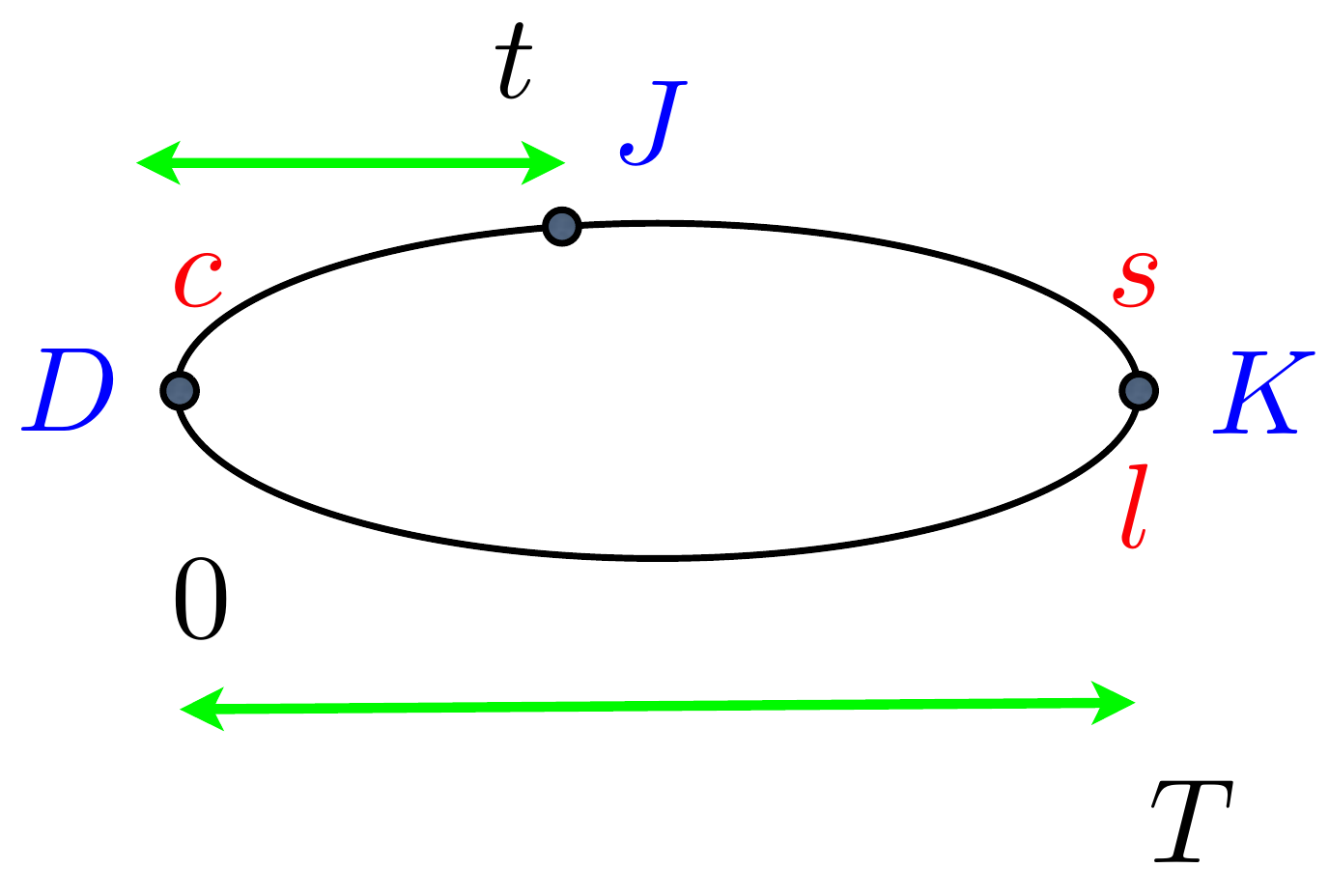}
  \caption{Schematic diagram of our three-point correlation function for current insertion $J$. }
  \label{fig:decay}
\end{figure}

Our goal is to extract scalar and temporal vector current matrix elements between $D$ and $K$ mesons 
for the determination of form factors using Eqs.~\eqref{eq:D2Kff} and~\eqref{eq:scff}. 
To do this we calculate three-point correlation functions on the lattice, as shown schematically in Figure~\ref{fig:decay}, 
constructed by multiplying together three valence quark propagators, obtained by solving the Dirac equation on 
the gluon field configuration. We use propagators for a $c$ quark, an $s$ quark and a `spectator' light quark, 
combined together with appropiate spins (implemented by a position-dependent phase for staggered 
quarks~\cite{Follana:2006rc}) to give a pseudoscalar meson at each end and an appropriate current operator at $J$. 
For computational cost it is most convenient to perform the calculation in the following way:
two of the propagators, $s$ and $l$, are generated from the same random 
wall source and the third quark propagator, the $c$, is an 
extended propagator using as a source the appropriate timeslice of the light quark propagator. 
In Figure~\ref{fig:decay} the $c\overline{l}$ pseudoscalar operator is placed at the origin and labelled by `$D$', 
the ground-state meson in that channel. Likewise the $s\overline{l}$ operator at $T$ is denoted by `$K$'. 
We calculate correlation functions from multiple different values of the origin timeslice (averaged over before fitting) 
to improve statistical 
errors. We also use multiple values of the time separation between $D$ and $K$, $T$, to improve the 
determination of the ground-state to ground-state matrix element. The $T$ values are listed 
in Table~\ref{tab:twists}.    

The extra `taste' degree of freedom for staggered quarks leads to some technical complications 
when constructing our meson 3-point correlation functions. We need to combine pseudscalar operators  
for the mesons at 0 and $T$ with either a scalar or temporal vector current operator at $t$. 
Staggered bilinears of different taste for a given spin are constructed with different 
point-splitting arrangements. 
Our preference is to use local operators because they are simple and most precise (since they do not incorporate gluon fields). A further advantage is that they have no tree-level discretisation errors.  
We then have to make sure that the three operators chosen have appropriate tastes; if not, the correlation 
function will be zero. The simplest way to test this is to write down the correlation 
function using naive quarks and apply the transformation to staggered quarks so that the correlation 
function factorises into a product of a color trace over a product of staggered quark propagators and a 
spin trace over a product of gamma matrices. The spin trace will be zero if tastes have been 
incorrectly combined~\cite{Follana:2006rc}. 

This means that we must use two different operators for the $c\overline{l}$ 
pseudoscalar meson depending on whether the current in the three-point correlation 
function is a scalar, $S=\bar{\psi}_s 1\otimes 1\psi_c$, or temporal vector, 
$V^{0}=\bar{\psi}_s\gamma^0\otimes\gamma^0\psi_c$. We give both operators here in 
their conventional `spin-taste' notation. The fact that the spin and taste gamma matrices 
are the same means that they are both local
(i.e. with $\psi$ and $\bar{\psi}$ fields at the same point). The operators are implemented for 
staggered quarks simply using a position-dependent patterning of $\pm 1$ instead of $\gamma$ matrices. 
In both cases we use an $s\bar{l}$ pseudoscalar 
operator for the $K$ meson of `Goldstone' form, i.e. 
$\bar{\psi}_l\gamma^5\otimes\gamma^5\psi_s$.     
For the $D$ meson we can use this same form,  
$\bar{\psi}_l\gamma^5\otimes\gamma^5\psi_c$,     
for correlation functions with the scalar current, since this is taste-singlet 
with a taste matrix of 1. 
Since the local temporal vector current has taste $\gamma^0$ we use a different, but still 
local, operator in its correlation functions. We distinguish this operator by denoting it by $\hat{D}$;  
$\hat{D}=\bar{\psi}_c\gamma^5\gamma^0\otimes\gamma^5\gamma^0\psi_l$. 
We also calculate two point correlation functions for the Goldstone pseudoscalar 
$K$, and the Goldstone and non-Goldstone $D$ bilinears detailed above. 
The $D$ meson masses for the Goldstone and non-Goldstone 
operators will not be the same but differ by a taste-splitting which is a discretisation
effect. These splittings are very small for heavy mesons 
such as the $D$~\cite{Follana:2006rc, Bazavov:2012xda}. We demonstrate that 
for this calculation in Appendix~\ref{App:corrfits}. 

We take the $D$ meson to be at rest and give spatial momentum to 
the $K$ meson so that we can map out the dependence of 
the form factors on $q^2$. We do this by using twisted boundary 
conditions~\cite{Guadagnoli:2005be} for the $s$-quark propagator. 
The twist is taken equally in all three spatial directions to generate a momentum
in the $(1,1,1)$ direction, minimising discretisation effects for a given 
value of $|\vec{p}_K|$. 
The twist angle, $\theta = |a\vec{p}_K|N_x/(\sqrt{3}\pi)$, where $N_x$ is the spatial 
extent of the lattice in lattice units. 
Different values of momentum were chosen so as to cover the full physical 
range of momentum transfer, $q$, and the twists used are listed in Table~\ref{tab:twists}.

We summarise below how the two-point correlation functions are built from 
quark propagators, $g_q(x_t,x_0)$, of flavour $q$ from point $x_0=(0,\vec{x}_0)$ 
to point $x_t=(t,\vec{x}_t)$. The two-point correlators are labelled by the ground-state 
meson in that channel    
\begin{equation}
  C_{D}(t)=\frac{1}{4}\sum_{\vec{x}_0,\vec{x}_t}\langle\text{Tr}[g_c^{\dagger}(x_t,x_0)g_l(x_t,x_0)] \rangle,
\end{equation}
\begin{equation}
  C_{\hat{D}}(t)=\frac{1}{4}\sum_{\vec{x}_0,\vec{x}_t}\langle(-1)^{\bar{x}_0^0+\bar{x}_t^0}\text{Tr}[g_c^{\dagger}(x_t,x_0)g_l(x_t,x_0)] \rangle,
\end{equation}
where $\bar{x}^{\mu}=\sum_{\nu\neq\mu}x^{\nu}$, and  
\begin{equation}
  C^{\vec{p}}_{K}(t)=\frac{1}{4}\sum_{\vec{x}_0,\vec{x}_t}\langle\text{Tr}[g_s^{\theta\dagger}(x_t,x_0)g_l(x_t,x_0)] \rangle .
\end{equation}
The factor of $1/4$ is the inverse of the number of staggered quark tastes. 
We sum over the spatial components of $x_t$ and $x_0$; the sum for $x_0$ is implemented using 
a random wall source. The $\langle\rangle$ denotes the average over gluon field 
configurations in an ensemble and the trace is over colour. 
$\theta$ denotes the twist that gives spatial momentum to the $s$ quark. 

Three-point correlation functions are built similarly~\cite{McLean:2019qcx,Parrott:2020vbe}
and labelled by the current operator
\begin{equation}
  C^{\vec{p}}_{S}(t,T)=\frac{1}{4}\sum_{\vec{x}_0,\vec{x}_t,\vec{x}_T}\langle\text{Tr}[g_c^{\dagger}(x_T,x_t)g_l(x_T,x_0)g^{\theta\dagger}_s(x_t,x_0)] \rangle,
\end{equation}
\begin{equation}
\begin{split}
    C^{\vec{p}}_{V^0}(t,T)&=\frac{1}{4}\sum_{\vec{x}_0,\vec{x}_t,\vec{x}_T}\langle(-1)^{\bar{x}_t^0+\bar{x}_T^0}\\
    &\times\text{Tr}[g_c^{\dagger}(x_T,x_t)g_l(x_T,x_0)g^{\theta\dagger}_s(x_t,x_0)] \rangle.
\end{split}
\end{equation}
In the next section we discuss how we fit these two- and three-point correlation functions to 
determine the $D$ to $K$ matrix elements and hence form factors. 

\subsection{Correlator Fits}
\label{sec:corrfits}

We perform a simultaneous multi-exponential fit to all of the two- and 
three-point correlation functions 
on each gluon field ensemble, using a standard Bayesian 
approach~\cite{Lepage:2001ym}\footnote{We 
use the corrfitter package~\cite{peter_lepage_2020_3707868,peter_lepage_2020_3715065,peter_lepage_2019_3563090} to do this.}.
The fit form that we use for the two-point correlator for meson $H$ is 
\begin{align}\label{Eq:2ptcorrfitform}
  C_H(t)&=\sum_{i=0}^{N_{\text{exp}}}\Big(|d_i^{H,n}|^2\, (e^{-E^{H,n}_i t} + e^{-E^{H,n}_i (N_t-t)}) \nonumber \\
  &-(-1)^{t/a}|d_i^{H,o}|^2\, (e^{-E^{H,o}_i t}+e^{-E^{H,o}_i (N_t-t)}) \Big) ,
\end{align}
where we include on the first line a tower of excited states of $H$ of energy $E^{H,n}_i$ and amplitude $d_i^{H,n}$ above the ground state ($i = 0$) generated by our lattice operator. 
Staggered quark operators also generate a tower of opposite parity states that oscillate in time 
and we also include such states in our fit (on the second line above) 
with their own amplitudes and energies, 
$d_i^{H,o}$ and $E_i^{H,o}$. 

Likewise the fit form for three-point correlators for current $J$ is: 
\begin{equation}\label{Eq:3ptcorrfitform}
\begin{split}
  &C^{\vec{p}}_J(t,T) = \sum_{i,j=0}^{N_{\text{exp}}} \Big( d_i^{K,n} J_{ij}^{nn} d_j^{\brachat{D},n}\, e^{-E^{K,n}_i t}\, e^{-E^{\brachat{D},n}_j (T-t)}\\
  &-(-1)^{(T-t)/a}\, d_i^{K,n}J_{ij}^{no}d_j^{\brachat{D},o}\, e^{-E^{K,n}_i t}\, e^{-E^{\brachat{D},o}_j (T-t)}\\
  &-(-1)^{t/a}\, d_i^{K,o} J_{ij}^{on} d_j^{\brachat{D},n}\, e^{-E^{K,o}_it}\, e^{-E^{\brachat{D},n}_j (T-t)}\\
  &+(-1)^{T/a}\, d_i^{K,o} J_{ij}^{oo} d_j^{\brachat{D},o}\, e^{-E^{K,o}_i t}\, e^{-E^{\brachat{D},o}_j (T-t)}\Big) .
\end{split}
\end{equation}
This includes the same towers of normal and oscillating states for $K$ and $D$ as those in 
two-point correlation functions. The only new parameters here are the three-point amplitudes, 
$J_{ij}$. To obtain these requires both two- and three-point correlator fits so that the 
$J_{ij}$ can be separated from the $d_i$ and $d_j$ amplitudes.  

The key parameters that we want to determine from these fits are the ground-state to 
ground-state amplitudes, $J_{00}^{nn}$, for the lattice temporal vector and scalar currents. 
We include the tower of excited states to remove contamination of excited states from the 
ground-state parameters and so that systematic errors from this contamination are fully included 
in the uncertainties on the ground-state parameters. 
Discarding data for $t < t_{\rm min}$ allows us to fit a finite number, 
$N_{\rm exp}$, of excited states, and $t_{\rm min}/a$ takes values in the range 2 to 5 for different correlators and different lattice spacings. 
Our fits use $N_{\rm exp}$ of 4 ($a = 0.15$ fm and 0.12 fm lattices) and 5 (finer lattices). 

Our fits use log-normal parameters to ensure non-negative amplitudes $d_i$ 
(because all of our two-point correlators have the same operator at source and sink) 
and energy differences between ordered states. 
We estimate priors for the ground state energies and amplitudes using the 
effective mass and effective amplitudes, as in~\cite{Parrott:2020vbe}, and give 
each a broad uncertainty (typically 5\%), ensuring that the final result of the fit is 
much more precisely determined (by at least an order of magnitude) than this prior.  
The ground-state energy in the oscillating channel is taken to be 0.4 GeV above the ground-state 
$D$ meson in the $D$ correlator and 0.25 GeV above the ground-state $K$ meson in the $K$ 
channel, using information from the Particle Data Tables~\cite{pdg}. The prior widths are 
typically taken as 20\% of the energy for the oscillating ground-state, again many times broader 
than the output from the fit. 
Likewise the priors for the ground-state to ground-state $J^{nn}_{00}$ 
are estimated from the three-point correlators by dividing through by the relevant two-point 
correlators and multiplying by their effective amplitudes. 
These priors are given an uncertainty of 20-50\% depending on the ensemble, again many 
times larger than the result from the fit.

For the $K$ mesons with non-zero momentum, 
we take priors for the ground-state energy and amplitude based on the priors for
the zero-momentum parameters and the dispersion relation. 
Denoting the prior for parameter $x$ as $\mathcal{P}[x]$ we use
\begin{equation}
  \begin{split}
    \mathcal{P}[aE^{K}_{0,\vec{p}}]&=\sqrt{(\mathcal{P}[aE^{K}_{0,\vec{0}}])^2+(a\vec{p})^2}\Big(1+\mathcal{P}[A]\Big(\frac{a\vec{p}}{\pi}\Big)^2\Big),\\
    \mathcal{P}[d^{K}_{0,\vec{p}}]&=\frac{\mathcal{P}[d^{K}_{0,\vec{0}}]}{(1+(\vec{p}/\mathcal{P}[E^{K}_{0,\vec{0}}])^2)^{1/4}}\Big(1+\mathcal{P}[B]\Big(\frac{a\vec{p}}{\pi}\Big)^2\Big).
  \end{split}
\end{equation}
We take priors for $A$ and $B$ as $0\pm1$. 

Priors for energy splittings between excited states are taken as 0.5 GeV with a 50\% uncertainty. 
Priors for excited state non-oscillating and all oscillating amplitudes 
are based on the size of ground-state amplitudes and generally given 100\% uncertainties. 
These are listed in Table~\ref{Tab:priorsforfit} in Appendix~\ref{App:corrfits} along 
with the priors for the remaining $J_{ij}^{kl}$.

\begin{figure}
\includegraphics[width=0.48\textwidth]{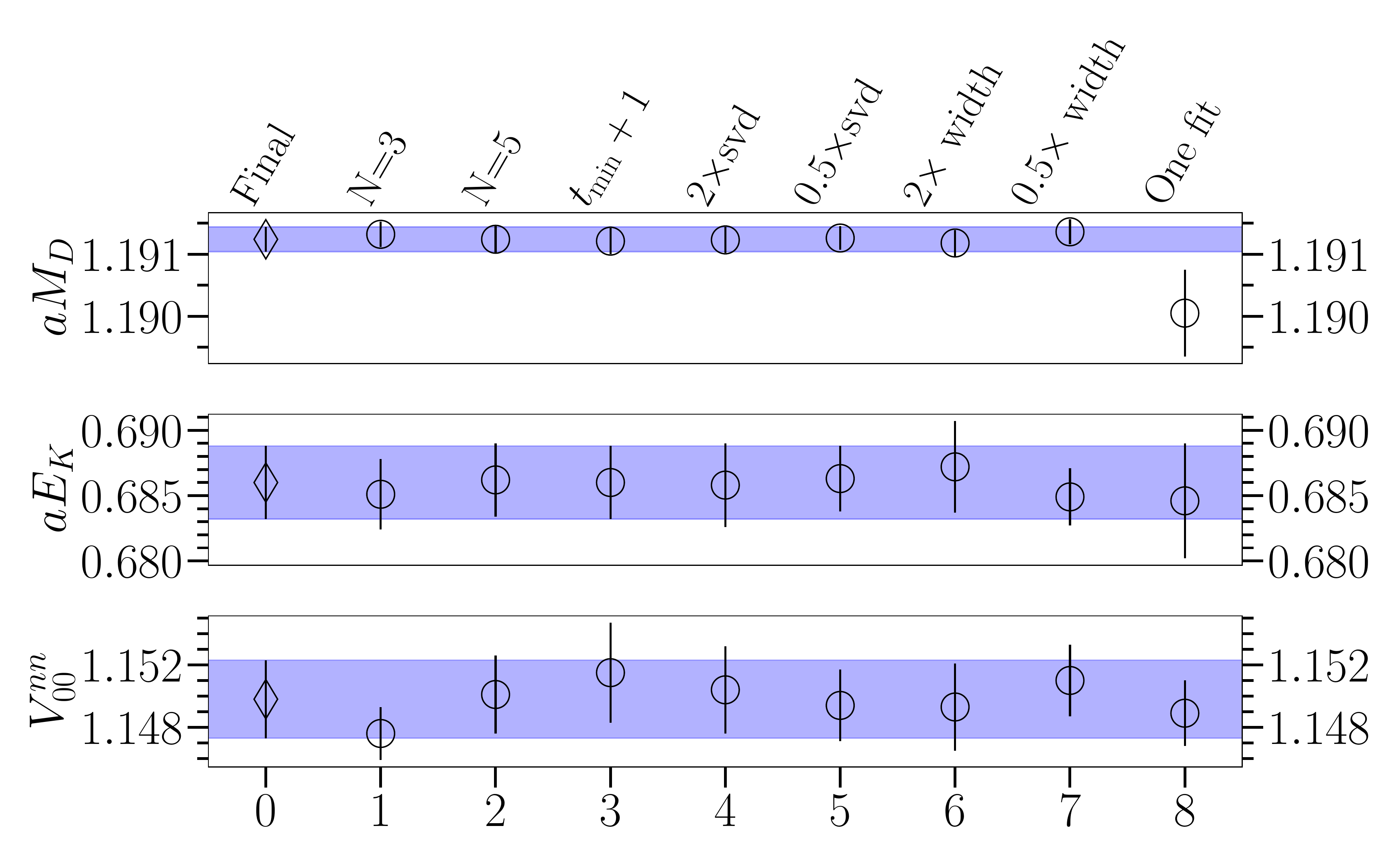}
\caption{Stability plot for our fit on the set 5 ($a = 0.12$ fm) lattice, with 
our preferred fit using $N=4$ exponentials, shown as the value at location 0. 
The different panels show (from the top) the mass of the $D$ (parameter $E_0^{D,n}$), 
the ground state energy of the $K$ (parameter $E_0^{K,n}$) 
with the largest twist for this set of 2.646, 
and current matrix element for the temporal vector current $V^{nn}_{00}$ 
(parameter $J_{00}^{nn}$) for twist 0.441. Tests 1 and 2 give the results 
from including one fewer and one more exponential respectively. 
Test 3 increases $t_\text{min}/a$ by 1 across the whole fit. 
Tests 4 and 5 double and halve the svd cut and tests 6 and 7 
double and halve all prior widths. 
The final test, 8, shows the results when the single correlator is fit on its own or, 
in the case of $V^{nn}_{00}$, just with the $D$ and $K$ two-point correlation functions required, 
rather than as part of one big simultaneous fit.}
\label{fig:corrstab}
\end{figure}

Since we have many correlators on each gluon field ensemble, the covariance matrix that must 
be inverted to minimise $\chi^2$ in our fits is very large. For a finite number of samples 
(gluon field configurations) there is a bias in the small eigenvalues of the covariance matrix  
that needs to be addressed in order to avoid underestimating uncertainties on the fit parameters; 
{see Appendix D of~\cite{Dowdall:2019bea} for a discussion of this}. 
We address this bias by applying a singular value decomposition (svd) cut 
on the eigenvalues using tools 
provided in our fitting package~\cite{peter_lepage_2020_3707868} for estimating 
an appropriate value. Using an svd cut leads to an artificial reduction in the $\chi^2$ value 
and so we implement additional `svd-noise'~\cite{Dowdall:2019bea,peter_lepage_2020_3707868} 
for a more reliable $\chi^2$ value. Our fit results are all based on fits for which this 
$\chi^2/\mathrm{dof}$ value is less than or close to 1.   

The results for the ground-state parameters for our preferred fits are given in 
Table~\ref{tab:fitresults} in Appendix~\ref{App:corrfits}. 

Figure~\ref{fig:corrstab} shows an example of tests of the stability of our correlator fits 
against a variety of changes. These tests are performed on all of our fits.  
We give further tests of our fit results in Appendix~\ref{App:corrfits}. 

Our fit results for the three-point amplitudes $J^{nn}_{00}$ are converted into 
the matrix elements we need in the following way: 
\begin{equation}\label{Eq:fitnormalisation}
  \bra{K}J\ket{\brachat{D}}=2Z_{\mathrm{disc}}\sqrt{M_{D}E_{K}}J^{nn}_{00},
\end{equation}
where $M_D$ is the Goldstone $D$ meson mass and $E_K$ the $K$ meson energy from 
the fit. 
We correct the normalisation for discretisation effects using the results of
~\cite{Monahan:2012dq}. $Z_{\mathrm{disc}}$ differs from 1 at $\mathcal{O}((am_c)^4)$ which is 
less than 1\% in all cases here; the values are given in Table~\ref{tab:ensembles}. 
For the temporal vector current the matrix element obtained above is 
$\bra{K}V^0_{\mathrm{latt}}\ket{\hat{D}}$. This needs to be normalised by 
multiplication by $Z_V$, which is determined using the matrix elements at 
zero-recoil and Eq.~\eqref{eq:zvcalc2}. The $Z_V$ values we obtain are listed in 
Table~\ref{tab:fitresults} in Appendix~\ref{App:corrfits}.     

\begin{figure}
\hspace{-30pt}
\begin{center}
\includegraphics[width=0.48\textwidth]{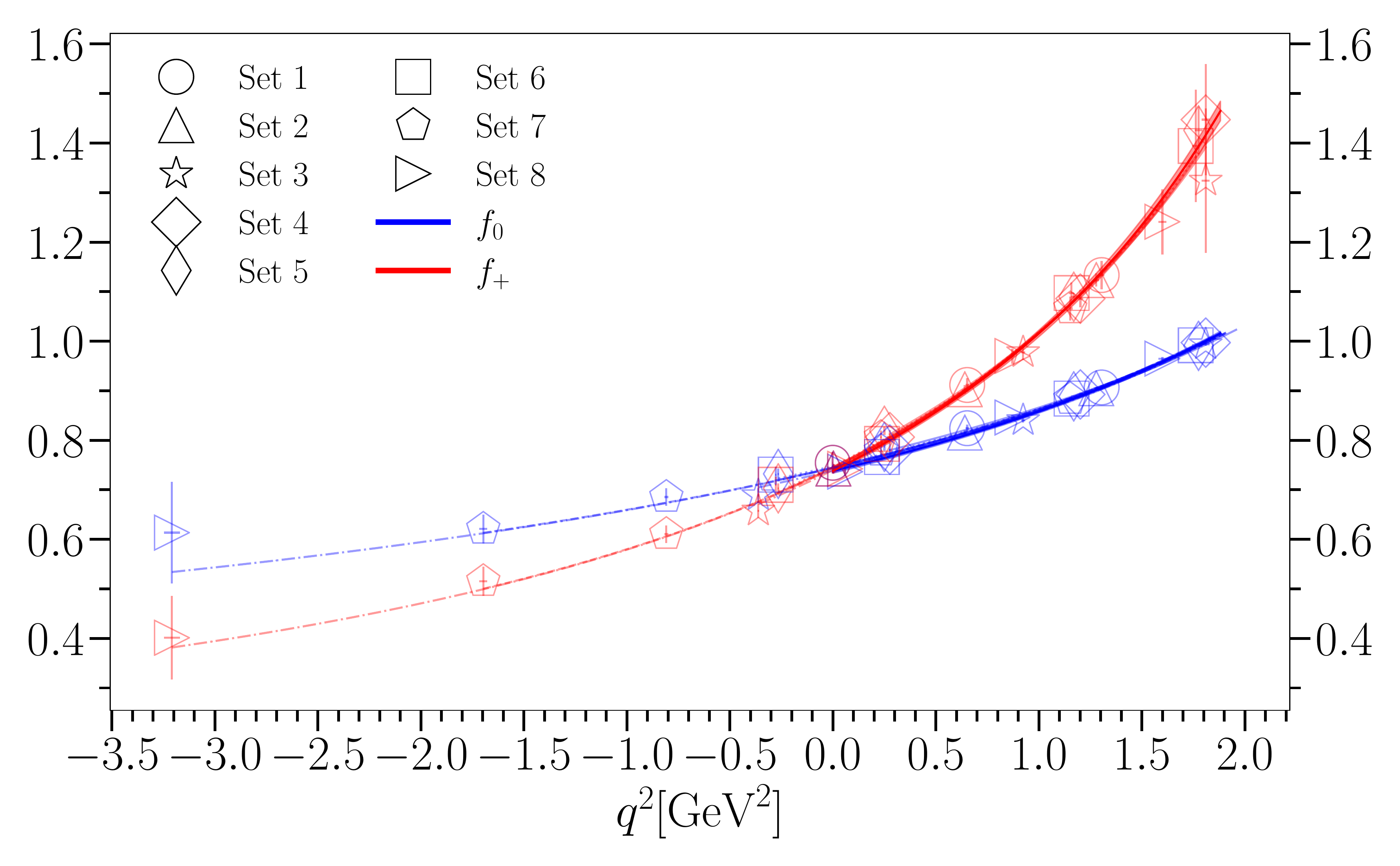}
\end{center}
\caption{$f_0$ and $f_+$ results on each of the 8 ensembles, marked by symbols as 
given in the legend. Our results cover the full physical $q^2$ range from $q^2=0$ to 
$q^2=(M_D-M_K)^2=1.88\,\mathrm{GeV}^2$.  
The solid blue and red curves correspond to our fit results for the form factors in the continuum 
limit, as described in Section~\ref{sec:ffphys}.
}
\label{fig:fdatabothinqsq}
\end{figure}

The vector and scalar form factors can now be determined from the 
matrix elements for the temporal vector and scalar currents on each gluon field ensemble using 
Eqs.~\eqref{eq:D2Kff} and~\eqref{eq:scff}. Our results for the form factors are 
listed in Table~\ref{tab:fitresults} in Appendix~\ref{App:corrfits} and plotted against $q^2$ 
in Figure~\ref{fig:fdatabothinqsq}. Little dependence on lattice spacing or sea light quark 
masses is visible. There are correlations between form factor 
values on a given field ensemble and these are captured in our correlator fits 
and passed on to the next stage of fitting. These correlations are sizeable between 
results for a given form factor ($f_+$ or $f_0$) 
at small values of the spatial momentum, close to zero-recoil. 
They are also substantial between $f_+$ and $f_0$ at large values of spatial momentum 
close to $q^2=0$.  

In the next section we discuss how we extrapolate our form factor results 
as a function of $q^2$ to the continuum limit.  

\subsection{Evaluating form factors at the physical point}
\label{sec:ffphys} 

Our results for the form factors at each value of $q^2$ on a given 
gluon field ensemble differ from the physical curve of $f(q^2)$ by 
discretisation effects and 
the mistuning of valence and sea quark masses. 
By fitting our results at multiple values of the lattice spacing and 
for multiple sea quark masses and allowing for valence quark mass mistuning 
we can account for both of these 
systematic effects. At the same 
time we interpolate in $q^2$ to obtain the physical form factor curves for 
the full kinematic range of $q^2$ values. 

Our preferred method for doing this is to extend the form factors to an analytic function 
in the complex $q^2$-plane and then map the physical region into a line inside 
the unit circle in $z$-space.  
This enables a simple fit and $a\rightarrow 0$ extrapolation in $z$-space and we can then 
transform back to $q^2$. 
We will describe that approach first in Section~\ref{sec:ffphysz}, along with a variety 
of tests of its robustness. 
In Section~\ref{sec:spline} we will compare results to a direct cubic spline fit in 
$q^2$-space. 

\subsubsection{Using a z-expansion}
\label{sec:ffphysz} 

\begin{figure}
\hspace{-30pt}
\begin{center}
\includegraphics[width=0.48\textwidth]{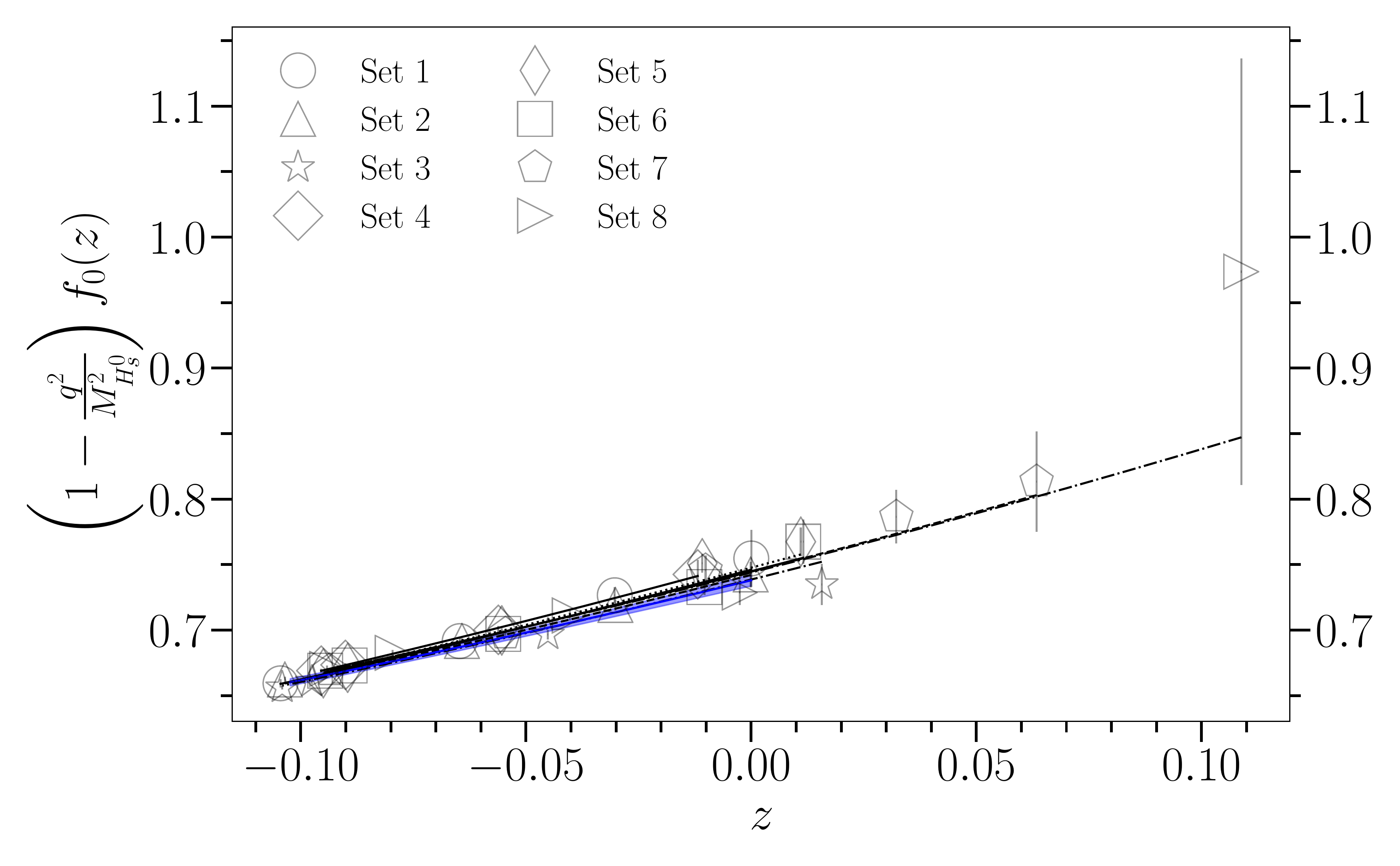}
\includegraphics[width=0.48\textwidth]{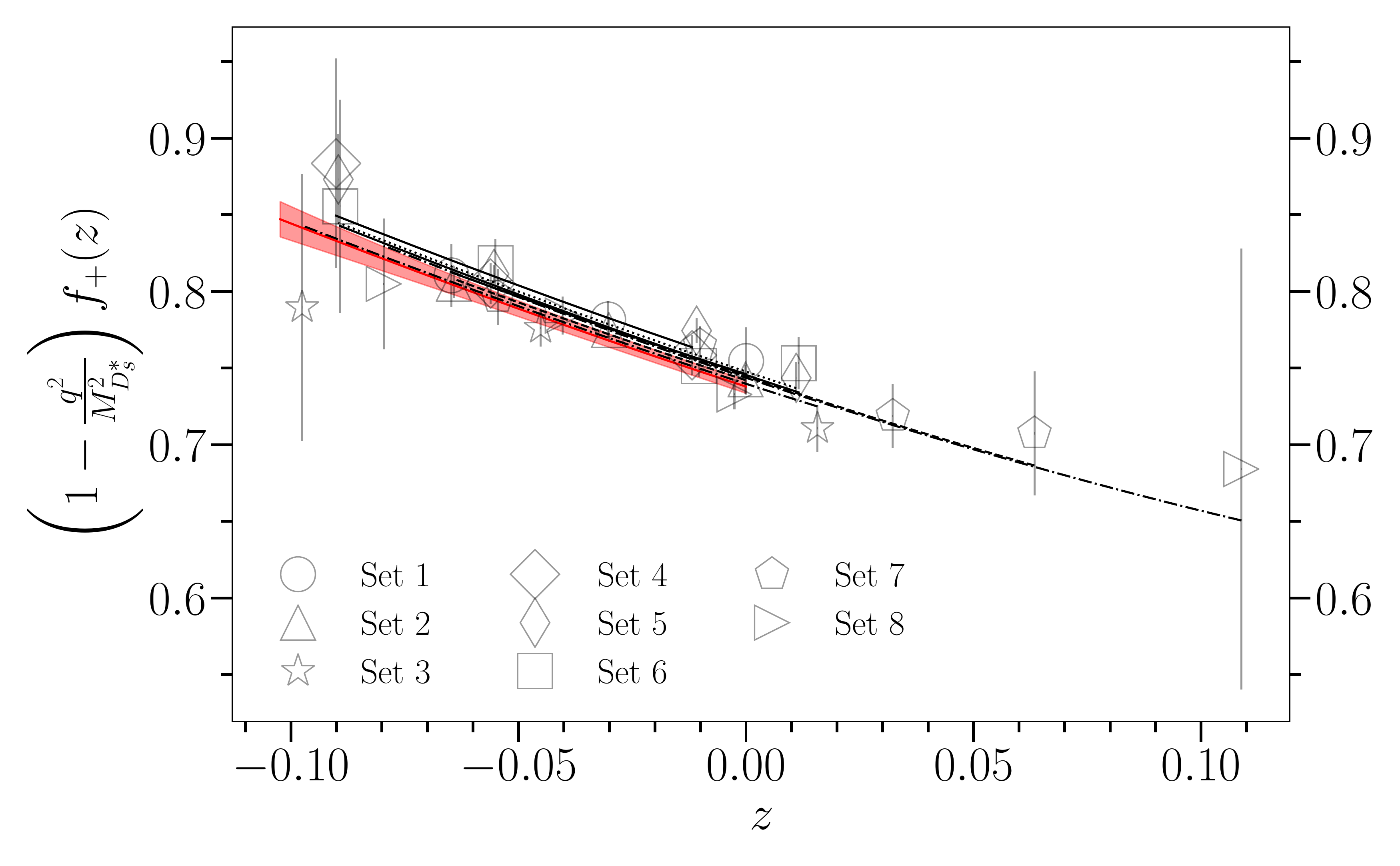}
\end{center}
\caption{Our lattice results for $f_0$ and $f_+$ on each of the 8 ensembles, plotted as a 
function of $z$ (Eq.~\eqref{eq:zspace}). In both cases the expected pole has been removed. 
The points plotted then correspond to the 
left-hand side of Eq.~\eqref{Eq:zexpansion}. 
The solid blue and red curves correspond to the fit 
described in the text evaluated in the continuum limit and with quark masses tuned to their physical 
values. The curves are plotted for the range in $z$ corresponding to the physical range 
in $q^2$. The black dashed lines give the fit results evaluated for each set of gluon field 
configurations. }
\label{fig:fbothnopoleinz}
\end{figure}

The physical region of $q^2$ values for the $D \rightarrow K$ form factors is from 
$q^2=0$ to $q^2_{\text{max}}=(M_D-M_K)^2$. In the larger complex $t=q^2$ plane we expect a branch cut 
to appear from $t =t_+=(M_D+M_K)^2$ upwards, corresponding to $DK$ production in the 
crossed channel. Since $M_D+M_K$ has the value 2.36 GeV, we also have two poles below $M_D+M_K$ 
in this channel, corresponding to the vector $D_s^*$ meson and the scalar $D_{s0}^*$. 
We can map the cut-$t$ plane into the interior of the unit circle in 
$z$-space using a standard mapping (see, for example~\cite{Hill:2006ub}):
\begin{equation}
\label{eq:zspace}
  z(q^2,t_0)=\frac{\sqrt{t_+-q^2}-\sqrt{t_+-t_0}}{\sqrt{t_+-q^2}+\sqrt{t_+-t_0}}.
\end{equation}
Here $t_0$ is the point mapping on to $z=0$. We take $t_0=0$ for simplicity but will 
show below that we get the same result using other values of $t_0$. 

Since the form factor, with sub-threshold poles removed, is analytic we can fit to a 
polynomial form in $z$-space, modified by terms to allow for 
lattice discretisation and quark mass-mistuning effects~\cite{Koponen:2013tua}. 
We use the Bourreley-Caprini-Lellouch (BCL) parameterisation~\cite{Bourrely:2008za},  
\begin{eqnarray}\label{Eq:zexpansion}
  (1-\frac{q^2}{M^2_{D_{s0}^*}})f_0(q^2)&=&(1+L(m_l))\sum_{n=0}^{N-1}a_n^0z^n\\
  (1-\frac{q^2}{M^2_{D_s^*}})f_+(q^2)&=&(1+L(m_l))\times \nonumber \\&&\sum_{n=0}^{N-1}a_n^+\Big(z^n-\frac{n}{N}(-1)^{n-N}z^N\Big) \nonumber.
\end{eqnarray}
We now describe each piece of this fit form in turn. 

The factors of $1-q^2/M^2$ on the left-hand side of Eq.~\eqref{Eq:zexpansion} remove the 
sub-threshold poles in the scalar and vector channels discussed above. 
The physical masses of the two mesons that appear are
well known from experiment~\cite{pdg}. 
It is convenient in our calculation to use pole masses that are related 
to our $D$ meson masses to minimise uncertainties from the lattice spacing. 
We therefore use two simple formulae for the pole masses in Eq.~\eqref{Eq:zexpansion}: 
\begin{eqnarray}
\label{eq:polemasses}
M_{D_{s0}^*} = M_D + \Delta_0; \\
M_{D_{s}^*} = M_D + \Delta_1 \, , \nonumber
\end{eqnarray} 
where $\Delta_0 = M^{\mathrm{phys}}_{D_{s0}^*} - M^{\mathrm{phys}}_D$ and 
$\Delta_1 = M^{\mathrm{phys}}_{D_{s}^*} - M^{\mathrm{phys}}_D$ using 
mass values from~\cite{pdg}. $M_D^{\mathrm{phys}}$ is the average of the 
experimental masses for $D^+$ and $D^0$.  
The $M_D$ values in Eq.~\eqref{eq:polemasses} correspond to those 
from our lattice QCD calculation and 
the $\Delta$ values are constructed so that the pole 
masses in Eq.~\eqref{Eq:zexpansion} 
are equal to the appropriate experimental masses when our lattice 
results are extrapolated to the physical point. 

Figure~\ref{fig:fbothnopoleinz} shows our results for the form factors with 
poles removed (i.e. the left-hand side of Eq.~\eqref{Eq:zexpansion}) as a 
function of $z$. We can see that the $z$-dependence is very benign, almost linear with 
opposite sign gradients for $f_+$ and $f_0$, 
and there are no large deviations for discretisation effects or mistuning of sea 
quark masses. This enables a simple fit in $z$-space. 

On the right-hand side of Eq.~\eqref{Eq:zexpansion} we have a polynomial 
expansion in $z$ multiplied by a term that includes a chiral logarithm, 
a function of the light quark mass.  
We discuss the logarithmic term 
below but first describe the polynomial expansion. 
We include $N$ powers of $z$ starting from $z^0$ and take each coefficient to 
be of the form
\begin{equation}\label{Eq:an}
  a_n^{0,+} = (1+\mathcal{N}^{0,+}_n)\times\sum^{N_{j}-1}_{j=0}d_{jn}^{0,+}\Big(\frac{am_c^{\text{val}}}{\pi}\Big)^{2j} \, .
\end{equation}
We take $N=N_{j}=3$ for our preferred fit and will show below that our fits are stable to a change in the number of terms. Eq.~\eqref{Eq:an} allows for discretisation effects in the coefficients 
of the $z$-expansion when $j$ is non-zero. For the HISQ action, discretisation 
effects appear as even powers of the inverse lattice cut-off $a/\pi$. 
We allow for discretisation effects that are set by the charm quark 
mass $m_c$  since that is the largest energy scale here. 
The coefficients that set the discretisation effects, $d^{0,+}_{jn}$ for $j>0$ take 
independent values for different values of $n$ to allow for $z$-dependent ($q^2$-dependent) 
discretisation effects. They also take independent values for $f_+$ and $f_0$. 
In the absence of discretisation effects we have the kinematic constraint that 
$f_+(0)=f_0(0)$. Since we are using $t_0=0$, we can easily enforce this 
constraint by setting $d_{00}^+=d_{00}^0$. 

The $\mathcal{N}^{0,+}_n$ term encodes (non-logarithmic) 
dependence on quark masses, again with independent coefficients for each value of $n$.  
\begin{eqnarray}
\label{eq:mistuning}
  \mathcal{N}_n^{0,+}&=c_{s,n}^{\text{val},0,+}\delta_s^{\text{val}}+c_{l,n}^{\text{val},0,+}\delta_l^{\text{val}} \nonumber \\
&\hspace{2em}+c_{s,n}^{\text{sea},0,+}\delta_s^{\text{sea}}+2c_{l,n}^{\text{sea},0,+}\delta_l^{\text{sea}} \nonumber \\
  &+c_{c,n}^{0,+}\Big(\frac{M_{\eta_c}-M_{\eta_c}^{\text{phys}}}{M_{\eta_c}^{\text{phys}}}\Big).
\end{eqnarray}
In the first four terms, 
\begin{equation}
\label{eq:deltadef}
\delta_q = \frac{m_q-m_q^{\mathrm{tuned}}}{10m_s^{\mathrm{tuned}}} 
\end{equation}
takes account of the mistuning of the light and strange valence and sea quarks, relative to the 
tuned $s$ quark mass. Dividing by $m_s^{\mathrm{tuned}}$ makes this a physical, scale-independent, 
ratio and the factor of 10 matches this approximately to the usual expansion parameter 
in chiral perturbation theory. We tune the $m_s$ mass using the mass of the artificial 
$s\overline{s}$ pseudoscalar meson, the $\eta_s$, 
whose mass can be determined in terms of those of the $\pi$ 
and $K$ mesons in lattice QCD~\cite{Davies:2009tsa,Dowdall:2013rya}.  
$m_s^{\text{tuned}}$ is obtained on each ensemble from~\cite{Chakraborty:2014aca}  
\begin{equation}\label{mstuned}
  m_s^{\text{tuned}}=m_s^{\text{val}}\Bigg(\frac{M_{\eta_s}^{\text{phys}}}{M_{\eta_s}}\Bigg)^2,
\end{equation}
with $M_{\eta_s}^{\text{phys}}$ = 0.6885(20) GeV~\cite{Dowdall:2013rya}. 
We then determine the tuned $l$ quark mass from this using~\cite{Bazavov:2017lyh}
\begin{equation}\label{mlmsrat}
  \frac{m_s^{\text{tuned}}}{m_l^{\text{tuned}}}=27.18(10).
\end{equation}
The final term in Eq.~\eqref{eq:mistuning} allows for mistuning of the valence $c$ quark mass. 
We take $M_{\eta_c}^{\text{phys}}$ equal to 2.9766 GeV, allowing for the fact that the 
$\eta_c$ mass determined from quark-line connected diagrams (only) on the lattice 
differs from the experimental value~\cite{pdg} by 7 MeV~\cite{Hatton:2020qhk}. 

\begin{figure}
\includegraphics[width=0.48\textwidth]{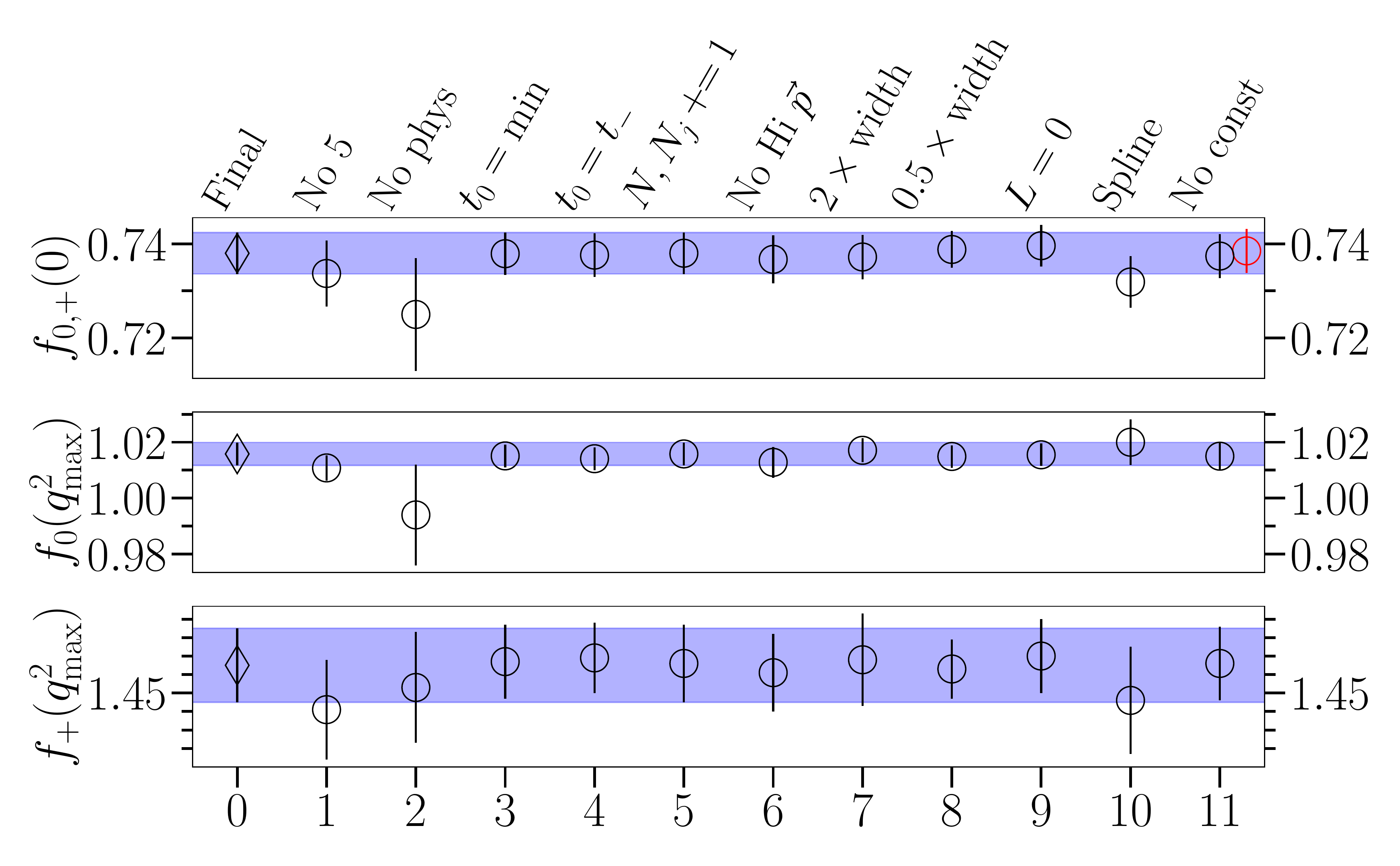}
\caption{Stability test of the $z$-expansion fit; 0 marks our final result. 
Test 1 removes all the results from gluon field configurations with $m_s/m_l=5$, so that only sets 1, 2 and 3 remain. Test 2 removes the results from sets 1, 2 and 3 and fits the others. 
Test 3 takes $t_0$ in the $q^2$ to $z$ mapping to the `minimum' 
prescription described in the text. 
Test 4 sets $t_0$ to $t_-$. Test 5 includes an extra term in the sums over $n$ up to 
$N$ and over $j$ up to $N_{j}$ (Eqs.~\eqref{Eq:zexpansion} and~\eqref{Eq:an}). 
Test 6 removes the highest momentum data point for each gluon field ensemble 
(and highest two on set 7 so that there are 
no results included with $q^2 < 0$). Test 7 doubles the width of all `$d$' priors 
(this decreased the Gaussian Bayes Factor), and 8 halves them. Test 9 sets the logarithmic 
factor $L(m_l)$ to zero (Eq.~\eqref{eq:logs}). Test 10 shows the results of a completely 
different kind of fit, a cubic spline fit in $q^2$ 
discussed in Section~\ref{sec:spline}. Test 11 removes the $f_0(0)=f_+(0)$ constraint, 
in this case the black point is $f_0(0)$ and the red is $f_+(0)$.}
\label{fig:extrapstab}
\end{figure}

Returning to Eq.~\eqref{Eq:zexpansion}, the first term on the right-hand side allows for the 
chiral logarithms expected from hard pion chiral perturbation theory~\cite{Bijnens:2010jg}. 
Following~\cite{Bouchard:2014ypa} 
 we include a chiral logarithm term multiplying the polynomial in $z$ for both $f_+$ and 
$f_0$. Because our light quark masses are small (with maximum $m_l/m_s=0.2$) the $K$ meson 
mass changes very little between different values of $m_l$. We therefore only include 
the chiral logarithm associated with the $\pi$ meson mass: 
  \begin{equation}
\label{eq:logs}
  L(m_l) =  - \frac{9g^2}{8}x_{\pi}\Big(\log x_{\pi}+\delta_{FV}\Big),
\end{equation}
where $x_{\pi}=M_{\pi}^2/\Lambda_{\chi}^2$, with $\Lambda_{\chi}$ the chiral scale of 
$4\pi f_{\pi}$. We rewrite $x_{\pi}$ in terms of quark masses as $m_l/(5.63m_s^{\mathrm{tuned}})$, 
using the ratio of $\Lambda_{\chi}$ to $M_{\eta_s}$ to evaluate the chiral logarithm 
accurately. 
$\delta_{FV}$ above is a finite-volume correction, calculated for each ensemble 
at the pion mass (See Eq.(47) of~\cite{Bernard:2001yj}). $\delta_{FV}$ has negligible effect in our fit.
We take the $DD^*\pi$ coupling, $g=0.570(6)$ from~\cite{Lees:2013uxa}. 
As shown in Eq.~\eqref{eq:mistuning} we include other terms in our fit, independently for 
each $z$-expansion coefficient, to allow for (analytic) dependence on 
$m_l$ from chiral perturbation theory. Our fit is not able to distinguish between linear 
and logarithmic dependence and so, as we will show below, gives the same result if the 
chiral logarithm of Eq.~\eqref{eq:logs} is dropped. We include it in our preferred fit, however.   

The priors on the $d_{0n}$ in Eq.~\eqref{Eq:an} that give the $z$-expansion coefficients in 
the continuum limit are taken to be $0\pm 2$. All other $d$ coefficients, that set the 
discretisation effects, are given prior 
$0\pm1$. The $c$ coefficients in Eq.~\eqref{eq:mistuning} that account for valence mass mistuning 
are given priors $0 \pm 1$; those that correspond to the smaller sea quark mass effects are given 
prior $0\pm 0.5$.  
An Empirical Bayes study~\cite{Lepage:2001ym} suggests that our priors are conservative. 

Our preferred fit, as described above, returns a $\chi^2/\mathrm{dof}$ of 0.67 with 64 
degrees of freedom. 
The stability of this fit against a variety of changes is demonstrated in Figure~\ref{fig:extrapstab}.  
We show the impact of omitting sets of lattice results, 
changing the numbers of terms in the $z$-expansion and the number of discretisation effects 
considered as well as doubling and halving the prior widths on all of the 
$d$ coefficients. Modifications to the fit in which we drop the logarithmic term 
of Eq.~\eqref{eq:logs} or remove the constraint that $f_+(0)=f_0(0)$ are tested.  
We also show the impact of changing $t_0$ from zero to 
the choice $t_0=t_+\Big(1-\sqrt{1-\frac{t_-}{t_+}}\Big)$, which minimises the maximum magnitude 
of $z$ as well as the choice $t_0=t_-\equiv(M_D-M_K)^2$. In both of these two cases 
we implement the constraint that $f_+(0)=f_0(0)$ by setting the difference between them 
equal to a parameter with prior $0\pm 1\times 10^{-6}$. These two different values of $t_0$ 
correspond to different ranges for the fit in $z$-space with the $q^2$ distribution 
mapped very differently into $z$-space. The good agreement is a strong validation of 
the $z$-expansion approach. In Section~\ref{sec:spline} we consider a completely different 
kind of fit, to cubic splines in $q^2$ space, and compare the results of that also in 
Figure~\ref{fig:extrapstab}.  
Our fit result is stable against all of these changes, although the uncertainties 
increase significantly if the lattice results for physical $m_l$ values (sets 1, 2 and 3)
are dropped. Dropping all of the lattice results for the unphysical $m_l$ values (sets 4--8) 
also increases the 
uncertainties but to a lesser extent. 
Note that dropping specific single lattices makes very little difference;  
the finest lattices (set 8) have almost no impact on the fit result. 

In the next section we compare our $z$-expansion fit to a fit in $q^2$-space using cubic splines. 

\subsubsection{Using a cubic spline in $q^2$}
\label{sec:spline}

\begin{figure}
\hspace{-30pt}
\includegraphics[width=0.48\textwidth]{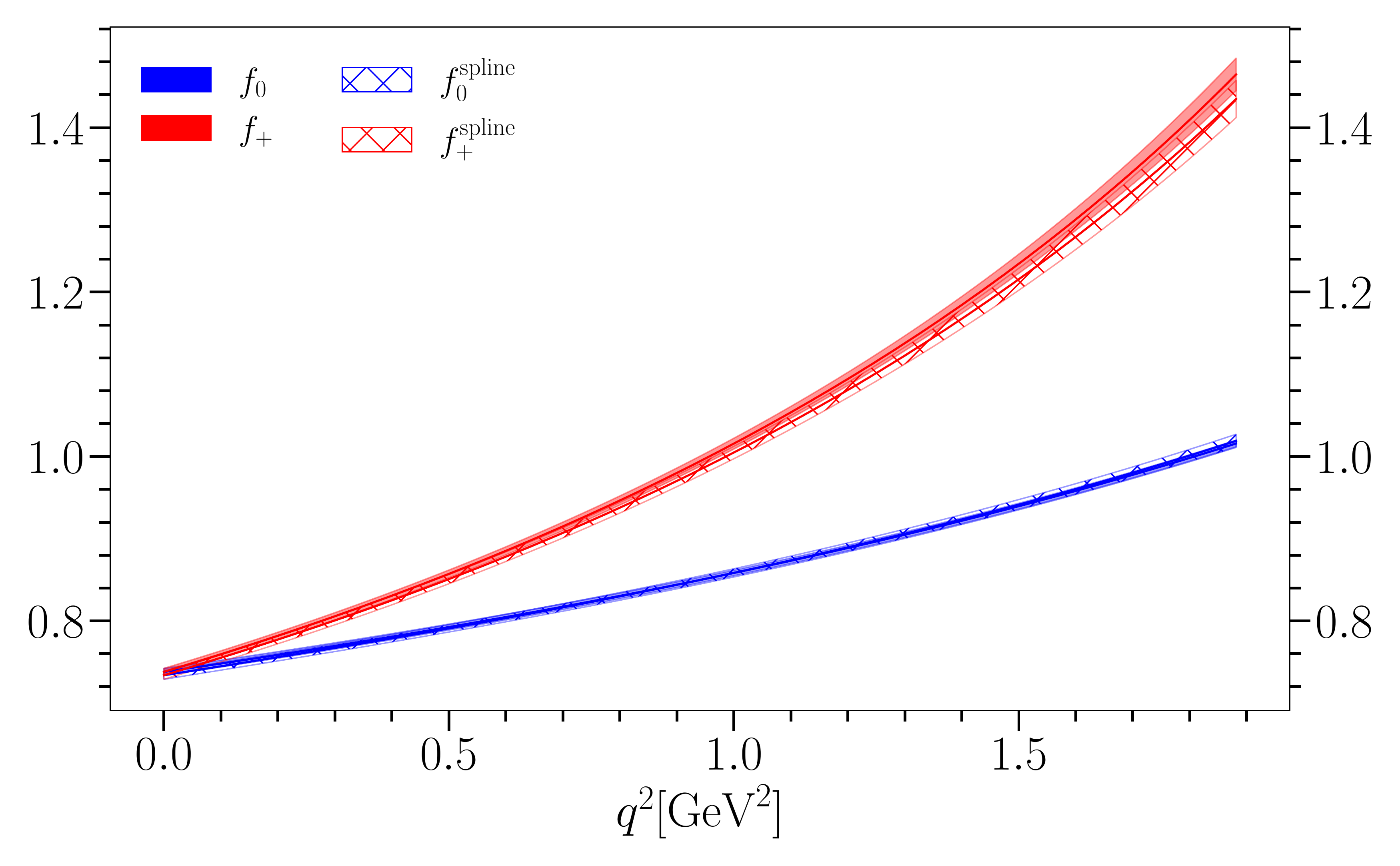}
\caption{A comparison of the $f_+$ and $f_0$ form factors, at the physical point ($a=0$ and physical 
quark masses), obtained from our preferred $z$-expansion fit of Section~\ref{sec:ffphysz} and from a cubic spline fit 
in $q^2$ of Section~\ref{sec:spline}. }
\label{fig:f0fpspline}
\end{figure}

There are choices to be made in implementing a $z$-expansion, 
from the choice of $t_0$ in the $q^2$ to $z$ mapping to the prefactors in 
front of the polynomial in $z$ 
(compare the form we use in Eq.~\eqref{Eq:zexpansion} to that used 
for the shape parameters in Eq.~\eqref{Eq:exp_zexpansion}). 
Here, since we have precise lattice QCD results over the full $q^2$ range 
of the decay, we can test a completely model-independent approach to 
the fit. 
Using cubic splines allows us to fit a very general function directly in $q^2$ space. 
We use the Steffen spline~\cite{Steffen:1990} to do this and denote each spline function, $g_i(q^2)$. 
After removing the expected pole, as described in Section~\ref{sec:ffphysz}, and including 
the chiral logarithm term of Eq.~\eqref{eq:logs}, we use a spline function $g_0$ to describe 
the physical dependence of each form factor on $q^2$ and further spline functions to account 
for discretisation and quark mass mistuning effects. The fit forms are given by:  
\begin{eqnarray}\label{Eq:spline-fit}
  (1-\frac{q^2}{M^2_{D_{s0}^*}})f_0(q^2)&=&(1+L(m_l))\times \\
&&\hspace{-4em}\left(g_0^0(q^2) + \sum_{j=1}^{N_j-1} \left[g_j^0(q^2)\left(\frac{am_c}{\pi}\right)^{2j} + \mathcal{N}^0\right]\right) ; \nonumber \\
  (1-\frac{q^2}{M^2_{D_s^*}})f_+(q^2)&=&(1+L(m_l))\times \nonumber \\
&&\hspace{-4em}\left(g_0^+(q^2) + \sum_{j=1}^{N_j-1} \left[g_j^+(q^2)\left(\frac{am_c}{\pi}\right)^{2j} + \mathcal{N}^+\right]\right) . \nonumber 
\end{eqnarray}
We take $N_j=2$ but taking $N_j=3$ gives no significant difference. 
For $\mathcal{N}$ we use further spline functions:
\begin{eqnarray}
\label{eq:Nspline}
  \mathcal{N}^{0,+}&=g_{s}^{\text{val},0,+}\delta_s^{\text{val}}+g_{l}^{\text{val},0,+}\delta_l^{\text{val}} \nonumber \\
&\hspace{2em}+g_{s}^{\text{sea},0,+}\delta_s^{\text{sea}}+2g_{l}^{\text{sea},0,+}\delta_l^{\text{sea}} \nonumber \\
  &+g_{c}^{0,+}\Big(\frac{M_{\eta_c}-M_{\eta_c}^{\text{phys}}}{M_{\eta_c}^{\text{phys}}}\Big).
\end{eqnarray}
The definitions of $\delta_l$ and $\delta_s$ are given in Eq.~\eqref{eq:deltadef}.   

All of the spline functions use the same four knots, positioned at $q^2$ values at either end 
of our range of results and with two values in between. 
This gives knot positions at \{-3.25, -1.5, 0.25, 2.0\} $\text{GeV}^2$. 
We take priors on the values of the spline functions at these knots. For $g_0^{0,+}$, 
which give 
the form factors in the continuum limit at physical quark masses, we take 0.75(15). 
This is informed by the range of the raw lattice results with pole removed 
(see Figure~\ref{fig:fbothnopoleinz}). The priors for the $g_j$, $g_s$ and $g_l$ 
are taken to be 0.0(5) and for the $g_c$ 0.0(1.0). 

The spline fit returns a $\chi^2/\mathrm{dof}$ value of 0.66 for 65 degrees of freedom. 
The form factors at the 
physical point can then be reconstructed from the $g_0(q^2)$ spline functions along with 
the $(1+L(m_l))$ and pole factors. A comparison of the form factors at the physical point 
with those from our $z$-expansion fit of Section~\ref{sec:ffphysz} is shown in 
Figure~\ref{fig:f0fpspline}. We see good agreement across the $q^2$ range.    
The cubic spline results are slightly less accurate (see also Figure~\ref{fig:extrapstab}) 
but the cubic splines also explore non-analytic functions of $q^2$ that we do not 
expect to contribute to the form factors. This is why we prefer the $z$-expansion fit results. 

\begin{figure}
\hspace{-30pt}
\includegraphics[width=0.48\textwidth]{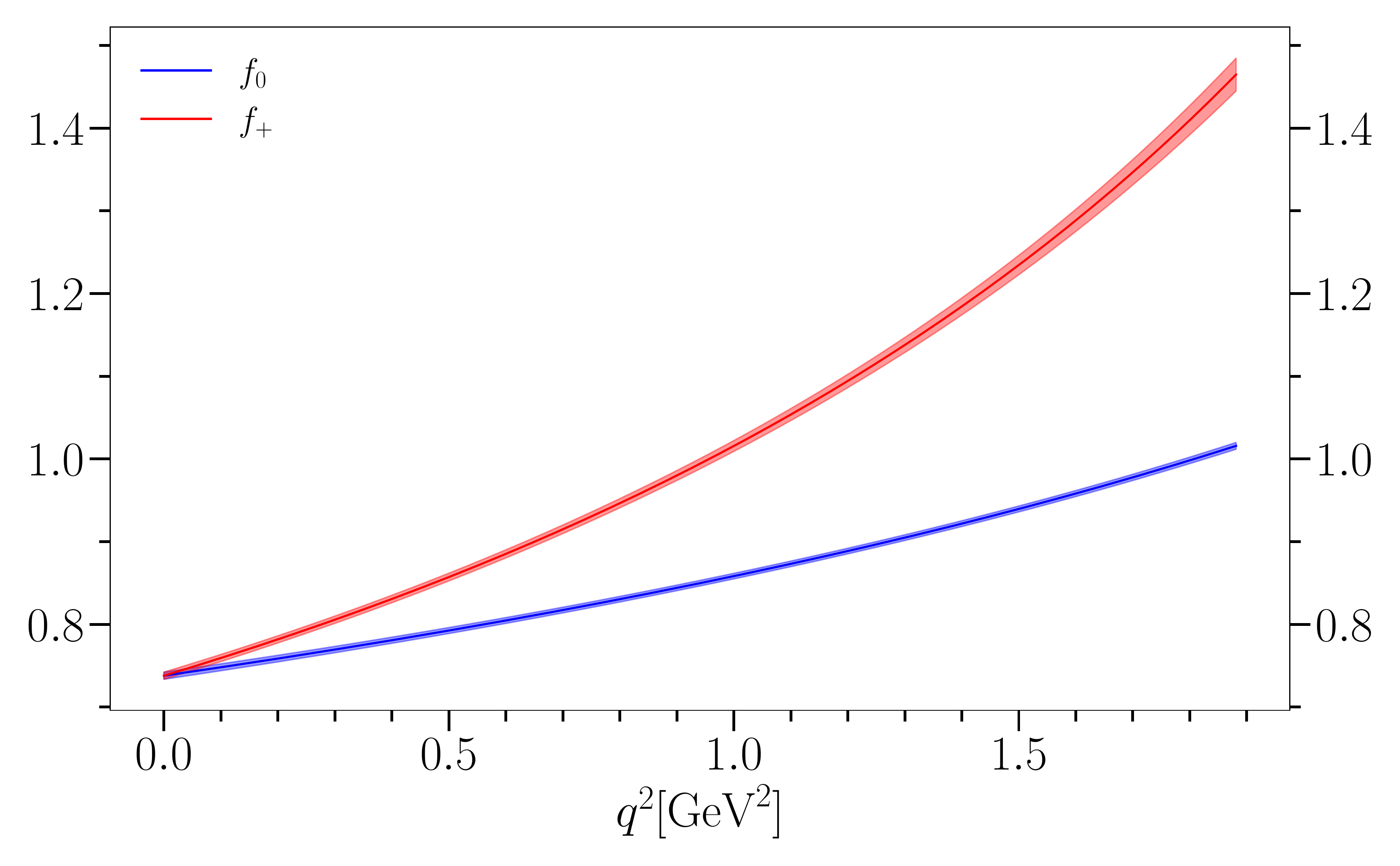}
\caption{Our results for the $f_+$ and $f_0$ form factors, at the physical point ($a=0$ and physical 
quark masses), as a function of squared momentum-transfer, $q^2$. }
\label{fig:f0fpinqsq}
\end{figure}

\begin{table*}
  \caption{Values and uncertainties for the fit coefficients $a_n^{0,+}$, 
pole masses, and chiral logarithmic term 
$(1+L(m_l))$ for the reconstruction of our form factors at the physical point as a function 
of $q^2$ from Eq.~\eqref{Eq:ff-cont}. The correlation matrix between these parameters is given 
below the row with their values. The pole masses are in GeV. The pole masses and $L(m_l)$ 
are very slightly correlated due to the way the fit function is constructed. 
These correlations are too small to have any meaningful effect on the fit, 
but we include them for completeness in reconstructing our results.}
  \begin{center} 
    \begin{tabular}{c c c c c c c c c}
      \hline
      $a_0^0$&$a_1^0$&$a_2^0$&$a_0^+$&$a_1^+$&$a_2^+$&$M^{\text{phys}}_{D^0_s}$& $M^{\text{phys}}_{D^*_s}$&$(1+L(m_l))$ \\ [0.5ex]
      0.7292(43)&0.825(80)&0.72(50)&0.7292(43)&-0.95(10)&1.1(1.3)&2.31780(50)&2.11220(40)&1.01200(26)\\ [1ex]
 \hline 

 1.00000&0.73103&0.51757&1.00000&0.29251&0.02299&-0.00023&-0.00005&-0.04904\\ [0.5ex]
 &1.00000&0.90723&0.73103&0.49742&0.01488&-0.01619&0.00001&-0.00795\\ [0.5ex]
 &&1.00000&0.51757&0.52335&0.00600&0.00368&0.00003&-0.00222\\ [0.5ex]
 &&&1.00000&0.29251&0.02299&-0.00023&-0.00005&-0.04904\\ [0.5ex]
 &&&&1.00000&0.49065&0.00007&-0.01488&0.00553\\ [0.5ex]
 &&&&&1.00000&0.00019&0.00362&-0.00017\\ [0.5ex]
 &&&&&&1.00000&-0.00000&-0.00000\\ [0.5ex]
 &&&&&&&1.00000&0.00000\\ [0.5ex]
 &&&&&&&&1.00000\\ [0.5ex]
      \hline
    \end{tabular}
  \end{center}
  \label{tab:ancoefficients}
\end{table*}

\section{Results for Form Factors}
\label{sec:results}

In Section~\ref{sec:ffphysz} we described how we fit the lattice form factor results, 
obtained at specific values of momentum for a set of lattice spacing values and quark masses, 
to a functional form (Eq.~\eqref{Eq:zexpansion}) that allows us to interpolate in $q^2$ and 
extrapolate to zero lattice spacing and physical quark mass values. To obtain the form factor 
at the physical point we set $\mathcal{N}_n$ and $a$ to zero in Eq.~\eqref{Eq:an}, 
so that $a_n^{0,+}=d_{0n}^{0,+}$. These values of $a_n$ are then substituted into 
Eq.~\eqref{Eq:zexpansion} with $L(m_l)$ evaluated for physical $m_l/m_s$ (Eq.~\eqref{mlmsrat}) 
and $\delta_{FV}$ set to zero.
$M_{D_s^*}$ and $M_{D_{s0}^*}$ take their experimental values~\cite{pdg} in the pole factors.   

In our lattice calculation we have degenerate $u$ and $d$ quarks with mass $m_l$ (for 
both valence and sea). 
Our physical point is defined as that where $m_l$ has a value equal to the physical 
average for $u$ and $d$ from Eq.~\eqref{mlmsrat}. 
We therefore do not distinguish between form factors for $D^0 \rightarrow K^-$ and 
$D^+ \rightarrow K^0$ decay. Our decay process is that for a $D$ with the average 
mass of $D^0$ and $D^+$ to that of a $K$ meson with the average mass of a $K^+$ and 
a $K^0$.  
When we determine $V_{cs}$ in Section~\ref{sec:vcs} we will include an uncertainty 
to allow for the fact that $m_u=m_d$ in our calculation.  

The form factors obtained in the continuum limit and with physical 
quark masses are plotted as a function of $q^2$ in Figure~\ref{fig:f0fpinqsq}.  

Table~\ref{tab:ancoefficients} gives the parameters needed to reconstruct our form factors 
at the physical point. As discussed above these are the $a_n^{0,+}$ coefficients of the 
$z$-expansion (Eq.~\eqref{eq:zspace}) in the $a \rightarrow 0$ limit with physical quark masses. 
The form factors are then reconstructed from
\begin{eqnarray}\label{Eq:ff-cont}
  f_0(q^2)&=&\frac{(1+L(m_l))}{(1-{q^2}/{M^2_{D_{s0}^*}})}\sum_{n=0}^{2}a_n^0z^n \, , \nonumber \\
  f_+(q^2)&=&\frac{(1+L(m_l))}{(1-{q^2}/{M^2_{D_{s}^*}})}\times \nonumber \\&&\sum_{n=0}^{2}a_n^+\Big(z^n+\frac{n}{3}(-1)^{n}z^3\Big) . 
\end{eqnarray}
Table~\ref{tab:ancoefficients} gives the coefficients and also their correlation matrix, including 
their correlations with $L(m_l)$ and the $D_s^*$ and $D_{s0}^*$ pole masses. 

\begin{figure}
\hspace{-30pt}
\includegraphics[width=0.48\textwidth]{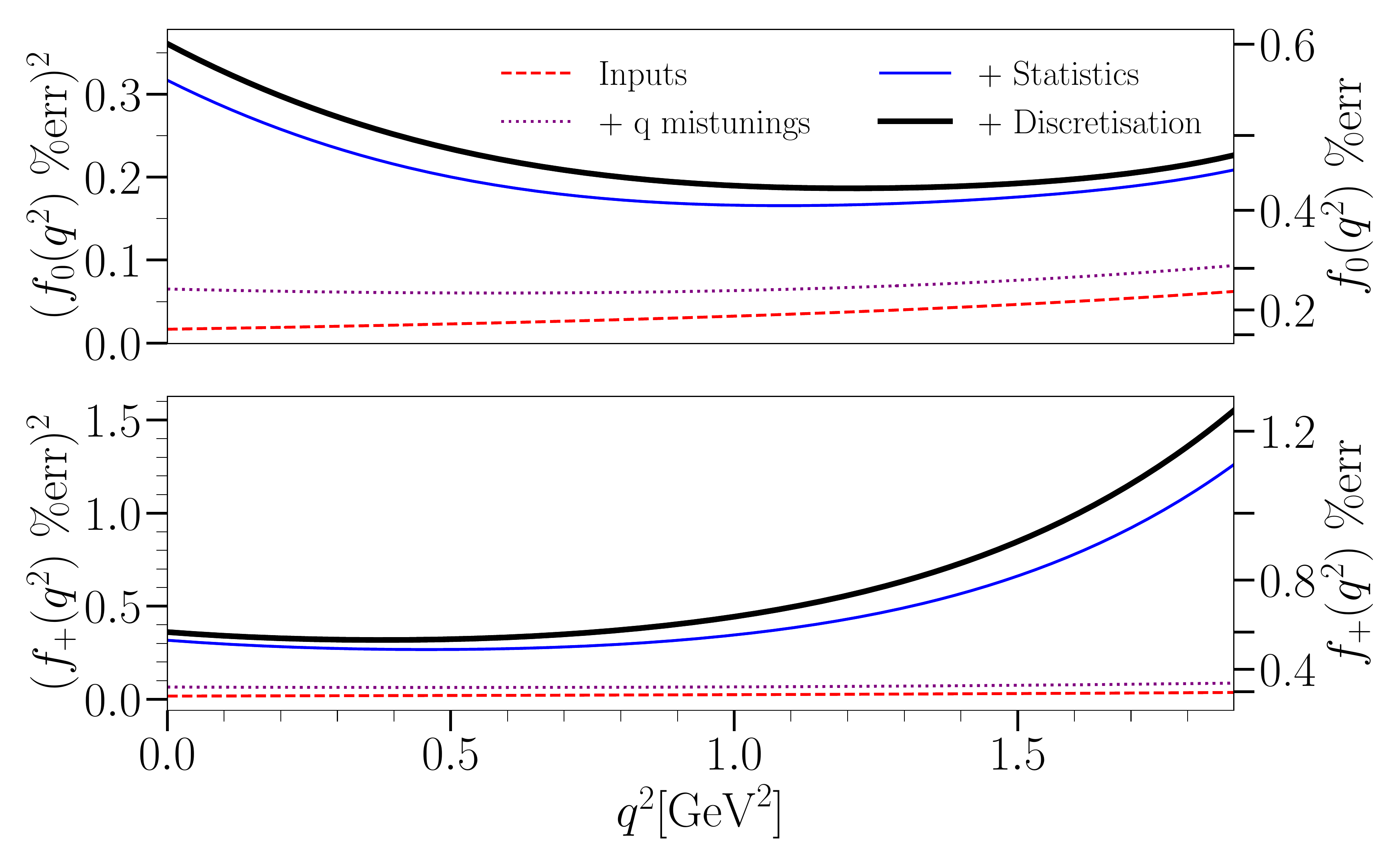}
\caption{Uncertainties for $f_0$ and $f_+$ (for $m_u=m_d$) as a function of $q^2$. 
The red line `Inputs' shows the uncertainties coming from fixed inputs, 
such as experimental meson masses used in the analysis. 
The purple line `q mistunings' adds in uncertainties arising from mistuning 
of valence and sea quark masses. The blue `Statistics' line further adds 
the statistical uncertainties from the lattice results (correlator fits). 
Finally, the black line (`Discretisation') gives the total uncertainty, now 
including the contribution from discretisation effects. 
These uncertainties add in quadrature, so we plot the squared percentage error 
and include an axis showing the corresponding percentage error on the right for clarity.}
\label{fig:f0fpluserr}
\end{figure}

Figure~\ref{fig:f0fpluserr} shows a breakdown of our errors as a function of $q^2$. 
We see that the total uncertainty is dominated by the statistical errors in the 
lattice QCD results. These can be reduced by collecting higher statistics, particularly 
on the finest lattice, set 8, where our statistical sample is not very large.  
The uncertainties are larger for $f_+$ than $f_0$; this is because of the way that 
the form factors are determined using Eqs.~\eqref{eq:D2Kff} and~\eqref{eq:scff}. 
The uncertainty for $f_+$ increases close to zero-recoil. This is because we have 
used the temporal vector current to determine $f_+$. Using a spatial vector current 
reduces this uncertainty, but it requires additional correlators to be calculated so 
we have not done that here. The region of $q^2$ close to zero-recoil is not important 
 for the determination of $V_{cs}$, as we shall see in Section~\ref{sec:vcs-diffrate}.  

\subsection{Comparison to previous results}
\label{sec:ffcompare}

\begin{figure}
\hspace{-30pt}
\includegraphics[width=0.48\textwidth]{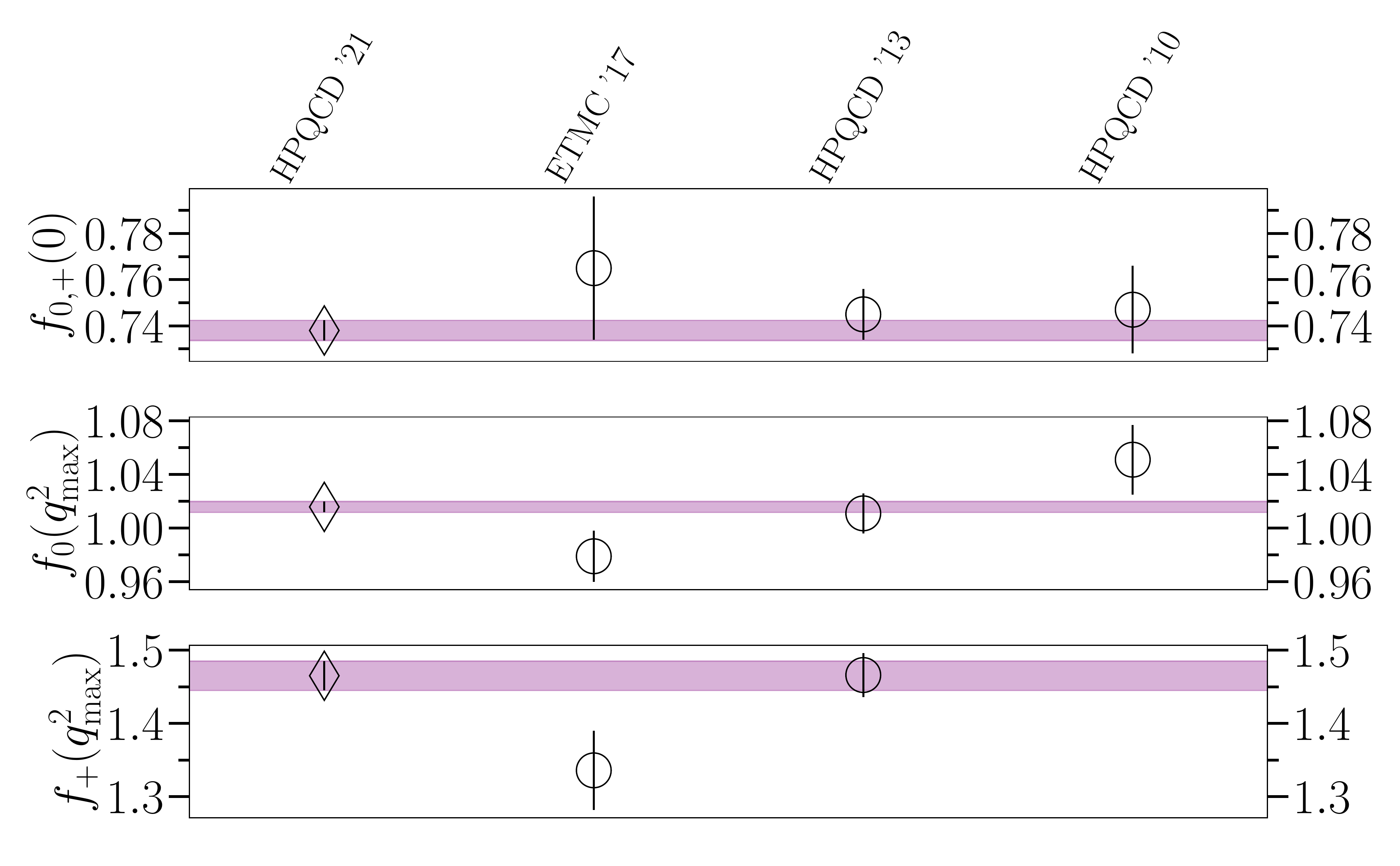}
\caption{Comparison of our lattice form factors at $q^2=0$ and $q^2_{\mathrm{max}}$ with 
earlier lattice QCD calculations. The points marked `HPQCD '10' are from~\cite{Na:2010uf}; 
the points marked `HPQCD '13' from~\cite{Koponen:2013tua} and the points marked `ETMC '17' from~\cite{Lubicz:2017syv, Riggio:2017zwh}.  
A preliminary analysis of the scalar form factor in~\cite{Li:2019phv} gives $f_0(0)=0.768(16)$, but we 
have not plotted that point. 
Our new results (Eq.~\eqref{eq:0maxvals}) are labelled `HPQCD '21' and demonstrate a significant 
improvement in uncertainty over earlier values.  
}
\label{fig:complat}
\end{figure}

We can compare our results for the $D \rightarrow K$ form factors to those
from earlier full lattice QCD calculations (all of which have $m_u=m_d$). 
In Figure~\ref{fig:complat} we show 
the comparison of results at the two ends of the physical $q^2$ range, $q^2=0$ and 
$q^2=q^2_{\mathrm{max}}=(M_D-M_K)^2$. For our results at $q^2_{\mathrm{max}}$ we 
use, as discussed above, $M_D=(M_{D^+}+M_{D^0})/2$ and $M_K=(M_{K^+}+M_{K^0})/2$.  
Previous results are from HPQCD:~\cite{Na:2010uf} calculating only the scalar form 
factor (in order to obtain the vector form factor at $q^2=0$ from 
$f_+(0)=f_0(0)$) and ~\cite{Koponen:2013tua} calculating the vector and scalar form 
factors across the full $q^2$ range of the decay. Both of these calculations were 
done on gluon field configurations that include 2+1 flavours of asqtad sea quarks. 
More recently ETMC has completed a calculation of the vector and scalar form factors 
across the full $q^2$ range using gluon field configurations with 2+1+1 flavours of 
twisted-mass sea quarks~\cite{Lubicz:2017syv, Riggio:2017zwh}. Our results here include 
2+1+1 flavours of HISQ sea quarks and are plotted as the leftmost results in 
Figure~\ref{fig:complat}. They show a significant improvement in uncertainty over the 
earlier results.  

Our results (plotted in Figure~\ref{fig:complat}) are
\begin{eqnarray}
\label{eq:0maxvals}
f_{+,0}(0) &=& 0.7380(44) \, , \nonumber\\
f_0(q^2_{\mathrm{max}}) &=& 1.0158(41) \, ,\nonumber \\ 
f_+(q^2_{\mathrm{max}}) &=& 1.465(20) \, . 
\end{eqnarray}
We observe a 2$\sigma$ tension with the results of~\cite{Lubicz:2017syv} at $q^2_{\mathrm{max}}$. 

\begin{figure}
\hspace{-30pt}
\includegraphics[width=0.48\textwidth]{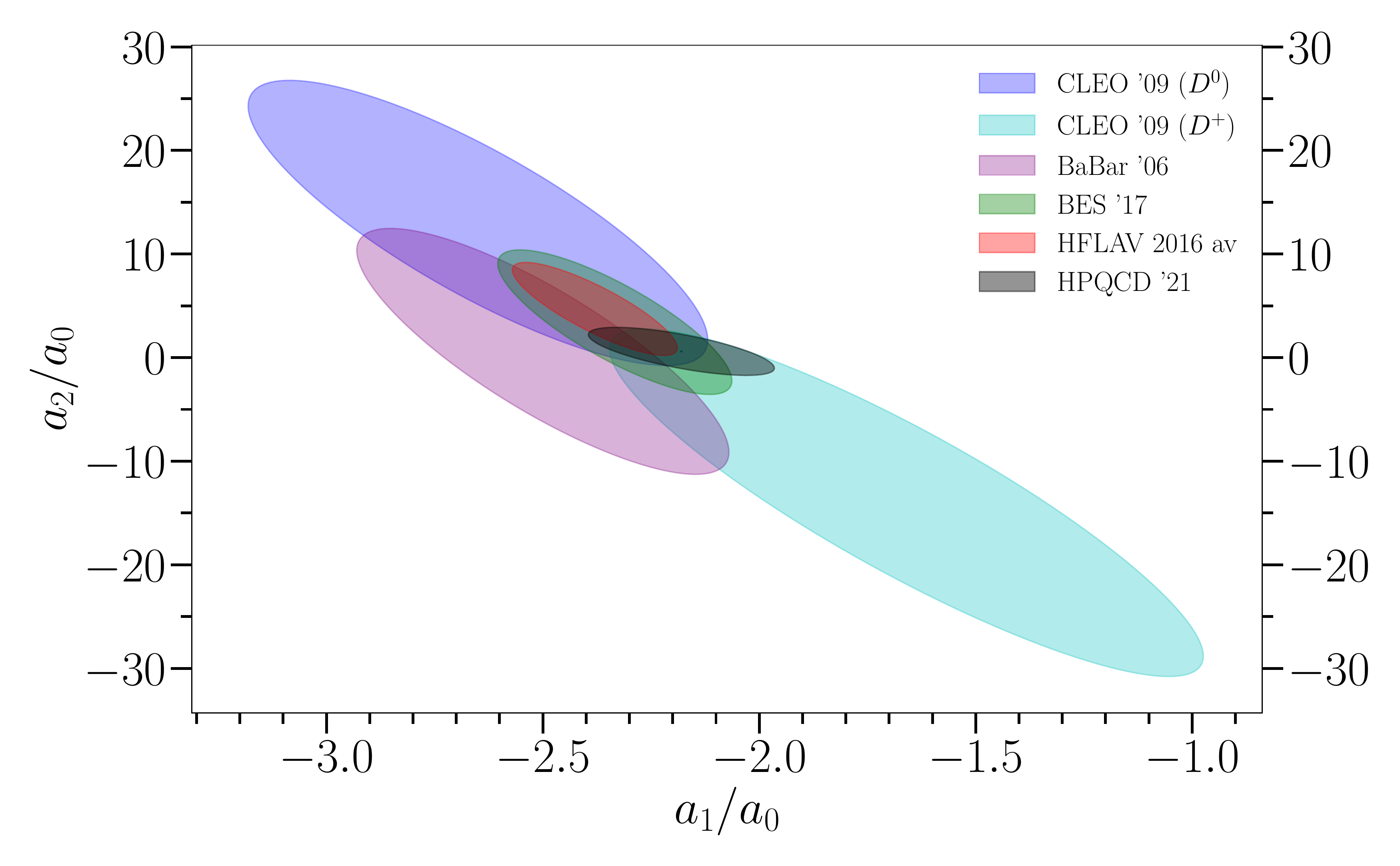}
\caption{Comparison of the shape of the vector form factor for $D \rightarrow K$ expressed 
in terms of ratios of the $z$-expansion coefficients $a_1$ and $a_2$ to $a_0$ 
for the fit form of Eq.~\eqref{Eq:exp_zexpansion}. 
Ellipses give the 68\% confidence limits ($\Delta \chi^2=2.3$). 
Experimental results are from~\cite{Besson:2009uv,Aubert:2007wg,Ablikim:2015ixa,Amhis:2016xyh}. 
CLEO results are for $D^0 \to K^-e^+\nu_e$ (dark blue) and $D^+ \to \bar{K}^0e^+\nu_e$ (light blue); 
all other experimental data is for $D^0 \to K^-e^+\nu_e$. 
The HFLAV experimental average~\cite{Amhis:2016xyh} 
is given as the red ellipse. Our results here are given by the black ellipse, 
showing good agreement. }
\label{fig:ellipse}
\end{figure}

Since the experimental differential rate for light leptons is proportional to the square of the 
vector form factor (Eq.~\eqref{eq:diffdecayrate}), the form factor shape can be determined 
from experiment. The experimental shape parameters come from a $z$-expansion fit but from a 
somewhat different one to the one that we have used here. To make a comparison we therefore 
need to fit our results in terms of the $z$-expansion used by the experiments. 
We do this by a `refitting' procedure that we describe in Appendix~\ref{App:expfitform}. 
The fit form used by the experiments is~\cite{Hill:2006ub} 
\begin{equation}\label{Eq:exp_zexpansion}
  f_+(q^2)=\frac{1}{z(q^2,t_0=M^2_{D_{s}^{*}})\phi(q^2)}\sum_{n=0}^{N-1}a_nz^n,
\end{equation}
where $\phi$ is an `outer function' given in Eq.~\eqref{eq:outerf} and $t_0$ is taken 
to be the value which minimises the maximum value of $z$ in the $q^2$ to $z$ 
mapping (Eq.~\eqref{eq:zspace}). 
The ratios $a_1/a_0$ and $a_2/a_0$ and their correlation coefficient 
then define the shape of the vector 
form factor. Experimental results from~\cite{Besson:2009uv,Aubert:2007wg,Ablikim:2015ixa} are plotted 
in Figure~\ref{fig:ellipse}.  

By fitting our form factors at the physical point (from Table~\ref{tab:ancoefficients}) 
to the form in Eq.~\eqref{Eq:exp_zexpansion} we obtain $a_1/a_0=-2.18(14)$ and 
$a_2/a_0=0.6(1.5)$ with a correlation coefficient of $\rho_{12}=-0.70$. 
As is clear from Figure~\ref{fig:ellipse} this agrees well with the experimental shape 
parameters, providing a good test of QCD. 
The HFLAV average~\cite{Amhis:2016xyh} of 
the shape parameters is more accurate than the individual experimental 
results giving $a_1/a_0=-2.38(13)$; $a_2/a_0=4.7(3.0)$ and $\rho_{12}=-0.19$. 
The agreement of our results with this average is particularly striking.

\subsection{Tests of lepton flavour universality}
\label{sec:lfu}

\begin{figure}
\hspace{-30pt}
\includegraphics[width=0.48\textwidth]{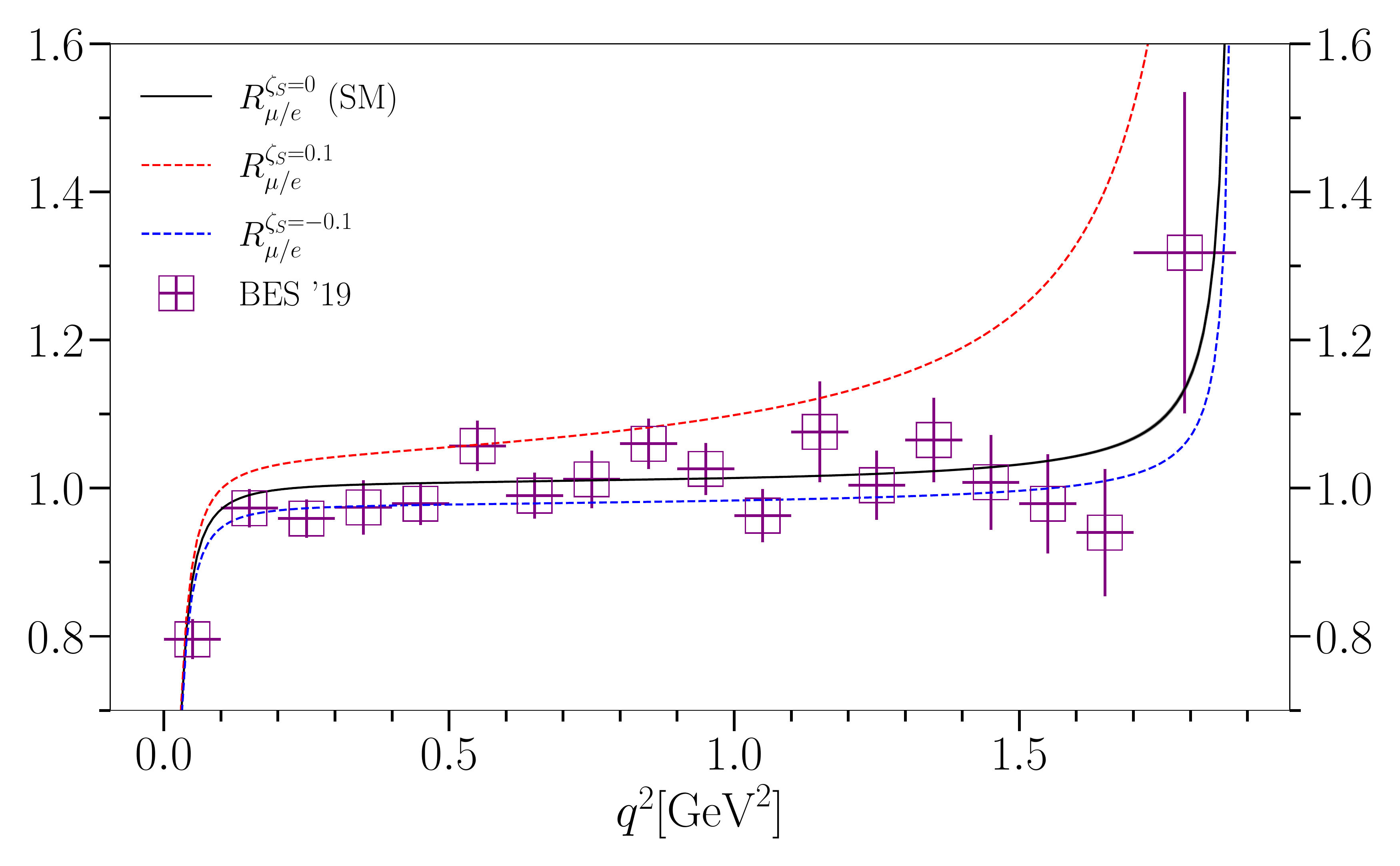}
\caption{Lepton flavour universality tests in $D \rightarrow K$ decay. The solid black 
curve as a function
of $q^2$ shows the 
Standard Model ratio of branching fractions for a muon in the final state to that for an electron 
obtained from our form factors using Eq.~\eqref{eq:diffdecayrate}. The width of the 
curve gives the 
(very small) uncertainty from our results. Possible QED effects are not included here. 
The points, with error bars, are from the BES experiment~\cite{Ablikim:2018evp}. For illustration the red and blue 
dashed lines show what the curve would look like in the presence of a new physics scalar 
coupling for the $\mu$ case (see Eq.~\eqref{eq:zeta} for definition of $\zeta_S$). }
\label{fig:Rmue}
\end{figure}

In the Standard Model the three charged leptons are copies of each other apart from 
having different masses. Hints are seen in experiment of violations of this 
lepton flavour universality in $B$ decays (for a review see~\cite{Bifani:2018zmi}) 
and this motivates a search for this also in $D$ decays~\cite{Fajfer:2015ixa,Fleischer:2019wlx}.  
We can only compare results with $\mu$ and $e$ in the final state for $D\rightarrow K$ 
decay because the production of $\tau$ leptons is kinematically forbidden. 
The BES experiment recently measured the ratio $R_{\mu/e}$ of branching fractions to $\mu$ and to $e$ 
as a function of $q^2$~\cite{Ablikim:2018evp}. We can calculate this ratio 
very accurately from our form factor results using Eq.~\eqref{eq:diffdecayrate},
because there is a lot of cancellation of uncertainties in the form factors in the ratio.  
If we ignore long-distance QED corrections (to be discussed below) we 
can compare the BES results to the curve derived from our form factors (solid 
black line) in Figure~\ref{fig:Rmue}. 
We see good agreement across the $q^2$ range. To quantify this 
agreement it would be necessary to have a correlation matrix for the experimental results. 
$R_{\mu/e}$ is smaller than 1 at small values of $q^2$, where the factor $(1-\epsilon)^2$ 
in Eq.~\eqref{eq:diffdecayrate} has 
an effect for the $\mu$. It is larger than 1 at large values of $q^2$
where the term containing the scalar form factor, $f_0$, contributes. 

The ratio of branching fractions to $\mu$ and to $e$, $R_{\mu/e}$, can be obtained by
integrating Eq.~\eqref{eq:diffdecayrate} from $q^2=m_{\ell}^2$ to $q^2_{\mathrm{max}}$. 
We take $q^2_{\mathrm{max}}$ from the $D$ and $K$ masses averaged over charged 
and neutral cases, although other choices make negligible difference. 
Our result for the ratio of branching fractions then has a 0.02\% uncertainty 
from lattice QCD. A larger source of uncertainty is the difference of long-distance 
QED corrections to the rate in the $\mu$ and $e$ cases. This could be a sizeable 
effect when there are electrically charged mesons in the final state, as in 
the BES experimental results which correspond to $D^0 \rightarrow K^-$ decay.  
Our result for $R_{\mu/e}$ is then 
\begin{equation}
\label{eq:rmue}
R_{\mu/e} = 0.9779(2)_{\mathrm{latt}}(50)_{\mathrm{EM}}, 
\end{equation}
allowing a 0.5\% uncertainty for the difference of QED corrections in 
the $D^0\rightarrow K^-$ case. 
Our $R_{\mu/e}$ agrees well with the BES result of 0.974(7)(12)~\cite{Ablikim:2018evp} 
but is much more accurate. 
We see some tension with the earlier ETMC result~\cite{Riggio:2017zwh} 
for this ratio using lattice QCD 
of $0.975(1)_{\mathrm{latt}}$. 

Violation of lepton flavour universality might be seen in comparison to 
the curve of Figure~\ref{fig:Rmue} 
with accurate 
enough experimental results, up to possible QED effects. 
We illustrate the impact of a new physics scalar coupling in 
the $\mu$ sector, $C_S^{(\mu)}$, with red and blue dashed lines. 
$C_S^{(\mu)}$ would multiply 
a new physics contribution to the effective Lagrangian consisting of a 
scalar $\overline{s}c$ current multiplying a $\overline{\nu}_{\mu}\mu$ current. 
Such a term affects the $D\rightarrow K$ differential rate, 
modifying the coefficient of the scalar form factor in Eq.~\eqref{eq:diffdecayrate} 
by a factor of $|(1+C_S^{(\mu)}q^2/(m_{\mu}(m_s-m_c)))|^2$~\cite{Fajfer:2015ixa} where 
$m_s$ and $m_c$ are the strange and charm quark masses. 
We show results for two possible 
real values of $C_S$ such that 
\begin{equation}
\label{eq:zeta}
\zeta_S\equiv\frac{C_S^{(\mu)}}{m_s-m_c} = \pm 0.1 \, \mathrm{GeV}^{-1}, 
\end{equation} 
which roughly encompass the range of variation of the central values of the 
BES data points from our Standard Model curve. 

\begin{figure}
\includegraphics[width=0.3\textwidth]{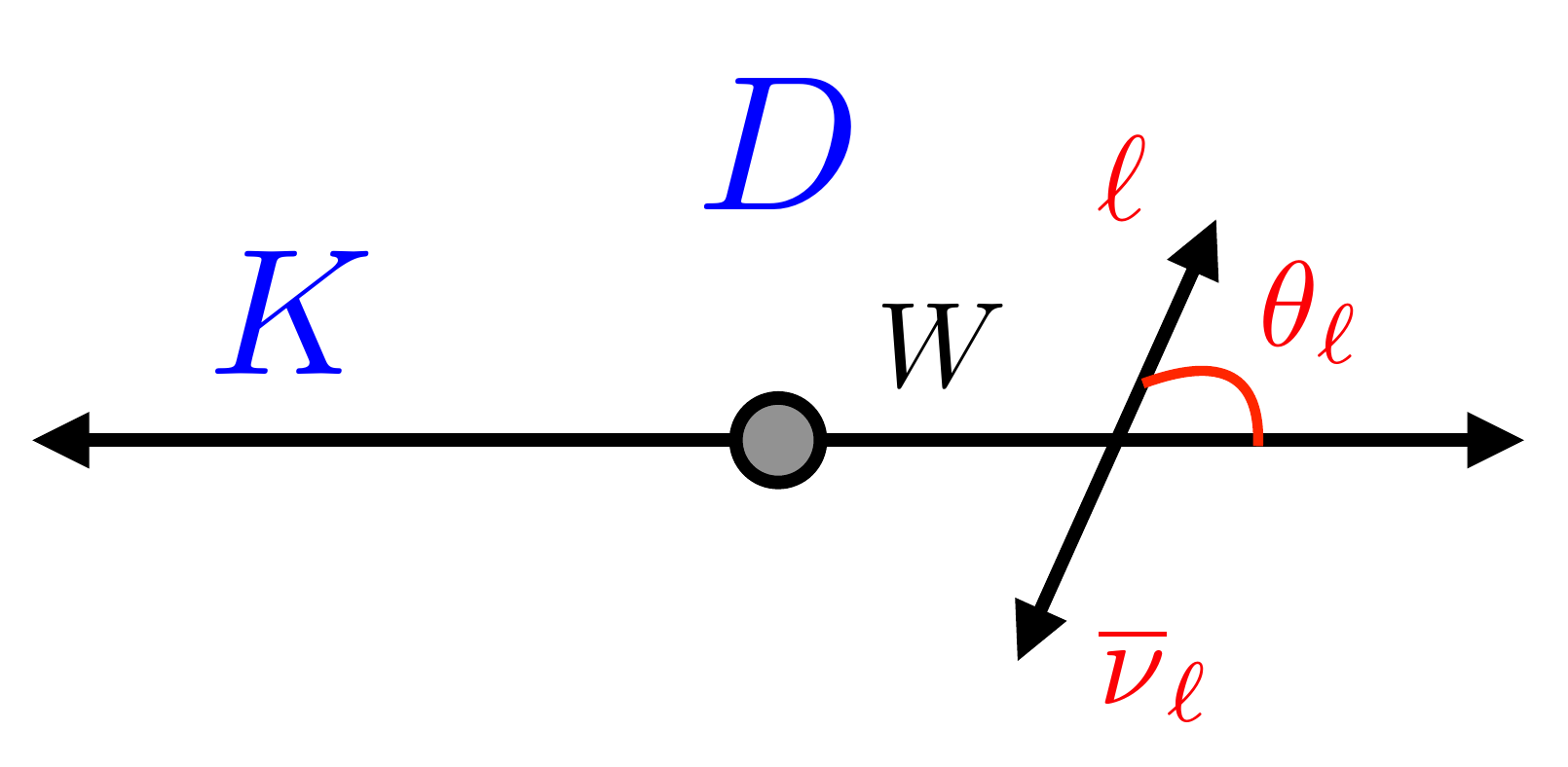}
\includegraphics[width=0.48\textwidth]{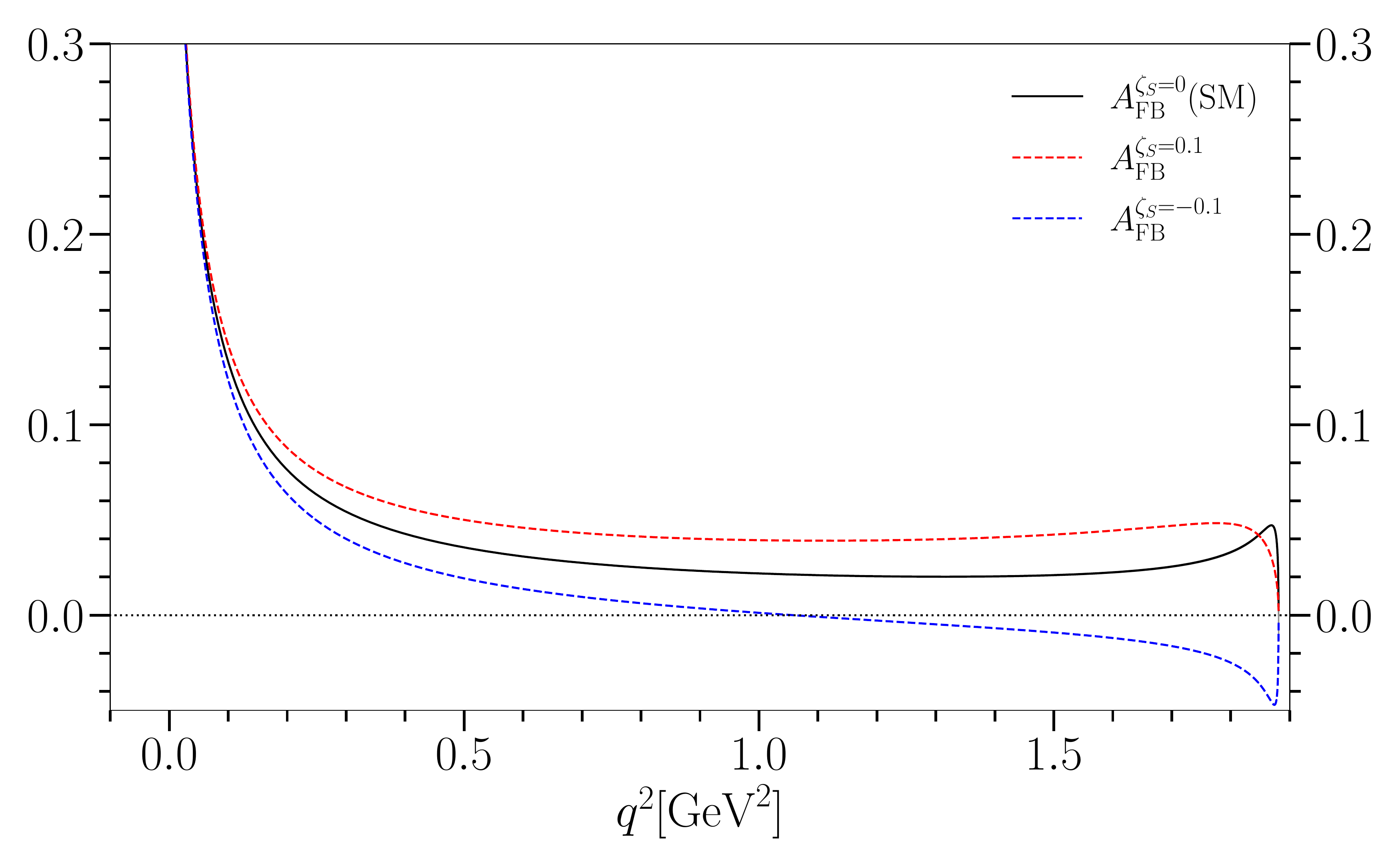}
\caption{The forward-backward asymmetry of the muon produced in $D \rightarrow K$ decay. 
This is defined with respect to the angle $\theta_{\ell}$ in the $W$ rest-frame indicated 
in the figure at the top. 
The solid black line shows the Standard Model result 
derived from our form factors, including the lattice QCD uncertainty but ignoring 
any uncertainty from possible QED corrections. 
For illustration the red and blue 
dashed lines show what the curve would look like in the presence of a new physics scalar 
coupling for the $\mu$ case (see text and Eq.~\eqref{eq:zeta}). }
\label{fig:AFB}
\end{figure}

Angular variables can also provide sensitive tests of the Standard Model and 
constraints on new physics. Figure~\ref{fig:AFB} 
plots the forward-backward asymmetry, $A_{FB}$, of the muon in $D \rightarrow K \mu \overline{\nu}$ 
decay as a function of $q^2$ in the Standard Model from our form factors (solid black line), 
ignoring possible QED corrections. This 
asymmetry is defined using the angle $\theta_{\ell}$ between the charged lepton momentum 
in the $W$ rest-frame and the $W$ momentum vector in the $D$ rest-frame. 
$\theta_{\ell}$ is shown in Figure~\ref{fig:AFB} and takes the range 0 to $\pi$.   
$A_{FB}$ is then defined as 
\begin{equation}
\label{eq:AFB}
A_{FB}^{(\ell)}(q^2) = -\frac{b_{\ell}}{d\Gamma^{(\ell)}/dq^2}
\end{equation}
where 
\begin{equation}
\label{eqbl}
\frac{d\Gamma^{(\ell)}}{dq^2d\cos\theta_{\ell}}=a_{\ell}(q^2) + b_{\ell}(q^2)\cos\theta_{\ell} + c_{\ell}(q^2)\cos^2\theta_{\ell}.
\end{equation}
$A_{FB}=0$ for massless leptons because only the helicity zero component of the $W$ can contribute. 
For massive leptons there is an interference term between scalar and vector form factor 
contributions~\cite{Fajfer:2015ixa}. Figure~\ref{fig:AFB} shows that this has a sizeable effect 
for muons in the final state, particularly close to $q^2=0$.  

$A_{FB}^{(\mu)}$ for $D\rightarrow K$ decay would be modified in the presence 
of a (real) scalar coupling from new physics 
because this affects the vector/scalar interference term. 
Figure~\ref{fig:AFB} shows the impact of two possible values of this coupling, 
as in Eq.~\eqref{eq:zeta} and Figure~\ref{fig:Rmue}. The impact of the new coupling is 
mainly at large $q^2$ values where $A_{FB}$ is small.  A positive value of $C_S^{(\mu)}$ 
(negative value of $\zeta_S$) can change the sign of $A_{FB}$ from that expected in 
the Standard Model at large $q^2$. 

\section{Determination of $|V_{cs}|$}
\label{sec:vcs}

Using the measured experimental rates for the $D\rightarrow K \ell \overline{\nu}$ decay we 
can determine the CKM element $|V_{cs}|$. The accuracy with which this can be done 
depends on the accuracy of both the experimental results and the accuracy of the lattice 
QCD form factors for the decay process. 
We show here that our improved form factor determination yields 
a significant improvement in the values of $V_{cs}$ obtained. We give three different methods 
for determining $V_{cs}$. Our preferred approach (Section~\ref{sec:vcs-diffrate}) 
is to use the experimental differential decay rate and Eq.~\eqref{eq:diffdecayrate}, integrated
over the $q^2$ bins used by the experiment~\cite{Koponen:2013tua}. 
This is the most direct approach, 
enabling use of the $q^2$ region where the experimental results 
are most accurate
and testing the $q^2$-dependence of the differential rate 
at the same time (although agreement here has already been 
demonstrated in Figure~\ref{fig:ellipse}). It requires experimental 
measurement of the differential 
rate with a covariance matrix for results in different bins and this is not always 
possible. We therefore also determine $V_{cs}$ in Section~\ref{sec:vcstotalbr} from 
the total rate, integrated 
over all $q^2$. In Section~\ref{sec:vcsfp0} we apply a third method that 
uses quoted experimental values from fitting the differential rate and  
extrapolating to $q^2=0$.  

Before giving details of these methods, we first discuss and estimate 
two further sources of systematic uncertainty beyond those of our calculated 
form factors and the experimental results: 

1. $m_u \ne m_d$. In determining $V_{cs}$ we will use our form factors obtained in QCD with $m_u=m_d$. 
The experimental results, however, correspond to the case with either valence $u$ quarks (for $D^0$ 
decay) or valence $d$ quarks (for $D^+$ decay). We therefore need to allow an uncertainty 
in our calculation for this mismatch. In determining the form factors at the physical point 
in Section~\ref{sec:ffphys} we set the physical value of the light quark mass, $m_l$, from 
Eq.~\eqref{mlmsrat}. We can test the effect of having a different light quark mass 
(corresponding to $u$ or $d$) by changing 
this condition. We take $m_d/m_u \approx 2$~\cite{pdg} so that 
$m_u/m_s \approx 2/(3\times 27.18)$ and $m_d/m_s \approx 4/(3\times 27.18)$ and compare 
to our original results using Eq.~\eqref{mlmsrat}. 
We find a change in our form factors of, at most, 0.15\%. Note that 
the calculation we really want to match to experiment changes only the light valence quark 
mass to $u$ or $d$, leaving the sea the same (with $u$ and $d$ quarks that match, to a linear 
approximation in quark mass, two quarks with mass $m_l$ equal to their average). 
To do this would require 
additional lattice calculations so here we simply take an additional uncertainty of 
0.15\% on our form factors (across the $q^2$ range) to account for this. This corresponds to 
0.25$\sigma$ at $q^2=0$ (see Eq.~\eqref{eq:0maxvals}). 

2. {\it QED}. Another issue that we must address in determining $V_{cs}$ is that of (long-distance) 
electromagnetic corrections. 
There are QED effects inside the mesons arising (mainly) from the valence quark electric 
charges. There are also effects from photon radiation, mainly from final-state interactions, that 
could be more sizeable for the case where a charged $K$ is produced. 
The experimental results include tests and corrections for radiated photons, 
to produce a photon-inclusive rate; 
this is typically done using PHOTOS~\cite{Barberio:1993qi} 
(see, for example, the discussions in~\cite{Aubert:2007wg, Besson:2009uv}). 
In Eq.~\eqref{eq:diffdecayrate} we include a factor of $(1+\delta_{EM})$ to allow 
for the effects of QED radiation as a $q^2$-independent uncertainty, since these 
effects have not been calculated. 
For $K \rightarrow \pi$ semileptonic decay, where the electromagnetic corrections have been 
calculated, results range from $\delta_{EM} \approx 0$ for neutral final state mesons to 
$\delta_{EM} \approx 0.7\%$~\cite{Antonelli:2010yf} for charged final 
state mesons and with small differences between $e$ and $\mu$ leptons in the final state. 
Here we will take independent uncertainties of 
$\delta_{EM}= \pm 1\%$ for the charged final state case ($D^0\rightarrow K^+$) and 
$\pm 0.5\%$ for the neutral final state case ($D^+\rightarrow K^0$). We will also take 
independent $\delta_{EM}$ for the case with final state $\mu$ from that with 
final state $e$, since these could differ. In our final 
result we will keep the QED uncertainty 
as a separate factor so that in the future it can be adjusted in the light of new information 
on these corrections (for example from lattice QCD+QED calculations~\cite{Sachrajda:2019uhh}). 

\begin{figure}
\hspace{-30pt}
\includegraphics[width=0.48\textwidth]{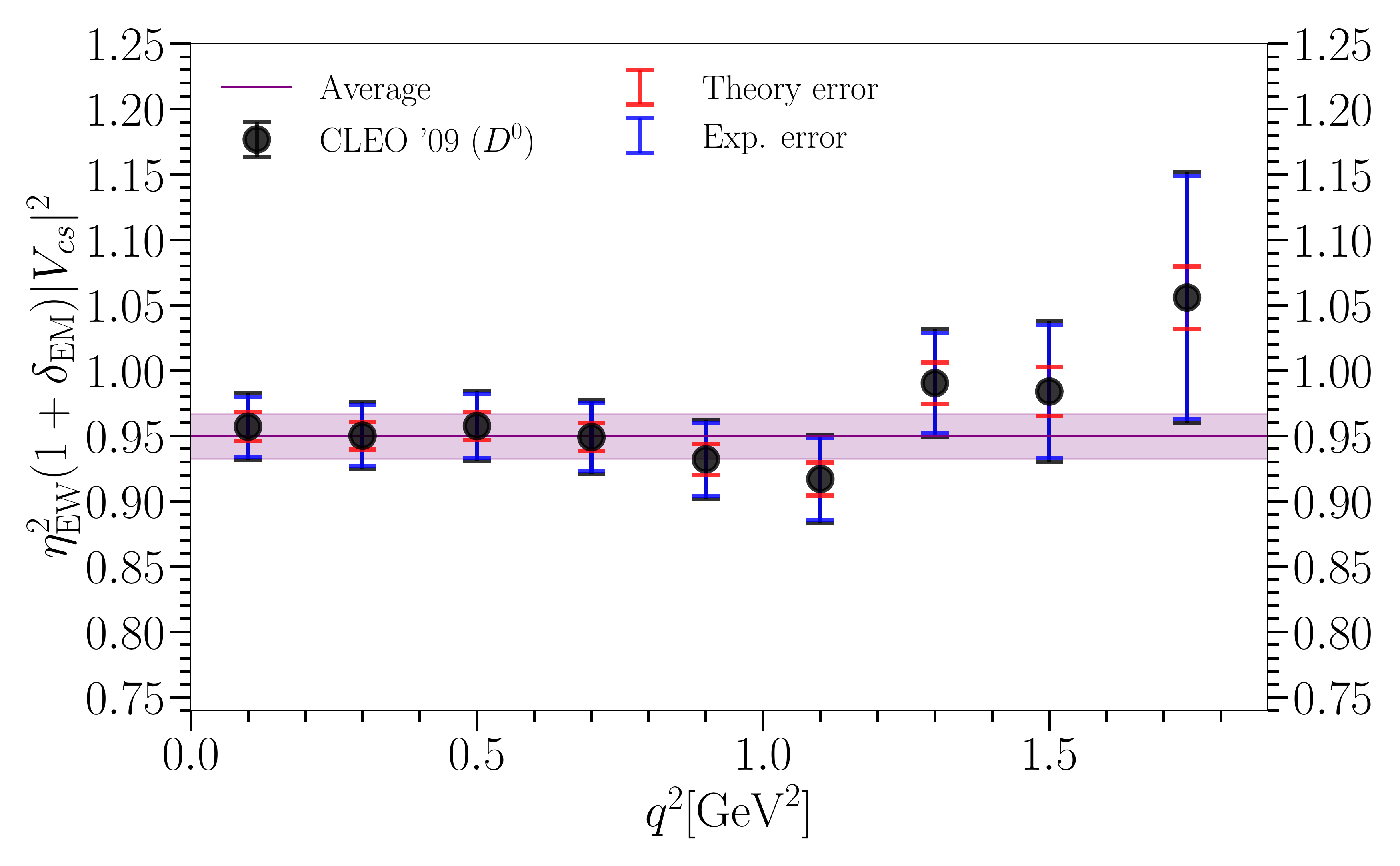}
\caption{Plot of the determination of $|\eta_{EW}V_{cs}|^2(1+\delta_{EM})$ 
per $q^2$ bin for CLEO $D^0$ results~\cite{Besson:2009uv}. 
The total uncertainty for each bin is given in black and this is broken down 
into experimental (blue) and theoretical (red) contributions, the latter coming from our form 
factors. Each data point is centred on the $q^2$ bin it corresponds to. Note that the uncertainties 
are correlated between $q^2$ bins. The purple band gives the weighted average 
for these data points, with all correlations included.}
\label{fig:Cleo1Vcsbybin}
\end{figure}

\begin{figure}
\hspace{-30pt}
\includegraphics[width=0.48\textwidth]{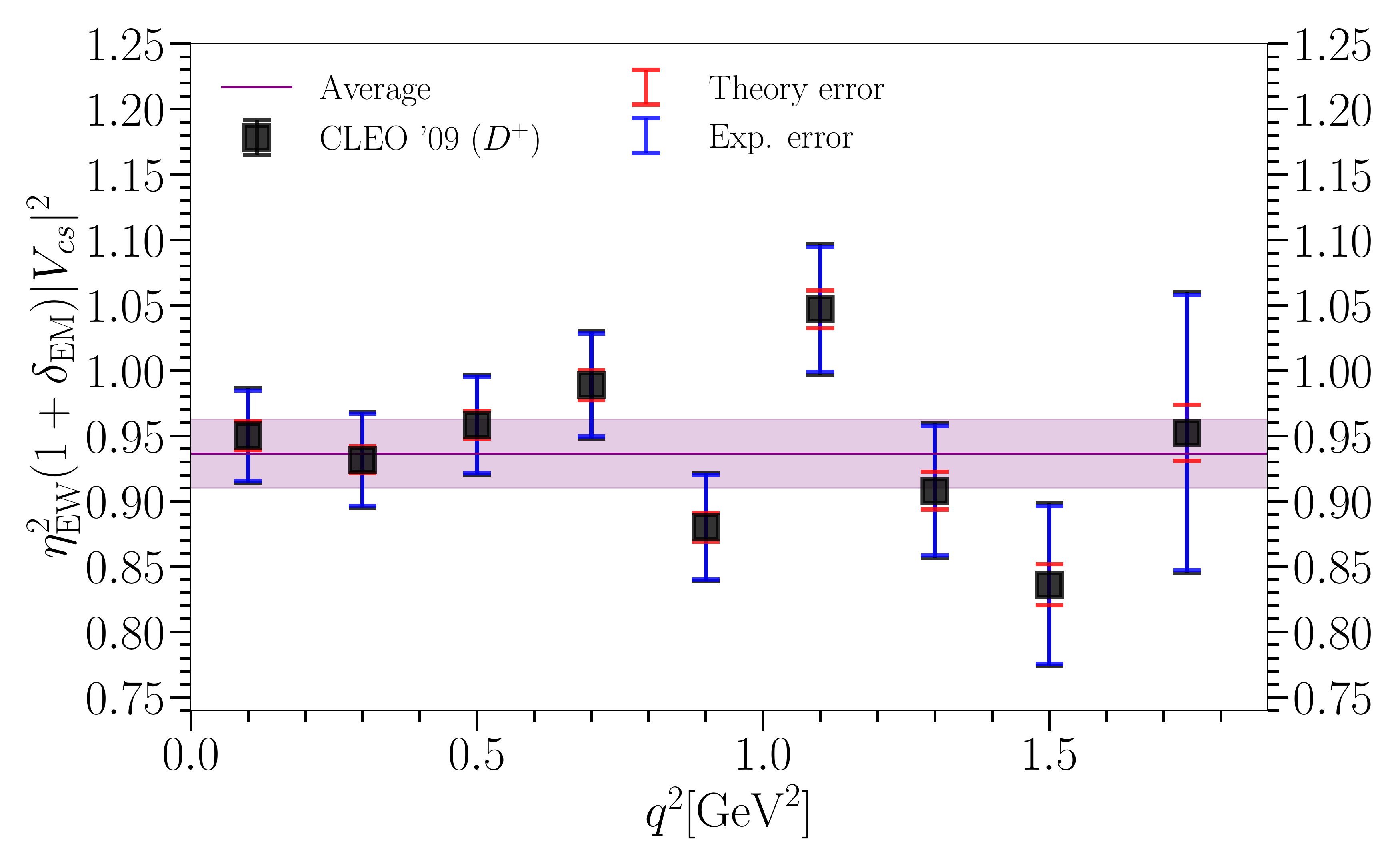}
\caption{Plot of the determination of $|\eta_{EW}V_{cs}|^2(1+\delta_{EM})$ 
per $q^2$ bin for CLEO $D^+$ results~\cite{Besson:2009uv}. 
The total uncertainty for each bin is given in black and this is broken down 
into experimental (blue) and theoretical (red) contributions, the latter coming from our form 
factors. Each data point is centred on the $q^2$ bin it corresponds to. Note that the uncertainties 
are correlated between $q^2$ bins. The purple band gives the weighted average 
for these data points, with all correlations included.}
\label{fig:Cleo2Vcsbybin}
\end{figure}

\begin{figure}
\hspace{-30pt}
\includegraphics[width=0.48\textwidth]{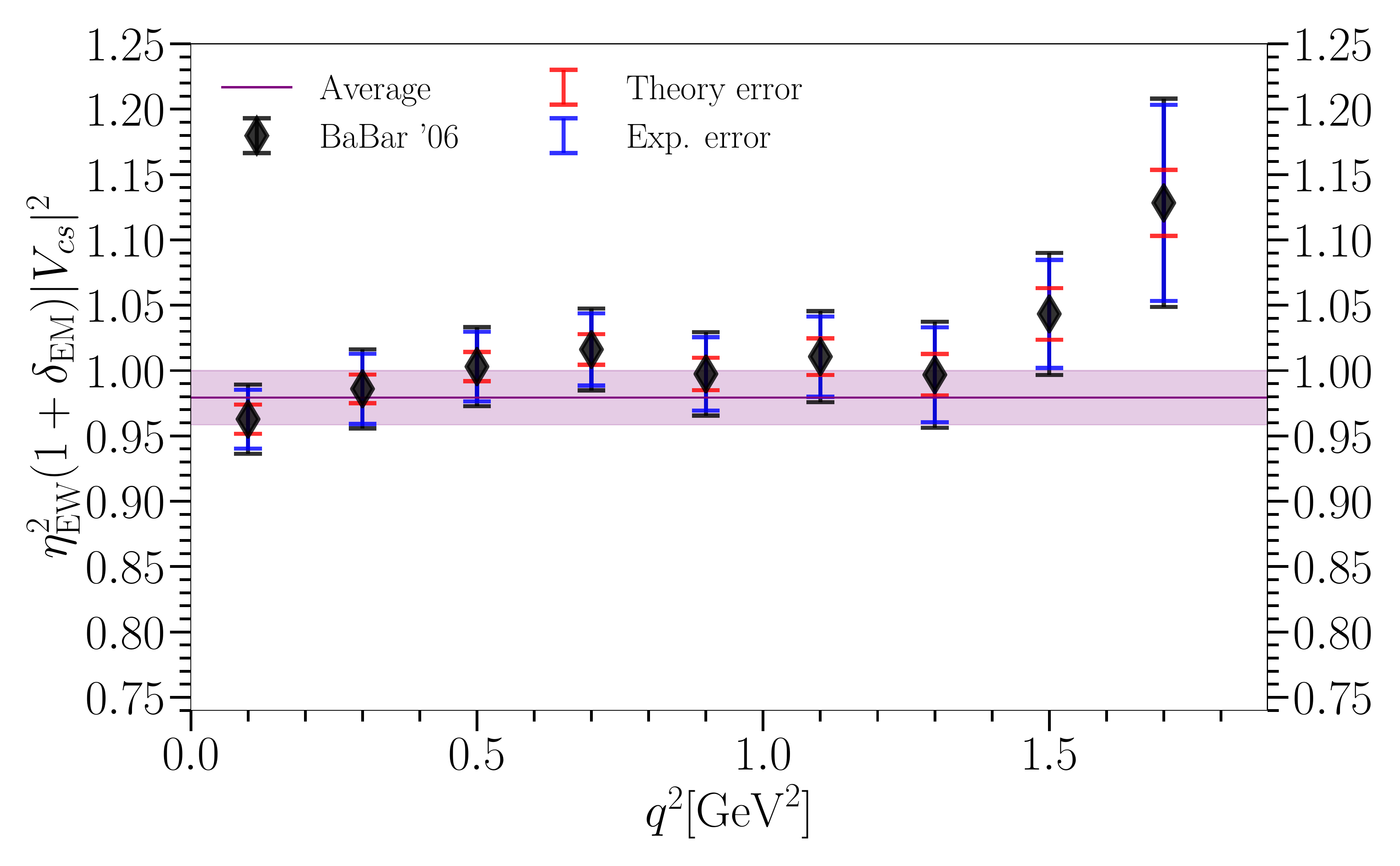}
\caption{Plot of the determination of $|\eta_{EW}V_{cs}|^2(1+\delta_{EM})$ 
per $q^2$ bin for BaBar $D^0$ results~\cite{Aubert:2007wg}. 
The total uncertainty for each bin is given in black and this is broken down 
into experimental (blue) and theoretical (red) contributions, the latter coming from our form 
factors. Each data point is centred on the $q^2$ bin it corresponds to. Note that the uncertainties 
are correlated between $q^2$ bins. The purple band gives the weighted average 
for these data points, with all correlations included.}
\label{fig:BaBarVcsbybin}
\end{figure}

\subsection{Using the differential rate}
\label{sec:vcs-diffrate}

We can use our form factor results across the full physical $q^2$ range 
to compare bin by bin in $q^2$ with experimental values of partial decay rates. 
For a given bin $(q^2_i,q^2_{i+1})$ the partial width is given by (from Eq.~\eqref{eq:diffdecayrate})
\begin{eqnarray}\label{eq:Vcs}
    \Delta_{i}\Gamma&=&\int^{q^2_{i+1}}_{q^2_i}\frac{d\Gamma}{dq^2}dq^2\\
    &=& \frac{G^2_F|\eta_{EW}V_{cs}|^2(1+\delta_{EM})}{24\pi^3}\times \nonumber \\ 
&&\int^{q^2_{i+1}}_{q^2_i}dq^2\bigg[ |\vec{p}_K|^3(1-\epsilon)^2(1+\frac{\epsilon}{2})|f_+(q^2)|^2 \nonumber \\
&& + |\vec{p}_K|(1-\epsilon)^2M_D^2\left(1-\frac{M_K^2}{M_D^2}\right)^2\frac{3\epsilon}{8}|f_0(q^2)|^2 \bigg] . \nonumber
\end{eqnarray}
The terms containing $\epsilon\equiv m_{\ell}^2/q^2$ have almost no impact here for 
either $\ell=e$ or $\ell=\mu$ but we include them nevertheless. 
We take $G_F=1.1663787(6)\times{}10^{-5}\mathrm{GeV}^{-2}$~\cite{pdg} from the muon lifetime 
and $\eta_{EW}=1.009(2)$ 
(Eq.~\eqref{eq:eta-ew-value}). 
$(1+\delta_{EM})$ allows for uncertainty from electromagnetic corrections, as discussed 
above. 
We perform the integral on the right-hand side of Eq.~\eqref{eq:Vcs} numerically for 
each $\Delta_{i}\Gamma$ matching those used in the experiment and carefully 
including the correlations of the form factor values between bins. 
As discussed above we use our form factors determined using $m_u=m_d$ and include 
an additional 0.15\% uncertainty to allow for variations between this and the experimental 
cases. For all of the kinematic factors in Eq.~\eqref{eq:Vcs} we use the experimental 
meson masses~\cite{pdg} for the charged or neutral meson cases as appropriate for that 
set of experimental data. 

Comparison to the experimental results enables us to determine 
$|\eta_{EW}V_{cs}|^2(1+\delta_{EM})$ for 
each bin and obtain a result as a weighted average across $q^2$ bins. 
We use experimental results for which a covariance matrix is provided for the partial rates 
between $q^2$ bins. We add covariance matrices for statistical and systematic uncertainties where 
they are provided separately (effectively adding the uncertainties in quadrature).  
In some cases an overall uncertainty on each bin is given along with the percentage 
breakdown into systematic and statistical uncertainty. We use this, along with the correlation 
matrices given, to obtain the separate covariance matrices and add them.

CLEO results are taken from~\cite{Besson:2009uv}, where both $D^0\rightarrow K^-e^+\nu_e$ and 
$D^+\rightarrow \overline{K}^0e^+\nu_e$ differential rates are measured and the correlations 
between them given. 
Partial rates were taken from Table V, and $\sigma_i^{\text{stat}}$, $\sigma_i^{\text{syst}}$ 
and their covariance matrices were calculated using these, the percentage error breakdowns 
in Tables VII and VIII and the correlation matrices in Tables XVI and XVII. 
These covariance matrices are then easily included in our calculation using the gvar 
package~\cite{peter_lepage_2020_3715065}. Our determination of $V_{cs}$ on a bin-by-bin 
basis is shown for the CLEO results in Figures~\ref{fig:Cleo1Vcsbybin} and~\ref{fig:Cleo2Vcsbybin}. 
The fit for the weighted average gives a $\chi^2/\mathrm{dof}$ of 0.64 in the $D^0$ case and 
1.7 in the $D^+$ case. In both cases there are nine degrees of freedom. 
The $q^2$ bins with the minimum total uncertainty 
are at the small $q^2$ end of the range, where the experiment is most accurate.  

BaBar results are taken from~\cite{Aubert:2007wg}; these are for 
the $D^0\rightarrow K^-e^+\overline{\nu}_e$ decay normalised by the branching fraction for 
$D^0\rightarrow K^-\pi^+$. 
Table II gives the normalised decay distribution and total correlation matrix. 
The leading diagonal values of the matrix give the $\sigma_i$. 
The distribution has been normalised so that the sum over all bins equals unity. 
A value is also given for
\begin{equation}
  R = \frac{\mathcal{B}(D^0\to K^-e^+\nu_e)}{\mathcal{B}(D^0\to K^-\pi^+)},
\end{equation}
which is included in the correlation matrix. Using this value, 
and multiplying by the global average for 
$\mathcal{B}(D^0\to K^-\pi^+)=0.03950(31)$~\cite{pdg}, we determine 
$\mathcal{B}(D^0\to K^-e^+\nu_e)$. This allows us to extract the branching fractions 
per bin from the decay distribution and convert these to partial rates by dividing by 
the $D^0$ lifetime $\tau_{D^0}=4.101(15)\times10^{-4}\,\text{ns}$~\cite{pdg}. 
We drop the largest $q^2$ bin from our weighted average fit (because it is equal to 
one minus the sum of the others from the normalisation constraint). We include 
the normalisation uncertainty after averaging to avoid normalisation bias.   
Our determination of $V_{cs}$ from the BaBar results is shown in Figure~\ref{fig:BaBarVcsbybin} 
and has a $\chi^2/\mathrm{dof}$ of 0.9 with nine degrees of freedom. 

\begin{figure}
\hspace{-30pt}
\includegraphics[width=0.48\textwidth]{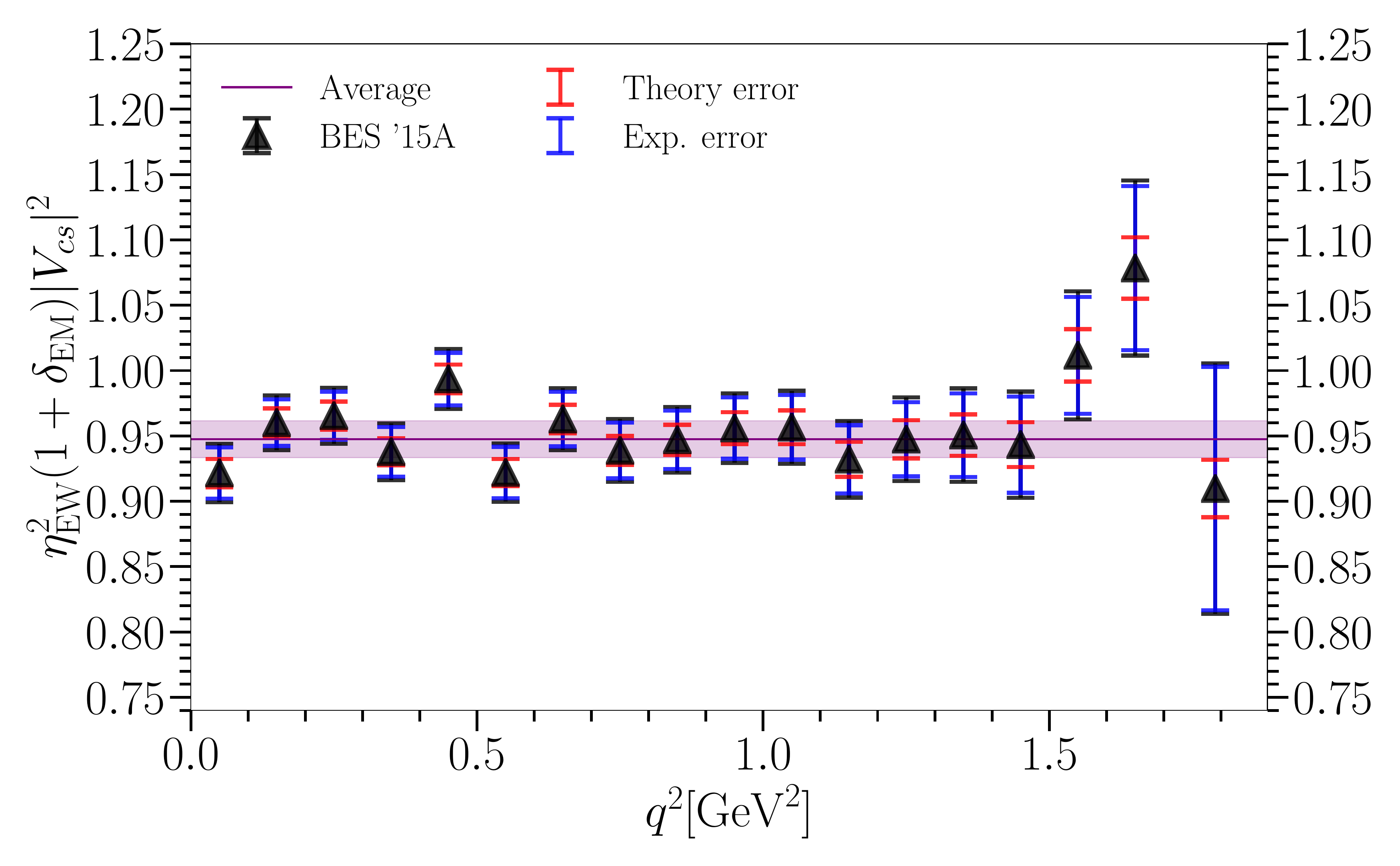}
\caption{Plot of the determination of $|\eta_{EW}V_{cs}|^2(1+\delta_{EM})$ 
per $q^2$ bin for BES $D^0$ results~\cite{Ablikim:2015ixa}. 
The total uncertainty for each bin is given in black and this is broken down 
into experimental (blue) and theoretical (red) contributions, the latter coming from our form 
factors. Each data point is centred on the $q^2$ bin it corresponds to. Note that the uncertainties 
are correlated between $q^2$ bins. The purple band gives the weighted average 
for these data points, with all correlations included.}
\label{fig:BESVcsbybin}
\end{figure}

\begin{figure}
\hspace{-30pt}
\includegraphics[width=0.48\textwidth]{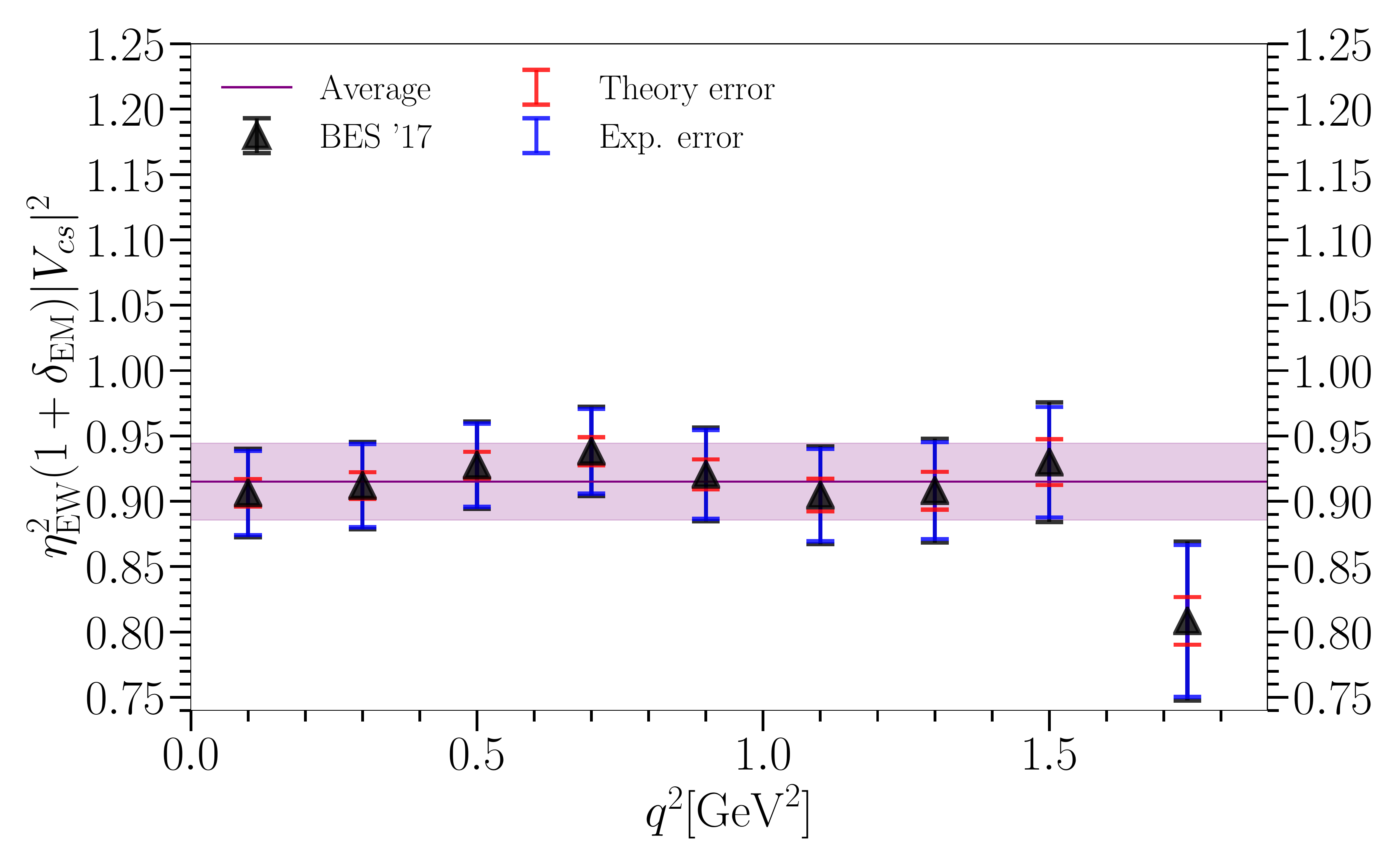}
\caption{Plot of the determination of $|\eta_{EW}V_{cs}|^2(1+\delta_{EM})$ 
per $q^2$ bin for BES $D^+$ results~\cite{Ablikim:2017lks}. 
The total uncertainty for each bin is given in black and this is broken down 
into experimental (blue) and theoretical (red) contributions, the latter coming from our form 
factors. Each data point is centred on the $q^2$ bin it corresponds to. Note that the uncertainties 
are correlated between $q^2$ bins. The purple band gives the weighted average 
for these data points, with all correlations included.}
\label{fig:BESDpVcsbybin}
\end{figure}

BES results are taken from~\cite{Ablikim:2015ixa} for the $D^0$ decay channel. The 
data can be found in Table V, and the breakdown of the percentage errors 
and correlation matrices for systematic and statistical uncertainty are given 
in Tables IX and XI.
BES results for the $D^+$ channel are given in~\cite{Ablikim:2017lks} (Table VI). 
Our determination of $V_{cs}$ on a bin-by-bin 
basis is shown for these two sets of BES results in 
Figures~\ref{fig:BESVcsbybin} and~\ref{fig:BESDpVcsbybin}, with $\chi^2/\mathrm{dof}$ 
1.1 ($\mathrm{dof}=18$) and 0.9 ($\mathrm{dof}=9$) respectively. 

\begin{figure}
\includegraphics[width=0.48\textwidth]{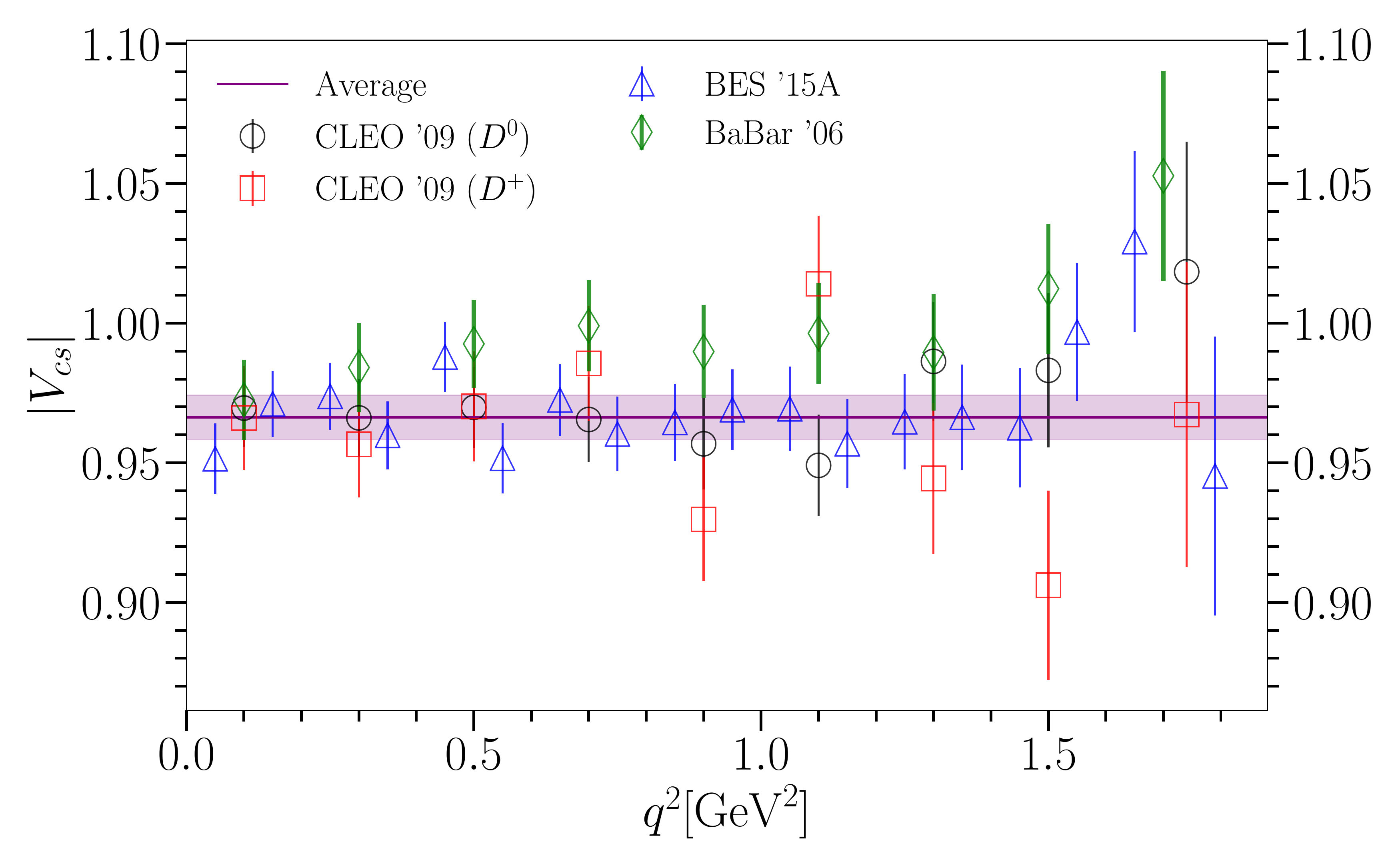}
\caption{Plot of $|V_{cs}|$ per bin for CLEO, BaBar and BES results 
from~\cite{Besson:2009uv,Aubert:2007wg,Ablikim:2015ixa}. Each data point is centred 
on the $q^2$ bin it corresponds to and the error bars plotted include the 
uncertainties from $\eta_{EW}$ and $\delta_{EM}$. 
The purple line and band give the result from 
our total weighted average for $|V_{cs}|^2$, 
with all correlations included. 
The width of the band includes the uncertainties from 
$\eta_{EW}$ and $\delta_{EM}$ as given in Eq.~\eqref{eq:vcs:dgamma}. 
}
\label{fig:Vcsbybin}
\end{figure}

\begin{figure}
\includegraphics[width=0.48\textwidth]{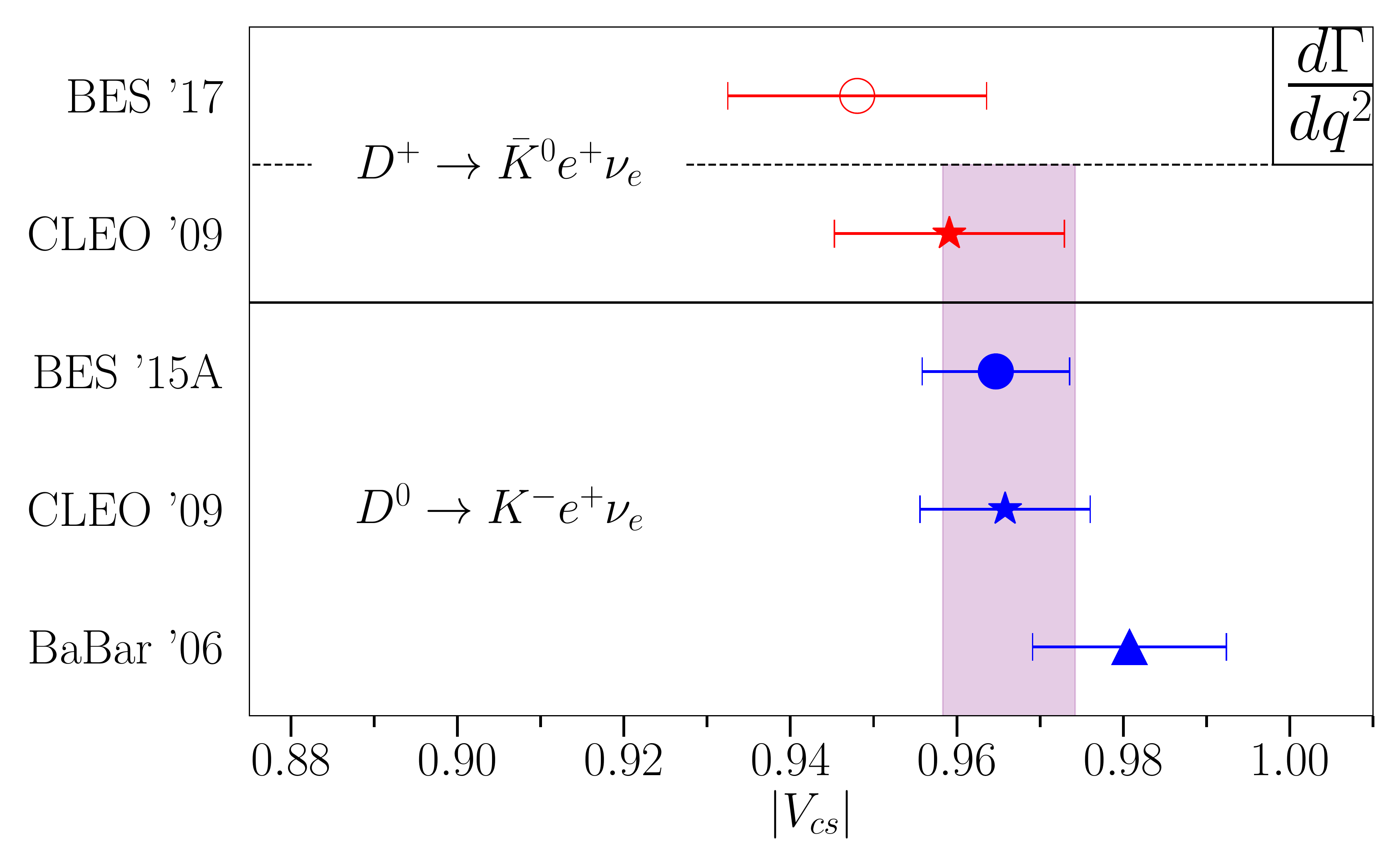}
\caption{Comparison plot of the determinination of $|V_{cs}|$ using the 
differential decay rate for CLEO, BaBar and BES results 
from~\cite{Besson:2009uv,Aubert:2007wg,Ablikim:2015ixa,Ablikim:2017lks} for $D^0$ and $D^+$ 
decays. 
The purple band gives the total weighted average for $V_{cs}$, not including 
the BES '17 result. 
The width of the band includes the uncertainties from 
$\eta_{EW}$ and $\delta_{EM}$ as given in Eq.~\eqref{eq:vcs:dgamma}. 
}
\label{fig:Vcsdgamma}
\end{figure}

The determinations of $V_{cs}$ from each experiment and each $q^2$ bin 
are plotted together as a function of $q^2$ 
in Figure~\ref{fig:Vcsbybin}. The weighted averaged results for $|V_{cs}|$ 
for each experiment are 
then compared in Figure~\ref{fig:Vcsdgamma}. The $|V_{cs}|$ result for 
each experiment is obtained by dividing the
square root of the weighted average 
of $|V_{cs}|^2\eta^2_{EW}(1+\delta_{EM})$ over 
the $q^2$ bins by $\eta_{EW}\sqrt{(1+\delta_{EM})}$. 

The results from each experiment are combined to give a total average for 
$|V_{cs}|$ which is shown by the purple band in Figures~\ref{fig:Vcsbybin} 
and~\ref{fig:Vcsdgamma}. 
Here we have assumed that correlations between the 
different experiments can be ignored. However, we do not include both sets 
of BES data, since the 
correlations between the two sets are not given. 
We include the more precise BES $D^0$ results~\cite{Ablikim:2015ixa} 
in Figure~\ref{fig:Vcsbybin} 
and drop the BES $D^+$ values.  
We note that in each $q^2$ bin 
the experimental error dominates over that from theory (our form factors). 
The fact that there are multiple sets of uncorrelated experimental 
results but only one set of lattice QCD form factors 
means that in the final average, however, the theory uncertainty dominates. 
We obtain a value of 
\begin{equation}
\label{eq:vcs:dgamma}
|V_{cs}|^{\text{d}\Gamma/\text{d}q^2} = 0.9663(53)_{\text{latt}}(39)_{\text{exp}}(19)_{\eta_{EW}}(40)_{\text{EM}}
\end{equation}
from using the binned differential rate. The fit to yield the average has 
a $\chi^2/\mathrm{dof}$ of 0.7 for 4 degrees of freedom.
The first uncertainty here is from our lattice QCD 
form factors, including an uncertainty for the fact that these are calculated for the $m_u=m_d$ case.
The second uncertainty comes from the experimental results. The third uncertainty is from $\eta_{EW}$ and the fourth
from long-distance QED corrections, amounting to 0.5\% in $V_{cs}$ for the case of a charged 
meson in the final state, and 0.25\% for a neutral meson in the final state, 
as discussed above. There is some sign in Figures~\ref{fig:Vcsbybin} and~\ref{fig:Vcsdgamma} 
that the central 
values of $V_{cs}$ for the results with a charged 
$K^-$ meson in the final state are slightly higher than those with 
a neutral $\overline{K}^0$ meson; this is 
consistent with what might be expected from QED effects if $\delta_{EM}>0$ 
but the uncertainties are too large 
for this to be clear. The fit to the average uses this information to 
arrive at the combined uncertainty from EM effects above.   

\subsection{Using the total branching fraction}
\label{sec:vcstotalbr}

\begin{table}
  \caption{ Total width for $D \rightarrow K$ semileptonic decay up to a factor 
of $|\eta_{EW}V_{cs}|^2(1+\delta_{EM})$ (see Eq.~\eqref{eq:diffdecayrate}), determined 
from our form factors. We give results for all 4 modes that we consider. They differ 
slightly in the parent and daughter meson masses and in the mass of the lepton in the final 
state; these affect the kinematic factors in the differential rate 
and the end-points of integration for 
the total width. These values can be combined with experimental values of the 
relevant branching fraction and $D$ meson lifetime to determine $|V_{cs}|$.  
}
  \begin{center} 
    \begin{tabular}{cc}
      \hline
& $\Gamma/(|\eta_{EW}V_{cs}|^2(1+\delta_{EM}))$ (${\mathrm{ns}^{-1}}$) \\
\hline 
\hline
$D^+\rightarrow \overline{K}^0\mu^+\nu_{\mu}$ &    88.30(99)       \\
$D^+\rightarrow \overline{K}^0e^+\nu_{e}$ &     90.3(1.0)      \\
$D^0\rightarrow K^-\mu^+\nu_{\mu}$ &    87.57(98)       \\
$D^0\rightarrow K^-e^+\nu_e$ &   89.5(1.0)        \\

      \hline
      \hline
    \end{tabular}
  \end{center}
  \label{tab:intrate}
\end{table}

We can also determine $V_{cs}$ from a comparison of theory and experiment 
for the total branching fraction for the semileptonic 
decay process. 
To obtain the total width, $\Gamma$, from the theory side we need to 
integrate Eq.~\eqref{eq:diffdecayrate} over the full physical $q^2$ range.  
The limits of integration use the experimental masses for the appropriate 
leptons and charged or neutral meson masses. 
Table~\ref{tab:intrate} gives our values for $\Gamma/(|\eta_{EW}V_{cs}|^2(1+\delta_{EM}))$ 
for each of the 4 modes we consider, i.e. charged and neutral $D$ meson decay 
to $e$ and $\mu$ in the final state. 

We convert the total width to a branching fraction using the experimental 
average values for the appropriate $D$ meson lifetime~\cite{pdg}. 
Comparison to experiment then yields a determination of $|V_{cs}|$. 
There are additional experimental results for the total 
branching fraction beyond 
those used in the determination of $|V_{cs}|$ from the differential decay rate 
in Section~\ref{sec:vcs-diffrate}. 
These come from Belle~\cite{Widhalm:2006wz} for $D^0$ decays to both 
$e$ and $\mu$ in the final state 
and from BES for $D^0$ decays to $\mu$ in the final state~\cite{Ablikim:2018evp} (discussed in 
Section~\ref{sec:lfu} in the context of tests of lepton flavour universality) 
and $D^+$ decays to $\mu$ in the final state~\cite{Ablikim:2016sqt}. 
There are also new results from BES~\cite{BESIII:2021mfl} for $D^0$ and $D^+$ decay to $e$ in the final state, using a new reconstruction method. 
In the summary of~\cite{BESIII:2021mfl} total branching fractions are quoted 
that are the average of the new results with their earlier values~\cite{Ablikim:2015ixa,Ablikim:2017lks}, 
accounting for correlations. 
It is these averages that we use in the following, denoting them as `BES21'. 
Note that there are then branching fraction results for 
all four possible modes for $D \rightarrow K$ decay. 

\begin{figure}
\includegraphics[width=0.48\textwidth]{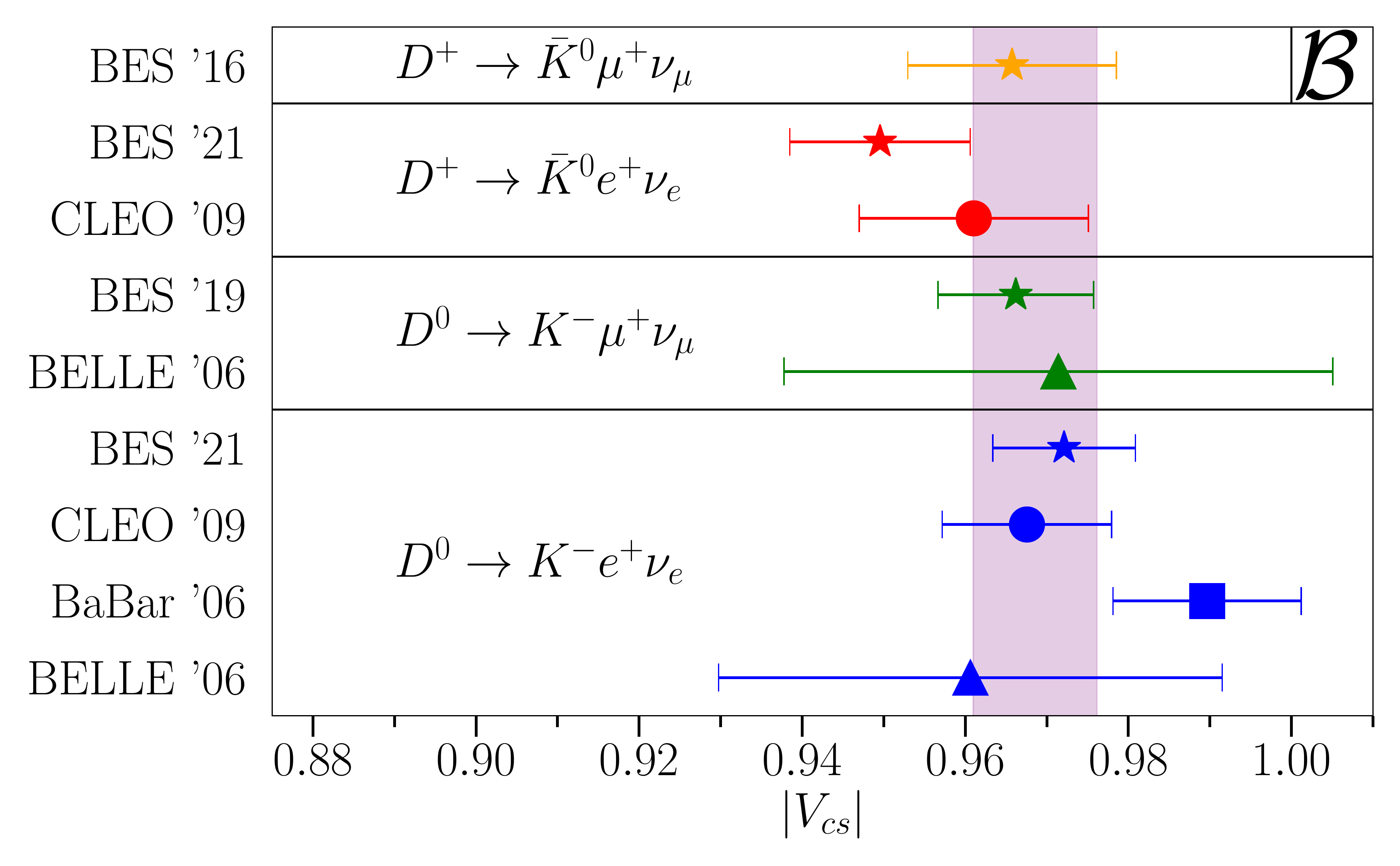}
\caption{Comparison plot of the determinination of $|V_{cs}|$ using the 
total branching fraction. Experimental results are 
from~\cite{Ablikim:2016sqt} for $D^+ \rightarrow \overline{K}^0 \mu^+\nu_{\mu}$,~\cite{BESIII:2021mfl,Besson:2009uv} for $D^+ \rightarrow \overline{K}^0 e^+\nu_e$, from~\cite{Ablikim:2018evp,Widhalm:2006wz} for $D^0\rightarrow K^-\mu^+\nu_{\mu}$ and from~\cite{BESIII:2021mfl,Besson:2009uv,Aubert:2007wg,Widhalm:2006wz} for $D^0\rightarrow K^-e^+\nu_e$    
decays. Note that the BES results for final state $e$ are the quoted averages 
for results from~\cite{BESIII:2021mfl} and~\cite{Ablikim:2015ixa,Ablikim:2017lks}.
The purple band gives the total average for $V_{cs}$, 
assuming 100\% correlation of systematic uncertainties for results from a 
given experiment. The width of the purple band encompasses all uncertainties, including those 
from $\eta_{EW}$ and $\delta_{EM}$.  
}
\label{fig:VcsB}
\end{figure}

Figure~\ref{fig:VcsB} shows the results of the determination of 
$V_{cs}$ using the total branching fraction for each experimental result. 
In fitting the experimental results to give a common (average) $|V_{cs}|$ value 
we have taken the systematic uncertainties 
for a given experiment to be 100\% correlated between the different results 
from that experiment. 
We obtain a final result for $V_{cs}$ from the total branching fraction of
\begin{equation}
\label{eq:vcs:B}
|V_{cs}|^{\mathcal{B}} = 0.9686(54)_{\text{latt}}(39)_{\text{exp}}(19)_{\eta_{EW}}(30)_{\text{EM}}.
\end{equation}
This fit has a $\chi^2/\mathrm{dof}$ of 1.7 for 9 degrees of freedom. 
Again the first uncertainty here is from our form factors (including 
an uncertainty from $m_u\ne m_d$), the second from 
the experimental results (including uncertainties from the $D$ meson lifetime), 
the third from $\eta_{EW}$ and the fourth 
is the uncertainty we allow for 
QED corrections from $\delta_{EM}$. $\delta_{EM}$ is taken as an independent uncertainty 
for the $e$ and $\mu$ cases and for charged and neutral mesons and the fit 
for the average constrains this uncertainty based on the data. 
The $\chi^2/\mathrm{dof}$ value drops to 1.4 if the BaBar result for 
$D^0 \rightarrow K^-$ is omitted from the fit; the average value 
obtained then falls by 0.35$\sigma$ (where $\sigma$ is the total uncertainty). 

\subsection{Using $f_+(0)$}
\label{sec:vcsfp0}

\begin{figure}
\includegraphics[width=0.48\textwidth]{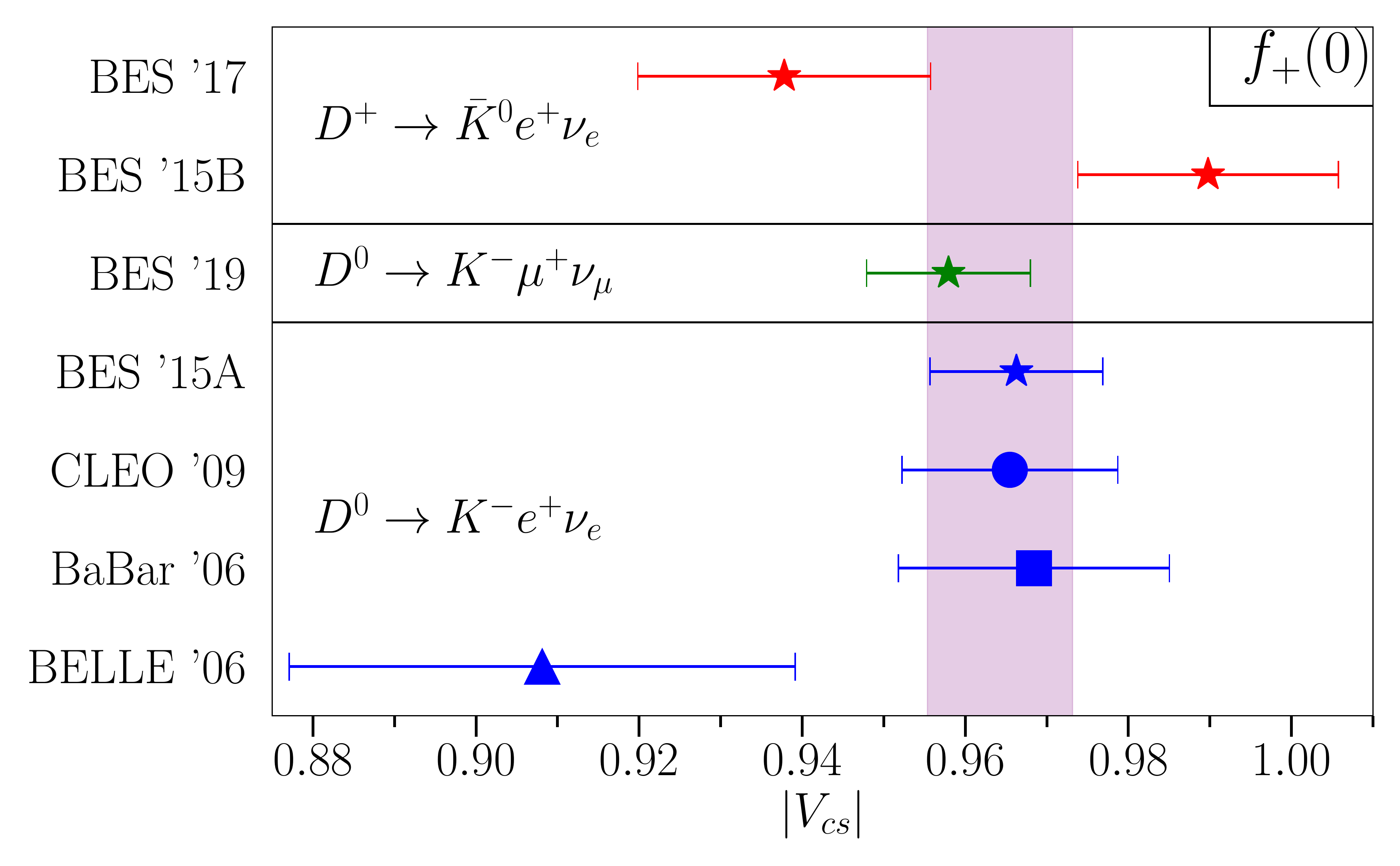}
\caption{Comparison plot of the determinination of $|V_{cs}|$ using the 
extrapolation of experimental results to $q^2=0$. 
Experimental results are from ~\cite{Ablikim:2017lks,Ablikim:2015qgt} for 
$D^+\rightarrow \overline{K}^0 e^+\nu_e$, from~\cite{Ablikim:2018evp} for 
$D^0 \rightarrow K^- \mu^+\nu_{\mu}$ 
and~\cite{Ablikim:2015ixa,Besson:2009uv,Aubert:2007wg,Widhalm:2006wz} for 
$D^0\rightarrow K^-e^+\nu_e$. 
The purple band gives the weighted average result for $V_{cs}$ 
obtained from the HFLAV weighted average~\cite{Amhis:2019ckw} 
of the experimental results but including a correction for 
$\eta_{EW}$ and an additional uncertainty from QED corrections 
(Eq.~\eqref{eq:vcs:f0}). 
}
\label{fig:Vcsf0}
\end{figure}

Following the approach for $K$ semileptonic decays, experimental groups 
have often provided results for the combination of $|V_{cs}|$ and form factor 
values at $q^2=0$ derived from fitting their differential decay rates.  
Simply dividing these results by the lattice QCD form factor result at 
$q^2=0$ can then give a determination of $V_{cs}$. However, what is 
generally quoted as a value for $|V_{cs}|f_+(0)$ is, in our notation using 
Eq.~\eqref{eq:diffdecayrate}, in fact
$|V_{cs}|f_+(0)\eta_{EW}\sqrt{(1+\delta_{EM})}$. 
Taking this into account, and using our $f_+(0)$ result from 
Eq.~\eqref{eq:0maxvals}, gives results for $V_{cs}$ from the 
experimental results available that are plotted in Figure~\ref{fig:Vcsf0}.  

To determine a weighted average for $V_{cs}$ from these values we 
take the HFLAV average~\cite{Amhis:2019ckw} for 
$|V_{cs}|f_+(0)\eta_{EW}\sqrt{(1+\delta_{EM})}$ 
(denoted $|V_{cs}|f_+(0)$ in~\cite{Amhis:2019ckw} and given as 0.7180(33)). 
This gives the purple band in Figure~\ref{fig:Vcsf0}. 
The value of $V_{cs}$ from this approach is then 
\begin{equation}
\label{eq:vcs:f0}
|V_{cs}|^{{f_+(0)}} = 0.9643(57)_{\text{latt}}(44)_{\text{exp}}(19)_{\eta_{EW}}(48)_{\text{EM}}.
\end{equation}
Again the last two uncertainties come from the uncertainty on $\eta_{EW}$ and 
QED corrections included in $\delta_{EM}$. Since the HFLAV average includes 
charged and neutral meson results and $\mu$ and $e$ final states, 
we take the largest uncertainty for $\delta_{EM}$ that we use here (1\% in the rate) 
and add this as a separate uncertainty. 

\section{Discussion: $V_{cs}$}
\label{sec:vcsdiscuss}

We have determined $|V_{cs}|$ in three different ways, with results given in 
Eqs.~\eqref{eq:vcs:dgamma},~\eqref{eq:vcs:B} and~\eqref{eq:vcs:f0}. 
The results vary in the experimental results that are included and the 
way in which the lattice QCD form factors enter the calculation. 
The agreement between the results is good, with the lowest ($V_{cs}^{f_+(0)}$)  
and highest ($V_{cs}^{\mathcal{B}}$) differing by 0.6$\sigma$. 
This is a good test, at this level of precision, that QCD gives 
the shape of the form factors seen in 
experiment (backing up Figure~\ref{fig:ellipse}). 
The uncertainties in each value are very similar, ranging from 0.8\% in 
both our preferred approach of $V_{cs}^{\text{d}\Gamma/\text{d}q^2}$ and 
for $V_{cs}^{\mathcal{B}}$ to 
0.9\% for $V_{cs}^{f_+(0)}$. This high accuracy is achievable because of good
statistical precision over a range of lattice spacing values and light quark 
masses, with accurately tuned $c$ and $s$ masses and weak current operators that 
are normalised fully nonperturbatively within the same calculation. 

\begin{figure}
\hspace{-30pt}
\includegraphics[width=0.48\textwidth]{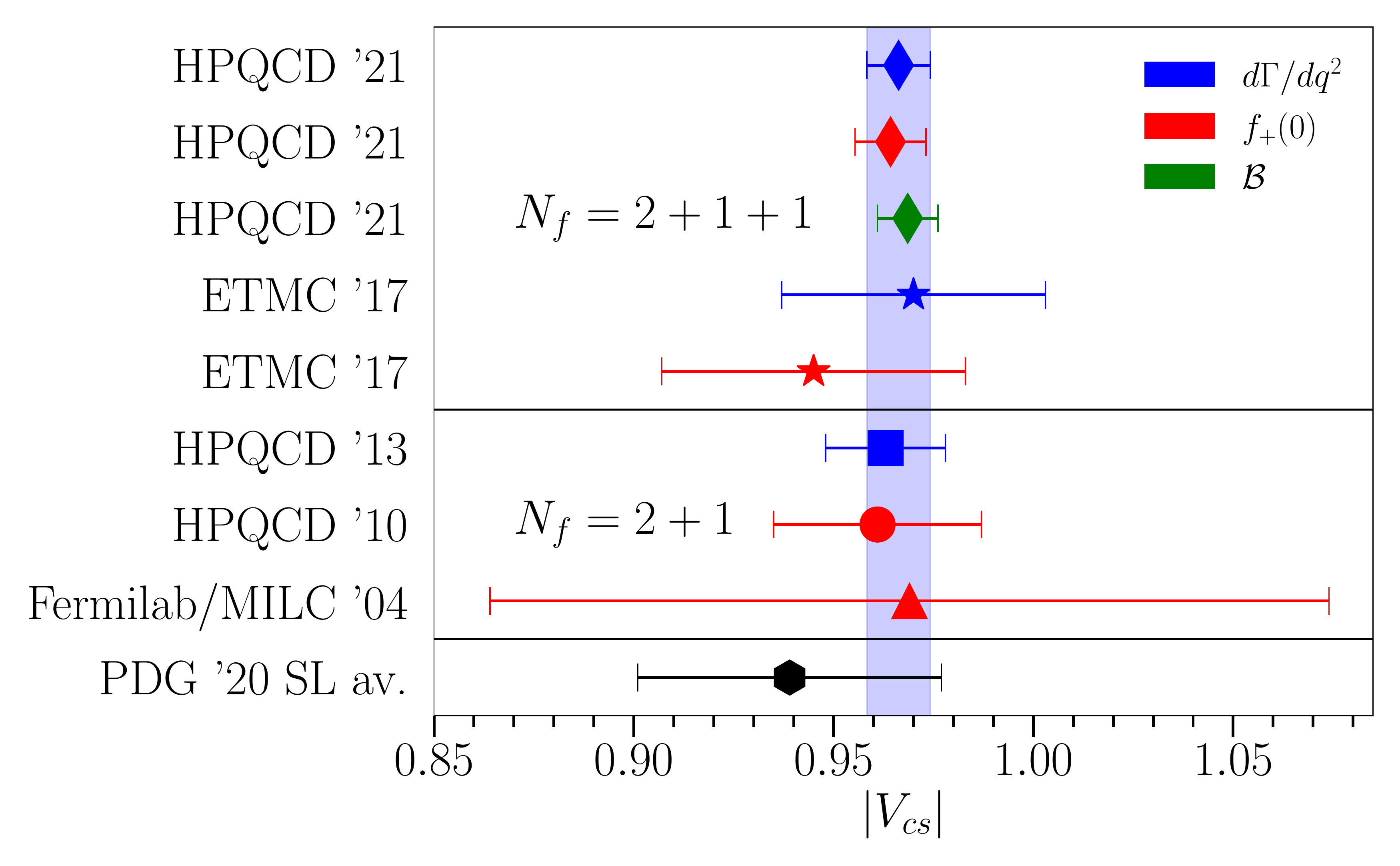}
\caption{Our $|V_{cs}|$ result compared with other $N_f=2+1+1$ and $N_f=2+1$ results using lattice QCD. 
Different symbols indicate different lattice calculations, whilst different
colours indicate the method used. 
Blue indicates use of the differential rate in $q^2$ bins, red 
indicates use of the $f_+(0)$ method and green indicates 
use of the total branching fraction for the decay.
Points marked `HPQCD '21' come from this work, `ETMC '17' is 
from~\cite{Lubicz:2017syv,Riggio:2017zwh}, `HPQCD '13' is 
from~\cite{Koponen:2013tua}, `HPQCD '10' is from~\cite{Na:2010uf} 
and `Fermilab/MILC '04' is from~\cite{Aubin:2004ej}. 
For comparison we give at the bottom the value currently quoted in the Particle Data Tables~\cite{pdg} 
from semileptonic $D \rightarrow K$ decay (Eq.~\eqref{eq:introvcs-sl}). 
The blue band carries our preferred result, 
$V_{cs}^{\text{d}\Gamma/\text{d}q^2}$, down the plot. 
}
\label{fig:Vcscomp}
\end{figure}

Figure~\ref{fig:Vcscomp} compares our new results for $V_{cs}$ to those from 
earlier full lattice QCD calculations. These go back to the Fermilab/MILC 
result of 2004~\cite{Aubin:2004ej}, completed before experimental results 
were available, using the clover action for $c$ quarks on gluon field 
configurations with $N_f=2+1$ flavours of asqtad sea quarks. 
The HPQCD results from 2010~\cite{Na:2010uf} and 
2013~\cite{Koponen:2013tua} use the HISQ action 
on the same gluon field configurations;
we build on these calculations with the improvements we have made here. 
The 2017 ETMC results~\cite{Lubicz:2017syv,Riggio:2017zwh} use the 
twisted mass formalism on gluon field configurations with $N_f=2+1+1$ 
flavours of sea quarks. We see good agreement between the results, including 
between those with $N_f=2+1$ and $N_f=2+1+1$ flavours. 

Our results show a significant improvement in uncertainty compared to these 
earlier values, being a factor of two more accurate than the previous 
best result from HPQCD in 2013. 
We note that the previous results set $\eta_{EW}$ to 1 and did not include 
an uncertainty to allow for long-distance QED effects on the experimental 
results.  

Our preferred result for $V_{cs}$ is 
\begin{equation}
\label{eq:finalvcs}
V_{cs} = 0.9663(80) ,
\end{equation}
from Eq.~\eqref{eq:vcs:dgamma}, adding uncertainties in quadrature. 
We can compare this to the result for $V_{ud}$ of 0.97370(14) from~\cite{pdg}. 
We see that $V_{cs}=V_{ud}$ within the 1$\sigma$ uncertainty in Eq.~\eqref{eq:finalvcs},
in good agreement with the expectation from the CKM matrix that this should be 
true up to effects of order $\lambda^4\approx 0.002$. 

We now compare our new result for $V_{cs}$ from semileptonic $D\rightarrow K$ decay 
to the value obtained from $D_s$ leptonic decay and look at the impact that our 
improved uncertainty has on our understanding of the unitarity of the CKM matrix.  

\begin{figure}
\includegraphics[width=0.48\textwidth]{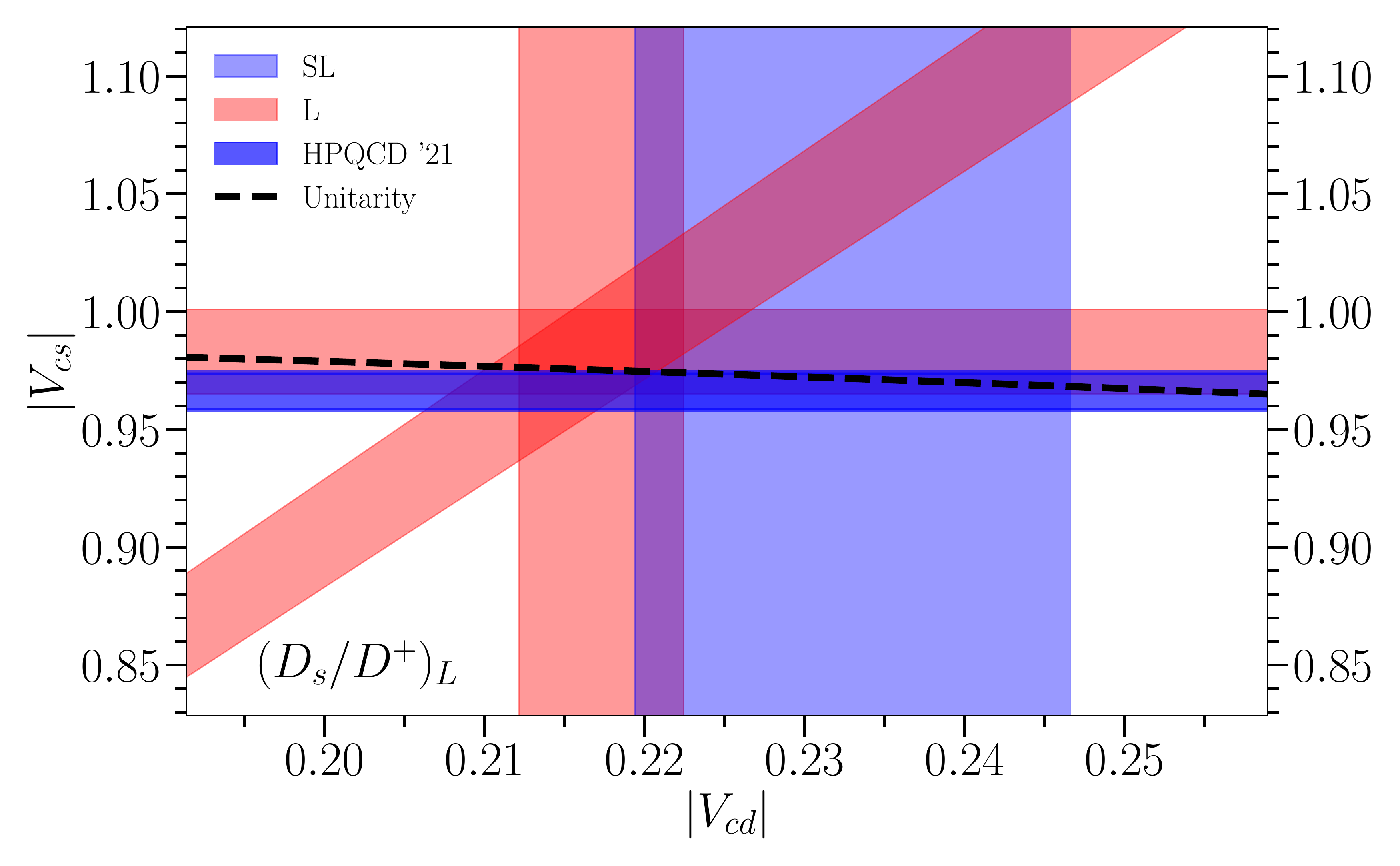}
\caption{A comparison of constraints on $V_{cs}$ and $V_{cd}$ with the expectation from 
CKM unitarity. 
Red bands show the $\pm 1\sigma$ range for the determination of $V_{cs}$ and $V_{cd}$
from leptonic decays of $D_s$ and $D^+$ combined with decay constants from lattice 
QCD. The diagonal red band is the constraint from the ratio of leptonic rates for 
$D_s$ and $D^+$ combined with the lattice QCD ratio of decay constants.  
The solid light blue band shows the result for $V_{cd}$ from 
the $D\rightarrow \pi \ell \overline{\nu}$ 
decay combined with lattice QCD form factor results. 
See the text for a discussion of the values used. 
The darker blue band shows our new determination here of $V_{cs}$ from 
$D\rightarrow K \ell \overline{\nu}$ with $\pm 1\sigma$ uncertainties. 
For comparison
the black dashed line gives the unitarity constraint curve 
of $|V_{cd}|^2 +|V_{cs}|^2+|V_{cb}|^2 = 1$. 
}
\label{fig:Vcscd}
\end{figure}

Figure~\ref{fig:Vcscd} plots the $\pm 1\sigma$ band for 
our determination of $V_{cs}$ from Eq.~\eqref{eq:finalvcs} as the darker blue band. 
This is compared to the result (red band) from $D_s$ leptonic decay of 0.983(18) 
from the `Leptonic decays of charged pseudoscalar mesons' 
review in~\cite{pdg}. This result uses lattice QCD results for the 
$D_s$ decay constant and includes uncertainties for 
$\eta_{EW}$ and long-distance QED effects. The `CKM Quark-Mixing Matrix' 
review gives a value of 0.992(12) but without including these effects. 
This value would then lie in the upper half of the $V_{cs}$ leptonic
band plotted in Figure~\ref{fig:Vcscd}.  
In either case it is clear that our new result for $V_{cs}$ is more 
accurate than that from leptonic decay and has a lower central value.

Figure~\ref{fig:Vcscd} also shows the constraints currently available 
on $V_{cd}$. The `CKM Quark-Mixing Matrix' review in~\cite{pdg} quotes a value 
for $V_{cd}$ from semileptonic $D \rightarrow \pi$ decay from combining 
experimental results with the form factor at $q^2=0$ determined in 
$N_f=2+1+1$ lattice QCD by ETMC~\cite{Lubicz:2017syv}. This gives 
$V_{cd}=0.2330(136)$. The value quoted in the same review from 
$D^+$ leptonic decays is 0.2173(51). This combines experimental 
results with the $D^+$ decay constant 
determined in $N_f=2+1+1$ lattice QCD by the Fermilab/MILC 
collaboration~\cite{Bazavov:2017lyh}. Another constraint follows from 
the ratio of $D_s$ to $D$ leptonic decay rates~\cite{Boyle:2018knm} combined with the 
ratio of $D_s$ and $D$ decay constants. Using ratios of $V_{cs}f_{D_s}$  
and $V_{cd}f_{D^+}$ averaged over experimental results from~\cite{pdg} 
and the lattice QCD result for $f_{D_s}/f_{D^+}$ 
from~\cite{Bazavov:2017lyh} gives the constraint $|V_{cd}|/|V_{cs}|=0.2209(56)$ 
if we assume that electromagnetic corrections to the leptonic rates will largely cancel. 

The black dashed line in Figure~\ref{fig:Vcscd} corresponds to the 
unitarity constraint $|V_{cd}|^2+|V_{cs}|^2+|V_{cb}|^2 = 1$. $V_{cb}$
has little impact on this curve; we use the average value of 0.0410(14) 
from~\cite{pdg}. Our result for $V_{cs}$ is in good agreement with the 
unitarity curve for values of $V_{cd}$ in the range given by the 
leptonic and semileptonic constraints. 

\begin{figure}
\includegraphics[width=0.48\textwidth]{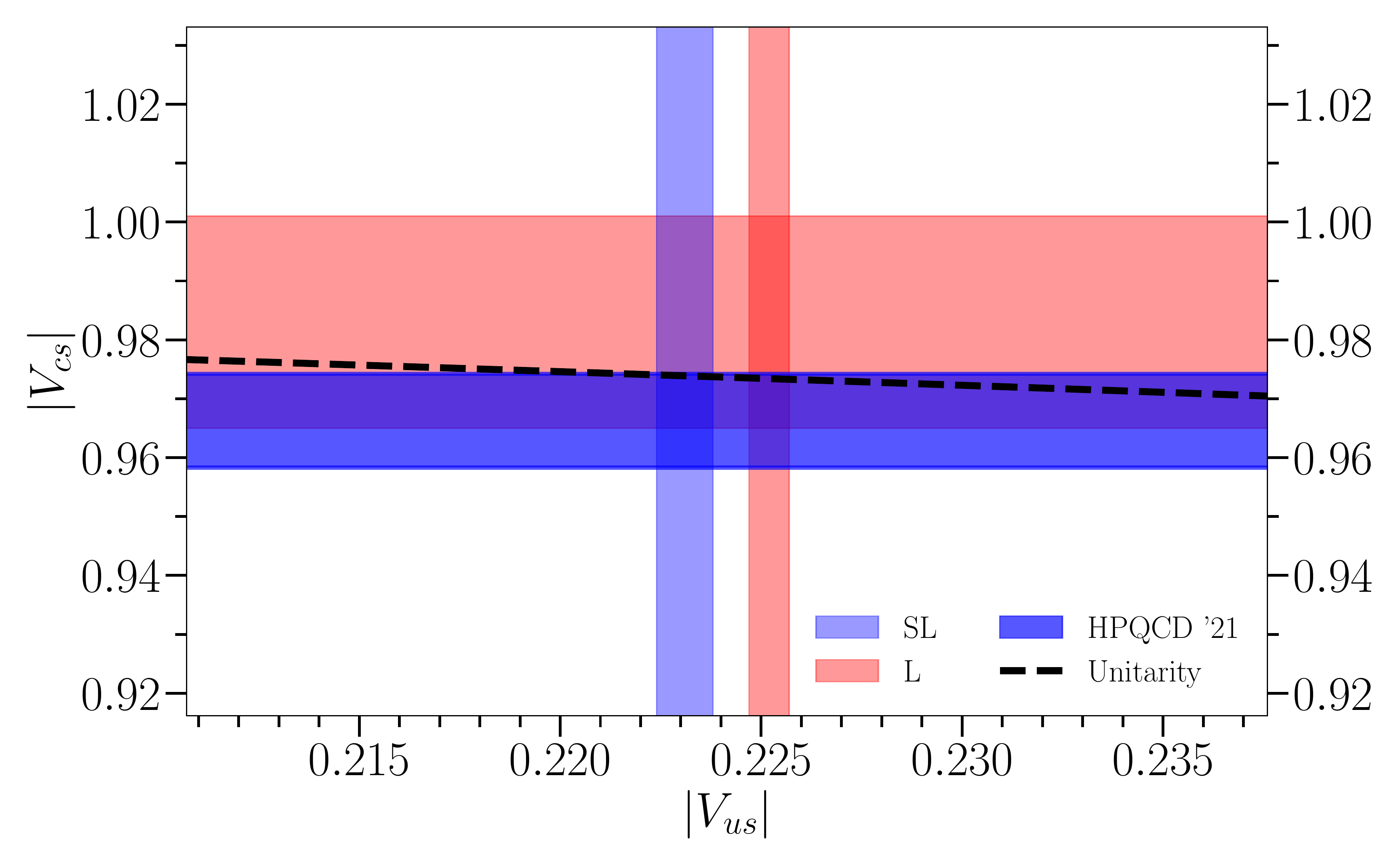}
\caption{A comparison of constraints on $V_{cs}$ and $V_{us}$ with the expectation from 
CKM unitarity. 
Red bands show the $\pm 1\sigma$ range for the determination of $V_{cs}$ and $V_{us}$
from leptonic decays of $D_s$ and $K^+$ combined with decay constants from lattice 
QCD. The light blue band shows the result for $V_{us}$ from $K\rightarrow \pi \ell \overline{\nu}$ 
decay combined with lattice QCD form factor results. 
See the text for a discussion of the values used. 
The darker blue band shows our new determination here of $V_{cs}$ from 
$D\rightarrow K \ell \overline{\nu}$ with $\pm 1\sigma$ uncertainties. 
For comparison 
the black dashed line gives the unitarity constraint curve, 
$|V_{us}|^2+|V_{cs}|^2+|V_{ts}|^2=1$.  
}
\label{fig:Vcsus}
\end{figure}

Figure~\ref{fig:Vcsus} gives the same picture for the $V_{us}$, $V_{cs}$, 
$V_{ts}$ column of the CKM matrix, showing constraints in the 
$|V_{cs}|-|V_{us}|$ plane. 
$|V_{cs}|$ values are as for Figure~\ref{fig:Vcscd} but plotted over a 
smaller range because of the higher accuracy of $|V_{us}|$ (we scale $x$ and $y$ 
axis ranges together). 

We take $|V_{us}|$ values from the  review `$V_{ud}$, $V_{us}$, Cabibbo angle and CKM 
unitarity' in~\cite{pdg}. This gives $|V_{us}|=0.2252(5)$ from leptonic decays of 
$K^+$ and 0.2231(7) from $K \rightarrow \pi$ semileptonic decay. The leptonic result 
uses an average~\cite{Aoki:2019cca} of lattice QCD results for the $K$ decay constants dominated 
by that from~\cite{Bazavov:2017lyh}. The semileptonic result uses an average~\cite{Aoki:2019cca} of 
lattice QCD results for the $K \rightarrow \pi$ form factor at $q^2=0$ 
from~\cite{Bazavov:2013maa,Carrasco:2016kpy}. The current most accurate lattice QCD results 
for the form factor are given in~\cite{Bazavov:2018kjg}.  

Figure~\ref{fig:Vcsus} shows the tension developing between leptonic and semileptonic 
determinations of $V_{us}$~\cite{Bazavov:2018kjg, pdg}. The black dashed line in 
the figure shows the unitarity constraint $|V_{us}|^2+|V_{cs}|^2+|V_{ts}|^2=1$. 
$|V_{ts}|$ has little impact on this curve; we use the current most accurate 
determination of $|V_{ts}| = 0.04189(93)$ from the measured oscillation rate of $B_s$ mesons~\cite{pdg} 
and HPQCD's lattice QCD determination~\cite{Dowdall:2019bea} of 
the matrix elements of the 4-quark operators 
that give the mass difference between the $B_s$ eigenstates. 
Our improved accuracy for $|V_{cs}|$, along with the unitarity curve, 
is not sufficient to distinguish between 
the two values for $|V_{us}|$.  

\section{Conclusions}
\label{sec:conclusions}
We have completed a detailed lattice QCD calculation of the scalar and vector form factors 
that parameterise the strong interaction effects in the $D\to{}K\ell\nu$ semileptonic 
decay process in the Standard Model. 
Our calculation covers the full physical range of momentum transfer. 
With high statistics on eight gluon field ensembles, three with physical light quarks, 
and a highly improved discretisation of QCD that allows nonperturbative normalisation of weak 
current operators, we have improved significantly on the precision of previous work. 

In Table~\ref{tab:ancoefficients} we give the parameters, and their uncertainties and correlation matrix, 
that enable our form factors to be reconstructed. 
We give our form factor values at $q^2=0$ and $q^2_{\text{max}}$ in Eq.~\eqref{eq:0maxvals}. 
Our lattice QCD calculations use $m_u=m_d=m_l$; we take an additional 0.15\% uncertainty on the 
form factors (uniformly in $q^2$) to allow for the impact of $m_u\ne m_d$ in the form factors 
when we compare to experimental results. 
Figure~\ref{fig:ellipse} compares the shape parameters for our form factors to those inferred
from the experimental differential rate, and shows good agreement.

In Section~\ref{sec:lfu} we give results for observables that allow tests of 
lepton flavour universality violation. These are the ratio of 
branching fractions for 
$D \rightarrow K \ell \nu$ decay for $\ell=\mu$ to that for $\ell=e$, $R_{\mu/e}$, 
and the lepton 
forward-backward asymmetry for the $\mu$ case (this quantity being very small for the $e$ case). 
We obtain (repeating Eq.~\eqref{eq:rmue})
\begin{equation}
\label{eq:rmue2}
R_{\mu/e} = 0.9779(2)_{\mathrm{latt}}(50)_{\mathrm{EM}} 
\end{equation}
in the Standard Model, including an uncertainty for QED corrections of 0.5\% for 
the $D^0\rightarrow K^-$ case being studied by BES~\cite{Ablikim:2018evp}. 
We show what the impact of a new physics coupling for muons 
could be in
Figures~\ref{fig:Rmue} and~\ref{fig:AFB}. 

Section~\ref{sec:vcs} gives our new determinations of $V_{cs}$ from combining
experimental measurements with our form factors. 
We give three different methods based on using the differential decay rate, binned in $q^2$, 
using the total branching fraction and using experimental results extrapolated to $q^2=0$. 
The results we obtain (repeating Eqs.~\eqref{eq:vcs:dgamma},~\eqref{eq:vcs:B} and~\eqref{eq:vcs:f0}) 
are
\begin{eqnarray}
\label{eq:vcs:list}
&&|V_{cs}|^{\text{d}\Gamma/\text{d}q^2} = 0.9663(53)_{\text{latt}}(39)_{\text{exp}}(19)_{\eta_{EW}}(40)_{\text{EM}} \nonumber \\
&&|V_{cs}|^{\mathcal{B}} = 0.9686(54)_{\text{latt}}(39)_{\text{exp}}(19)_{\eta_{EW}}(30)_{\text{EM}} \nonumber \\
&&|V_{cs}|^{{f_+(0)}} = 0.9643(57)_{\text{latt}}(44)_{\text{exp}}(19)_{\eta_{EW}}(48)_{\text{EM}}.
\end{eqnarray}
Our preferred result is the top one; adding uncertainties in quadrature this gives 
(repeating Eq.~\eqref{eq:finalvcs})
\begin{equation}
\label{eq:finalvcs2}
V_{cs} = 0.9663(80) .
\end{equation}
This total 0.83\% uncertainty is a significant improvement (by a factor of two) on the 
previous most accurate result~\cite{Koponen:2013tua}. The uncertainty is reduced by a factor 
of four over the value from~\cite{pdg} quoted in Eq.~\eqref{eq:introvcs-sl} 
in Section~\ref{sec:intro}. This is the first time that 
a direct determination of $V_{cs}$ has been accurate enough to see a significant 
difference (over $4\sigma$) from 1. 

As discussed in Section~\ref{sec:intro} the limitation on the determination of 
$V_{cs}$ from semileptonic decays (unlike for leptonic decay processes) was 
the accuracy of the lattice QCD calculation. Improving the accuracy of the form 
factors has then allowed us to leverage a significant improvement in the outcome 
for $V_{cs}$. There is still room for further improvement, as can be seen 
in Eq.~\eqref{eq:vcs:list}. The lattice QCD uncertainty is still larger than 
that from experiment, but not by a large margin, so a reduction in the experimental 
uncertainty would also help.   
A significant source of uncertainty is from long-distance QED corrections to the 
$D \rightarrow K $ semileptonic process. Improved understanding of these is 
needed and new methods in lattice QCD+QED may help here~\cite{Sachrajda:2019uhh}. 
Further improvements would include lattice calculations with $m_u \ne m_d$. 

In Table~\ref{tab:intrate} we give the integrated total rates calculated from 
our form factors, $\Gamma/(|\eta_{EW}V_{cs}|^2(1+\delta_{EM}))$, for $D\rightarrow K$ 
semileptonic decay for the four different meson charge and lepton modes we consider here. 
These can be used with improved experimental determinations of the total branching 
fractions to improve $|V_{cs}|$ in the future, even if an improved determination of the 
differential rates is not available. 

Finally, we update the second row and column unitarity tests using our new value for 
$V_{cs}$ in Eq.~\eqref{eq:finalvcs2} and results for other elements as given 
in Section~\ref{sec:vcsdiscuss} and plotted in Figures~\ref{fig:Vcscd} and~\ref{fig:Vcsus}.
For the second row, using $V_{cd}=0.2173(51)$ from leptonic $D^+$ decays 
and $V_{cb}=0.0410(14)$~\cite{pdg} we have
\begin{eqnarray}
|V_{cd}|^2 + |V_{cs}|^2 + |V_{cb}|^2 &=& \\
&&0.9826(22)_{V_{cd}}(155)_{V_{cs}}(1)_{V_{cb}}. \nonumber
\end{eqnarray}
For the second column, using a weighted average of leptonic and semileptonic values 
of $V_{us}$ of 0.2245(4)~\cite{pdg} and $V_{ts}=0.04189(93)$~\cite{Dowdall:2019bea} gives
\begin{eqnarray}
|V_{us}|^2 + |V_{cs}|^2 + |V_{ts}|^2 &=& \\ 
&&0.9859(2)_{V_{us}}(155)_{V_{cs}}(1)_{V_{ts}} \, . \nonumber 
\end{eqnarray}
Both are in good agreement with the value of 1 for unitarity. Since the total 
uncertainty on the unitarity relation depends mainly on that from $V_{cs}$, 
our new result for $|V_{cs}|$ has enabled a very substantial improvement over earlier results, 
giving a total uncertainty on the unitarity tests of 1.6\%.

\subsection*{\bf{Acknowledgements}} 
We are grateful to the MILC collaboration for the use of
their configurations and code. 
We thank R. Briere, A. Davis, J. Harrison, D. Hatton and M. Wingate for useful discussions. 
This work used the DiRAC Data Analytic system at the University of Cambridge, operated by the University of Cambridge High Performance Computing Service on behalf of the STFC DiRAC HPC Facility (www.dirac.ac.uk). This equipment was funded by BIS National E-infrastructure capital grant (ST/K001590/1), STFC capital grants ST/H008861/1 and ST/H00887X/1, and STFC DiRAC Operations grant ST/K00333X/1. DiRAC is part of the National E-Infrastructure.
We are grateful to the Cambridge HPC support staff for assistance.
Funding for this work came from the Gilmour bequest to the University of Glasgow, 
the Isaac Newton Trust, the Leverhulme Trust ECF scheme, 
the National Science Foundation and the 
Science and Technology Facilities Council.

\begin{appendix}

\section{Correlator Fits: further details and results}
\label{App:corrfits}

The fits to the correlators that we calculate in lattice QCD are described in Section~\ref{sec:corrfits}. Here we give more details of prior choices for the fit parameters, give the table of results 
for ground-state parameters and illustrate some of the tests of the fit results. 

Section~\ref{sec:corrfits} discusses how the priors for ground-state energies and 
two- and three-point amplitudes can 
be estimated from the correlators. Table~\ref{Tab:priorsforfit} gives the prior values that we 
use for excited state amplitudes for non-oscillating and oscillating states. 
It also lists the priors for the three-point parameters $J_{00}^{pq}$ 
(see Eq.~\eqref{Eq:3ptcorrfitform}) for the case where 
$pq$ includes oscillating states. The priors for $J_{ij}^{pq}$ when 
$ij \ne 00$ are 0.0(5) in all cases.  

\begin{table}[t]
  \caption{Priors used in the fit on each set. $d^{H}_{i\neq{}0}$ ($H=D/K$) indicates 
the amplitudes for normal and oscillating $D$ mesons and for 
normal $K$ mesons. $d^{K,o}_i$ is the amplitude for oscillating $K$, 
which we expect to be smaller because the oscillation vanishes at 
zero momentum when the quark masses are the same. 
Parameters denoted $S$ and $V$ refer to the $J_{ij}^{kl}$ parameters for the 
scalar and temporal vector currents respectively. 
Columns 4 and 5 then give the priors for the ground-state to ground-state 
parameter cases where at least one of the states is an oscillating state. 
For the cases where at least one state is an excited state, 
$\mathcal{P}[S_{ij\neq{}00}^{kl}]=\mathcal{P}[V_{ij\neq{}00}^{kl}]=0.0(5)$ in all cases.}
  \begin{center} 
    \begin{tabular}{c c c c c c}
      \hline
      set & $\mathcal{P}[d^{D}_{i\neq{}0}]$ & $\mathcal{P}[d^{K,o}_i]$  &$\mathcal{P}[S_{00}^{kl\neq{}nn}]$&$\mathcal{P}[V_{00}^{kl\neq{}nn}]$   \\ [0.5ex]
      \hline
      1 & 0.15(20) &0.05(5)&0.0(1.0)&0.0(1.0) \\ [1ex]
      2 & 0.15(10) &0.05(5)&0.0(1.0)&0.0(1.0) \\ [1ex]
      3 & 0.10(10) &0.05(5)&0.0(1.5)&0.0(1.5) \\ [1ex]
      4 & 0.20(20) &0.05(5)&0.0(1.5)&0.0(1.5) \\ [1ex]
      5 & 0.20(20) &0.03(3)&0.0(1.0)&0.0(1.0) \\ [1ex]
      6 & 0.10(10) &0.05(5)&0.0(1.5)&0.0(1.5) \\ [1ex]
      7 & 0.05(5)  &0.02(2)&0.0(1.0)&0.0(2.0) \\ [1ex]
      8 & 0.08(10) &0.01(2)&0.0(1.0)&0.0(1.5) \\ [1ex]

      \hline
    \end{tabular}
  \end{center}
  \label{Tab:priorsforfit}
\end{table}

\begin{figure}
\hspace{-30pt}
\includegraphics[width=0.48\textwidth]{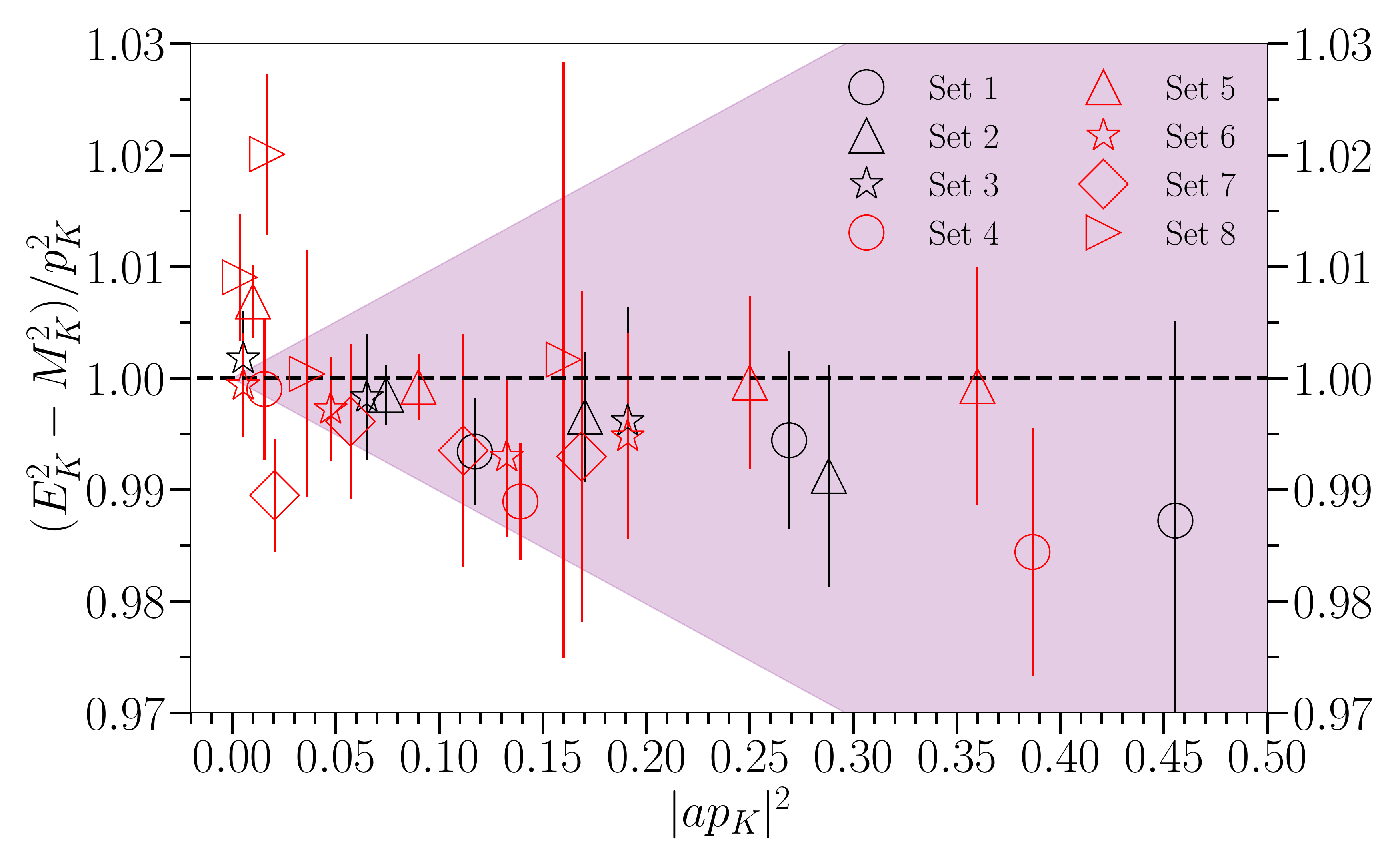}
\caption{For each ensemble, we plot the ratio $(E_{K}^2-M_{K}^2)/|\vec{p}_{K}|^2$ from 
our fit results against $|a\vec{p}_{K}|^2$ to check that the $K$ meson energy 
agrees with that expected from the spatial momentum 
given to the meson in the lattice calculation. 
The points for gluon field configurations with physical sea $u/d$ quark mass are in black. 
The ratio agrees with the expected value of 1 throughout the range of momenta 
and lattice spacing values. The purple wedge shows $1\pm |a\vec{p}_K/\pi|^2$.
}
\label{fig:speedoflight}
\end{figure}

Table~\ref{tab:fitresults} gives the ground-state parameters from our preferred fit 
to the correlators for each gluon field ensemble. 
Columns 8 and 9 of the table give the results for the 
scalar and vector form factors determined from the matrix elements 
as described in Section~\ref{sec:formalism}. These are given as a function of $q^2$ 
in lattice units where $q^2$ is determined from the $D$ and $K$ meson masses and the 
input lattice spatial momentum for the $K$. The results for the form factors on a 
given gluon field ensemble are correlated through our fit. 
We preserve those correlations through to the next stage of the fit where 
we determine the physical curve with uncertainty bands for $f_+(q^2)$ and $f_0(q^2)$, as described in 
Section~\ref{sec:ffphys}.  

\begin{table*}
  \caption{Ground-state parameters determined from our correlator fits for each gluon field ensemble.Columns 3 and 5 give the ground-state $D$ Goldstone 
meson mass and $K$ energy in lattice units, for the $q^2$ value given in lattice units in Column 4. 
Columns 6 and 7 give the matrix elements between $D$ and $K$ of the local scalar current and the local temporal vector current ({\it before} multiplication with $Z_V$ given in column 10 (and determined from Eq.~\eqref{eq:zvcalc2}). Columns 8 and 9 give the scalar and vector form factors (determined from Eqs.~\eqref{eq:D2Kff} and~\eqref{eq:scff}). }
  \begin{center} 
    \begin{tabular}{c c c c c c c c c c}
      \hline
      set &  $am^{\text{val}}_{c}$& $aM_{D}$ &$(aq)^2$&$aE_{K}$ &$\bra{K}S\ket{D}$&$\bra{K}V^0_{\mathrm{latt}}\ket{\hat{D}}$&$f_0(q^2)$&$f_+(q^2)$&$Z_V$\\ [0.5ex]
      \hline
\hline 
1&0.8605&1.44857(46)&1.1443(10)&0.37886(17)&2.524(13)&1.792(16)&1.0236(49)&&1.0440(87)\\ [1ex]
 &&&0.76263(88)&0.51059(13)&2.236(12)&1.605(14)&0.9066(46)&1.133(29)&\\ [1ex]
 &&&0.38113(75)&0.64227(10)&2.033(18)&1.480(21)&0.8243(72)&0.912(14)&\\ [1ex]
 &&&-0.00042(62)&0.773970(85)&1.861(54)&1.425(59)&0.755(22)&0.755(22)&\\ [1ex]
 \hline 
 2&0.643&1.15450(30)&0.72338(50)&0.303983(49)&2.1519(74)&1.4643(83)&1.0240(31)&&1.0199(54)\\ [1ex]
 &&&0.48244(44)&0.408334(36)&1.9015(60)&1.3104(68)&0.9049(26)&1.123(13)&\\ [1ex]
 &&&0.24166(38)&0.512611(29)&1.713(10)&1.193(11)&0.8154(49)&0.9029(90)&\\ [1ex]
 &&&0.00092(32)&0.616870(24)&1.561(21)&1.093(22)&0.7428(98)&0.7430(98)&\\ [1ex]

 \hline 
 3&0.433&0.83391(27)&0.37853(32)&0.218659(54)&1.6558(46)&1.0625(46)&1.0151(24)&&1.0056(43)\\ [1ex]
 &&&0.35885(32)&0.230461(51)&1.6207(44)&1.0435(47)&0.9935(24)&1.32(15)&\\ [1ex]
 &&&0.18311(26)&0.335833(35)&1.3707(71)&0.8967(80)&0.8403(44)&0.977(15)&\\ [1ex]
 &&&-0.07179(18)&0.488663(24)&1.123(24)&0.769(25)&0.689(15)&0.658(14)&\\ [1ex]

 \hline 
4&0.888&1.49339(36)&1.16033(71)&0.41621(17)&2.5532(58)&1.868(10)&1.0147(20)&&1.0370(52)\\ [1ex]
 &&&1.10578(70)&0.43447(16)&2.5090(54)&1.8355(97)&0.9971(18)&1.45(11)&\\ [1ex]
 &&&0.73402(62)&0.55894(13)&2.2457(94)&1.661(12)&0.8925(37)&1.086(18)&\\ [1ex]
 &&&0.16891(50)&0.748140(96)&1.966(30)&1.523(36)&0.781(12)&0.806(14)&\\ [1ex]

 \hline 
 5&0.664&1.19124(20)&0.73636(31)&0.333122(92)&2.1645(32)&1.5025(35)&1.0086(13)&&1.0233(21)\\ [1ex]
 &&&0.70138(30)&0.347804(88)&2.1261(30)&1.4756(33)&0.9906(12)&1.426(48)&\\ [1ex]
 &&&0.46203(27)&0.448268(68)&1.9069(40)&1.3325(41)&0.8885(18)&1.0896(75)&\\ [1ex]
 &&&0.09875(22)&0.600748(51)&1.688(16)&1.187(15)&0.7867(75)&0.8191(86)&\\ [1ex]
 &&&-0.10483(19)&0.686198(45)&1.571(23)&1.109(21)&0.732(11)&0.7029(97)&\\ [1ex]

 \hline 
 6&0.449&0.86434(23)&0.38678(26)&0.24243(10)&1.6898(37)&1.1173(47)&1.0100(19)&&1.0005(39)\\ [1ex]
 &&&0.36829(26)&0.253122(99)&1.6594(36)&1.0978(46)&0.9918(19)&1.39(11)&\\ [1ex]
 &&&0.24226(23)&0.326027(77)&1.4791(59)&0.9796(73)&0.8841(35)&1.097(21)&\\ [1ex]
 &&&0.04974(18)&0.437394(57)&1.282(11)&0.875(12)&0.7661(64)&0.7926(73)&\\ [1ex]
 &&&-0.05805(16)&0.499751(50)&1.222(27)&0.879(34)&0.731(16)&0.710(16)&\\ [1ex]

 \hline 
 7&0.274&0.56711(21)&0.16562(16)&0.160142(78)&1.1898(39)&0.7371(45)&1.0074(30)&&0.9940(56)\\ [1ex]
 &&&0.10378(14)&0.214663(59)&1.0429(45)&0.6534(54)&0.8830(37)&1.066(24)&\\ [1ex]
 &&&0.02098(11)&0.287665(44)&0.920(14)&0.584(18)&0.779(12)&0.806(13)&\\ [1ex]
 &&&-0.072829(79)&0.370376(34)&0.809(21)&0.516(22)&0.685(18)&0.610(18)&\\ [1ex]
 &&&-0.152857(51)&0.440934(29)&0.733(35)&0.497(33)&0.621(29)&0.515(29)&\\ [1ex]

 \hline 
 8&0.194&0.42167(21)&0.09183(12)&0.118624(77)&0.9325(40)&0.5501(44)&1.0109(38)&&0.9929(74)\\ [1ex]
 &&&0.07976(11)&0.132947(69)&0.8900(36)&0.5284(43)&0.9648(35)&1.241(66)&\\ [1ex]
 &&&0.043459(98)&0.175987(52)&0.7804(44)&0.4683(48)&0.8460(46)&0.970(15)&\\ [1ex]
 &&&0.002959(79)&0.224011(41)&0.6795(88)&0.4182(94)&0.7366(96)&0.7425(99)&\\ [1ex]
 &&&-0.1600027(18)&0.417246(22)&0.566(95)&0.338(72)&0.61(10)&0.401(84)&\\ [1ex]
 
 \hline
 \hline
    \end{tabular}
  \end{center}
  \label{tab:fitresults}
\end{table*}

\begin{figure}
  \hspace{-30pt}
  \includegraphics[width=0.48\textwidth]{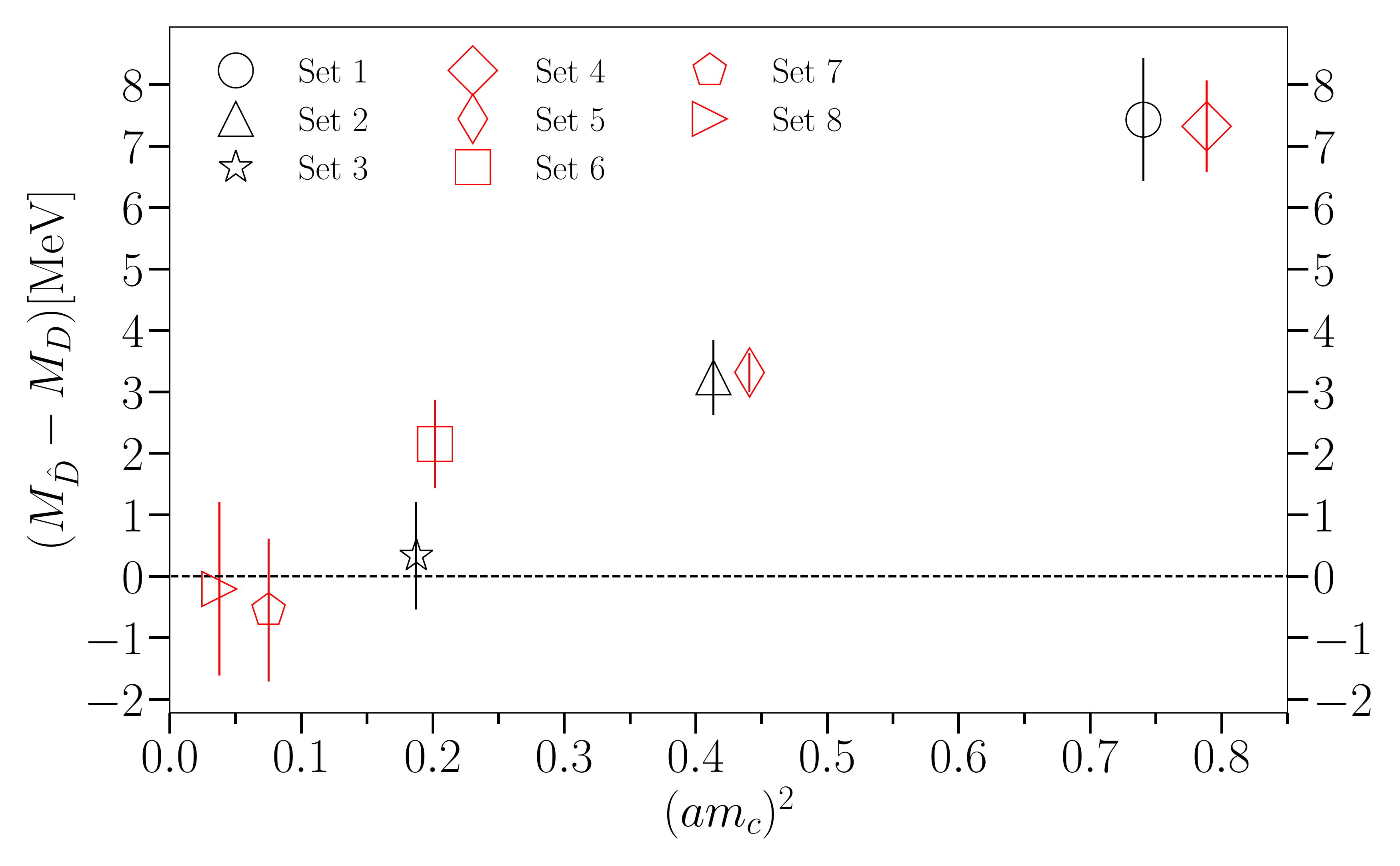}
  \caption{The difference between the non-Goldstone $\hat{D}$ and Goldstone $D$ meson masses, from our fit 
results, as a function 
of lattice spacing. The points in black are for gluon field configurations with 
 physical $u/d$ sea quark mass. The results show clearly that the splitting is a discretisation 
effect and is only a few MeV even on the coarsest lattices. }
  \label{fig:gold-non-split}
\end{figure}

\begin{figure}
  \hspace{-30pt}
  \includegraphics[width=0.48\textwidth]{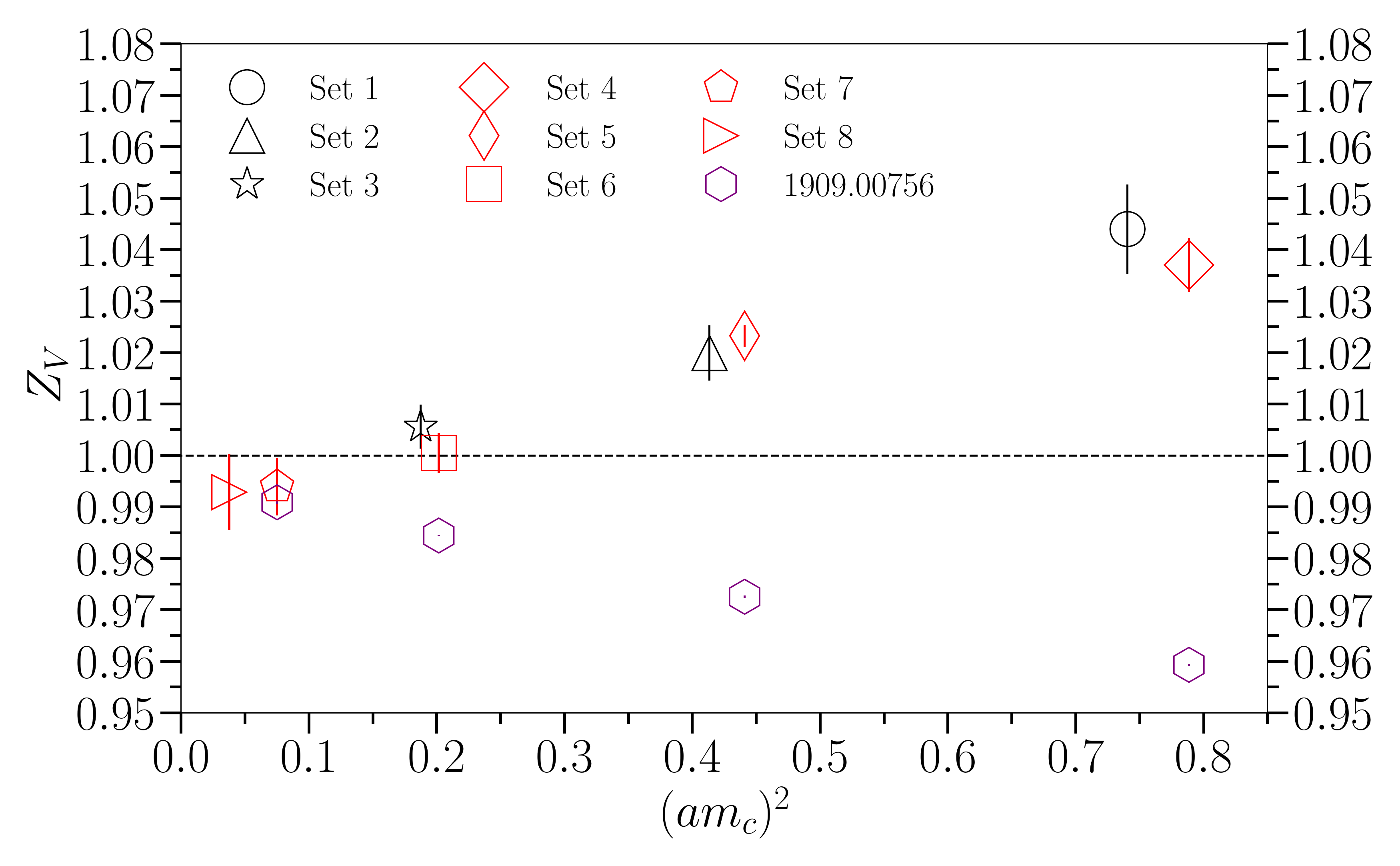}
  \caption{The renormalisation factor for the local temporal vector current, $Z_V$, 
plotted as a function of lattice spacing. The points in black correspond to 
gluon field configurations with physical $u/d$ sea quark mass. The purple hexagons give results 
for $Z_V$ values for the local vector current determined in a symmetric momentum-subtraction 
scheme on the lattice~\cite{Hatton:2019gha}. The two sets of $Z_V$ values differ at finite 
lattice spacing by discretisation effects. }
  \label{fig:ZV}
\end{figure}

A test of our fit results, plotted in Figure~\ref{fig:speedoflight}, is to 
work out the `speed of light' from the energy and mass of the $K$ meson at 
the different values of spatial momenta that we use. Our results show no 
significant disagreement with the result of one expected from relativity 
at the $\approx$1\% level of our 
statistical errors in this quantity. There is no sign of significant discretisation effects. 

Figure~\ref{fig:gold-non-split} shows the mass difference between the Goldstone 
$D$ meson and the nonGoldstone $D$ meson (denoted $\hat{D}$) that we use in the 
temporal vector three-point correlation functions. We see that the difference in mass 
is, as expected, a discretisation effect, vanishing as $a \rightarrow 0$. 
This shows that any effects in our form factors from this mass difference are easily taken care 
of in the discretisation effects that we allow in our extrapolation of the form factors 
to the $a \rightarrow 0$ limit. 

Figure~\ref{fig:ZV} plots our results for the renormalisation factor for the temporal 
vector current, $Z_V$. This is determined from the matrix elements of the scalar and 
temporal vector currents when both the $D$ and $K$ are at rest (zero recoil) from 
Eq.~\eqref{eq:zvcalc2}. Since this renormalisation constant matches the lattice regularisation 
of QCD to that in the continuum for a current with no anomalous dimensions, 
it takes the form of a perturbative 
series in $\alpha_s$, up to discretisation effects.  
Our results for $Z_V$ are very similar, not surprisingly, to those determined for 
the $c\overline{s}$ temporal vector current in $B_c \rightarrow B_s$ decays 
in~\cite{Cooper:2020wnj}. In that paper a comparison was made to the results for 
an $s\overline{s}$ current in~\cite{Chakraborty:2017hry} where $Z_V$ was shown to have 
the expected behaviour. The comparison in~\cite{Cooper:2020wnj} shows that 
the results for $Z_V$ for $c\overline{s}$ and $s\overline{s}$ 
differ only by discretisation effects.  

Another way to determine $Z_V$ is using a symmetric 
momentum-subtraction scheme, known as RI-SMOM, on 
the lattice. In Figure~\ref{fig:ZV} we compare results for $Z_V$ for the local 
vector current determined this way from~\cite{Hatton:2019gha}, taking values at $\mu=2$ GeV.   
These $Z_V$ values differ from the ones used here by discretisation effects. 
Hence, as in the paragraph above, we conclude that using a different prescription for $Z_V$ would give the 
same results in the continuum limit.  

\begin{figure}[t]
\includegraphics[width=0.48\textwidth]{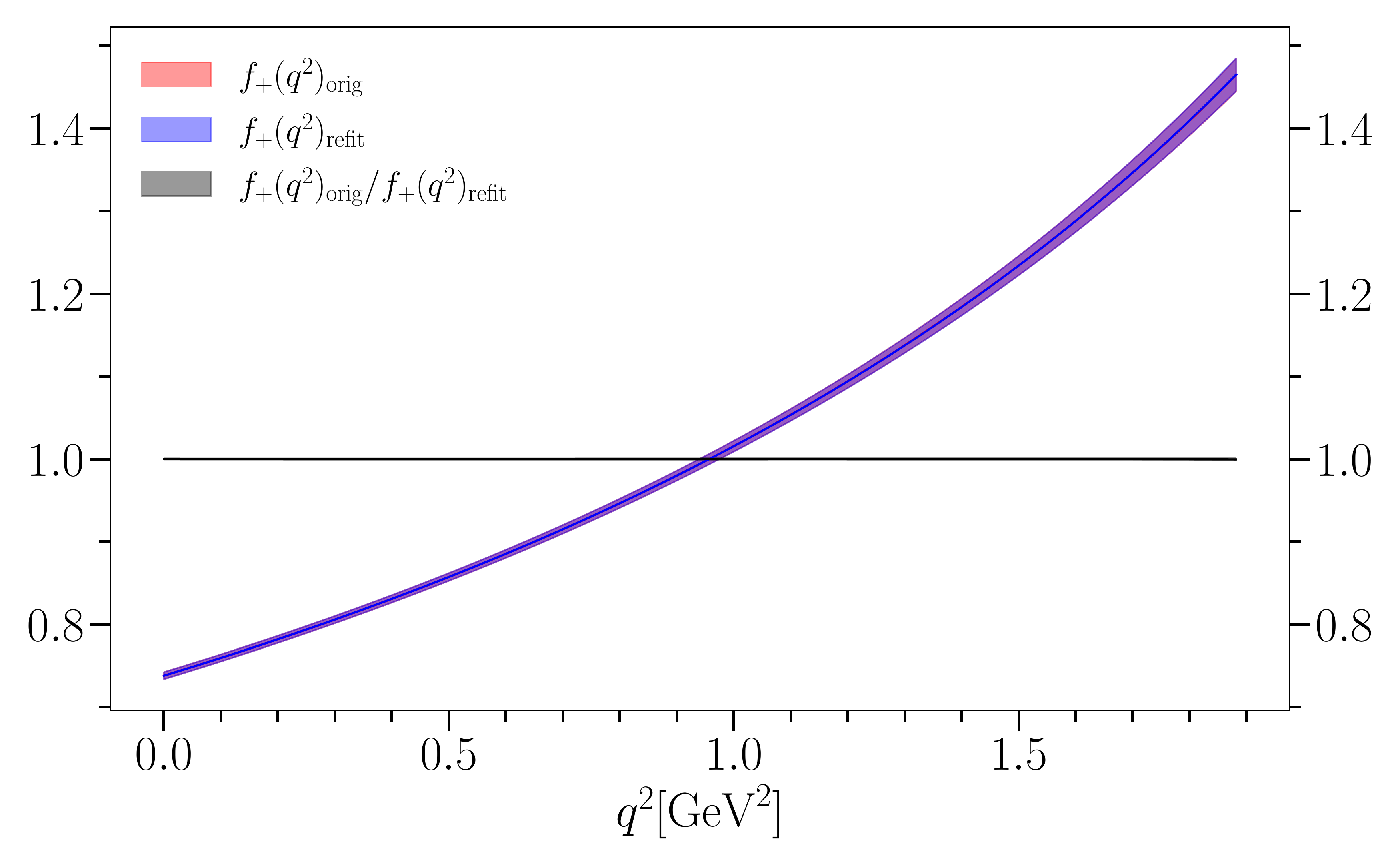}
\caption{The original $f_+$ form factor, as well as the result using the refitting procedure 
described here. The black line (and grey uncertainty band, barely visible) 
is the ratio of the two. 
We see that the refitting reproduces the original form factor and its uncertainty accurately. }
\label{refitfig}
\end{figure}

\section{Obtaining parameters for the $z$-expansion fit form used by experiments}\label{App:expfitform}

We can compare the shape of our vector form factor to that inferred from the experimental 
differential rate by comparing the parameters obtained from $z$-expansion fit. 
To do this we must use the same $q^2$ to $z$ mapping and the same form for the 
$z$-expansion as that used by the experiments. This form is 
\begin{equation}\label{Eq:exp_zexpansion2}
  f_+(q^2)=\frac{1}{z(q^2,t_0=M^2_{D_{s}^{*}})\phi(q^2)}\sum_{n=0}^{N-1}a_n^+z^n,
\end{equation}
where the outer function,
\begin{equation}\label{eq:outerf}
  \begin{split}
  \phi(q^2,t_0) &= \sqrt{\frac{\pi}{3}}m_c\Bigg(\frac{z(q^2,0)}{-q^2}\Bigg)^{5/2}\Bigg(\frac{z(q^2,t_0)}{t_0-q^2}\Bigg)^{-1/2}\\
  &\times\Bigg(\frac{z(q^2,t_-)}{t_--q^2}\Bigg)^{-3/4}\frac{t_+-q^2}{(t_+-t_0)^{1/4}}.
  \end{split}
\end{equation}
The $q^2$ to $z$ mapping (see Eq.~\eqref{eq:zspace}) uses $t_0=t_+(1-(1-t_-/t_+)^{1/2})$ 
(for $t_{+/-}=(M_D\pm M_K)^2$). This is the prescription that minimises the maximum value of $z$ over 
the $q^2$ range of the decay. The parameter $m_c = 1.25\mathrm{GeV}$.

We apply the fit form of Eq.~\eqref{Eq:exp_zexpansion2} to our form factors at the physical point, 
generating synthetic data from Table~\ref{tab:ancoefficients}. 
We used 20 evenly spaced points but changing the number of points makes no difference. 
This gives us the parameters $a_n^+$ for this fit form, along with their correlation matrix 
and these are the values plotted in Figure~\ref{fig:ellipse}. 

Figure~\ref{refitfig} compares our original vector form factor and the refitted one and also plots the 
ratio of the two. This confirms that our refitting process does not change the form factor or its 
uncertainty, but is simply a convenient way to determine the parameters of 
Eq.~\eqref{Eq:exp_zexpansion2} for comparison to experiment.

\end{appendix}

\bibliography{DKpaper}
\end{document}